\documentclass[11pt,a4paper]{article}
\usepackage[makeroom]{cancel}
\usepackage{latexsym}
\usepackage{longtable}
\usepackage{epsfig}
\usepackage{graphicx,bbm,psfrag}
\usepackage[normalem]{ulem}
\usepackage[usenames,dvipsnames]{xcolor}
\usepackage{amssymb,amsmath,amsthm,booktabs,mathtools,soul,cancel}
\usepackage{bm}
\usepackage{dutchcal}
\usepackage{hyperref}
\hypersetup{
    colorlinks,
    citecolor=red,
    linkcolor=blue,
}
\numberwithin{equation}{section}

\newtheorem{rema}{Remark}[section]

\newcommand{\orcid}[1]{\href{https://orcid.org/#1} 
  {\includegraphics[width=10pt]{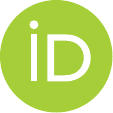}}}

\newcommand{\bc}{\begin{center}}
\newcommand{\ec}{\end{center}}
\def\ba#1{\begin{array}{#1}\displaystyle}
\newcommand{\ea}{\end{array}}

\newcommand{\beq}{\begin{equation}}
\newcommand{\eeq}{\end{equation}}
\newcommand{\beqa}{\begin{eqnarray}}
\newcommand{\eeqa}{\end{eqnarray}}
\newcommand{\no}{\nonumber}
\newcommand{\n}{\nonumber\\}
\newcommand{\bi}{\begin{itemize}}
\newcommand{\ei}{\end{itemize}}

\def\lt#1{\left#1}
\def\rt#1{\right#1}
\def\t#1{\tilde{#1}}
\def\h#1{\hat{#1}}
\def\b#1{\bar{#1}}
\def\frc#1#2{\frac{#1}{#2}}

\def\bs#1{\boldsymbol{#1}}

\newcommand{\p}{\partial}

\newcommand{\ket}{\rangle}

\newcommand{\R}{{\mathbb{R}}}

\newcommand{\ep}{\epsilon}

\newcommand{\dd}{{\rm d}}

\DeclareMathOperator{\sgn}{sgn}

\begin{document}

\begin{center}
{\Large {\sc Hamiltonian formulation and aspects of integrability\\[0.1cm] of generalised hydrodynamics}}

\vspace{1cm}

{\large Thibault Bonnemain\orcid{0000-0003-0969-2413}$^{1,2}$, Vincent Caudrelier\orcid{0000-0003-0129-6758}$^3$, Benjamin Doyon\orcid{0000-0002-5258-5544}$^1$}

\vspace{0.2cm}
$^1$Department of Mathematics, King’s College London, Strand, London WC2R 2LS, UK\\
$^2$Univ. Lille, CNRS, UMR 8523 - PhLAM -
Physique des Lasers Atomes et Mol\'ecules, \\ 
F-59 000 Lille, France\\
$^3$School of Mathematics, University of Leeds, Leeds LS2 9JT, UK\\
\ec

\begin{abstract}
Generalised Hydrodynamics (GHD) describes the large-scale inhomogeneous dynamics of integrable (or close to integrable) systems in one dimension of space, based on a central equation for the fluid density or quasi-particle density: the GHD equation. We consider a new, general form of the GHD equation: we allow for spatially extended interaction kernels, generalising previous constructions. We show that the GHD equation, in our general form and hence also in its conventional form, is Hamiltonian. This holds also including force terms representing inhomogeneous external potentials coupled to conserved densities. To this end, we introduce a new Poisson bracket on functionals of the fluid density, which is seen as our dynamical field variable. The total energy is the Hamiltonian whose flow  under this Poisson bracket generates the GHD equation. The 
fluid density depends on two (real and spectral) variables and the GHD equation can be seen as a $2+1$-dimensional classical field theory.  {In its $1+1$-dimensional reduction corresponding to the case without external forces,} we further show  the system admits an infinite set of conserved quantities that are in involution for our Poisson bracket, hinting at integrability of this field theory.

\end{abstract}

\tableofcontents

\section{Introduction}\label{intro}

Hydrodynamic equations of Euler type are hyperbolic equations that emerge at large scales of space and time in many-body systems \cite{spohn2012large}. They describe the propagation of local relaxation -- the separation between a slow, emergent dynamics and fast projection of local observables onto conserved quantities. In one dimension of space, the situation of interest here, they take the local ``conservation form"
\begin{equation}\label{basicconservation}
    \partial_t \mathtt q_i + \partial_x \mathtt j_i = \mathcal F_i
\end{equation}
where $i$ parametrises the set of admitted local conservation laws, and $\mathcal F_i$ represent the contributions from external force fields (which naturally break the conservation laws). The fluxes $\mathtt j_i$ and force fields contributions $\mathcal F_i$ are functions of the conserved densities $\mathtt q_i$ only (the equations of state), and are determined by thermodynamic considerations, such as entropy maximisation. For instance, the standard Euler equations for Galilean fluid, and Relativistic hydrodynamics, are of this type.

In one dimension of space, many-body systems often display the property of integrability \cite{calogero2012integrability,faddeev2007hamiltonian}. In such cases, there are infinitely many conservation laws, and the universal theory for their emergent Euler-scale hydrodynamics is Generalised Hydrodynamics (GHD) \cite{doyon2020lecture,spohn2023hydrodynamic}. This recovers the Euler-scale hydrodynamic equations for hard rods \cite{spohn2012large,Doyon_2017} and soliton gases \cite{el2021soliton,bonnemain2022generalized,suret2023solitonRev} obtained earlier, but applies more universally, including in classical and quantum systems of interacting particles, spin chains, and quantum field theory; see the reviews in \cite{Bastianello_2022}.

GHD recasts the infinite set of (broken) conservation laws \eqref{basicconservation} into a family parametrised by a continuous ``spectral" parameter $\theta$ instead of the discrete index $i$; we denote the conserved densities in real space, spectral space and time as $\rho(x,\theta,t)$. The spectral parameter enumerates the possible asymptotic objects of the corresponding scattering theory (particles, solitons, etc.), including their momentum and possible internal states. The precise set in which $\theta$ lies depends on the model, but in many simple situations it is a subset of $\R$, or $\R$ itself, and represents the asymptotic momenta. In these cases, the coordinates $x,\theta$ span an effective ``spectral phase space", on which $\rho$ is a density\footnote{It may be defined physically by a time-of-flight thought experiment: $\rho(x,\theta)\dd x\dd\theta$ is the number of particles with asymptotic spectral parameters in $[\theta,\theta+\dd\theta]$, that are asymptotically observed if the fluid element on $[x,x+\dd x]$ is extracted and let to expand in the vacuum.}. The inclusion of external force fields coupled to conserved densities was achieved in \cite{doyon2017note}, where it was shown that GHD takes the form
\begin{equation}\label{ghdintro}
    \p_t \rho + \p_x (v^{\rm eff}\rho)
    + \p_\theta (a^{\rm eff}\rho) = 0.
\end{equation}
Here $v^{\rm eff}$ and $a^{\rm eff}$ are appropriate functionals of $\rho(x,\cdot,t)$, and the last term on the left-hand side is the contribution from force fields. Other types of forces have been studied \cite{PhysRevLett.122.240606,bastianello2019generalized}, which we will not consider here. GHD can therefore be understood, at this level of generality, as hydrodynamics for a {\em two-dimensional fluid}, with a conserved fluid density. It generically has extended interaction range on the spectral component $\theta$ of phase space, which represents scattering between asymptotic objects of different spectral parameters. In systems of non-interacting particles, the spectral phase space is the true phase space of classical mechanics, and the GHD equation is the  {(single-particle)} Liouville equation, or collision-less Boltzmann equation, as arises from basic kinetic theory. GHD generalises this to integrable interactions.

What are the intrinsic properties of the equation \eqref{ghdintro}, as a dynamical system? In particular, a natural question in hydrodynamics is that of the Hamiltonian structure: is there a Poisson bracket for the conserved densities seen as the dynamical variables, such that the total fluid energy is a Hamiltonian generating the fluid equations? The standard Euler equations, in any dimension, do have a Hamiltonian structure, this is well documented \cite{ZakharovHamil1,ZakharovHamil2,OLVER1982233,MorrisonHamil} and has important consequences \cite{MorrisonHamil}, see also the recent study \cite{10.21468/SciPostPhys.14.5.103} in two spatial dimensions. When it comes to GHD, it is known that, without external force fields, so-called ``polychromatic reductions" have Hamiltonian structures, and even are integrable \cite{el2011kinetic,Bulchandani_2017,vergallo2023hamiltonian,vergallo2024hamiltonian}. Regarding the non-reduced, force-less GHD equation, although integrability has not been addressed, solutions by quadrature to the initial value problem are known \cite{bulchandani2017solvable,DOYON2018570}, and with external force fields, stationary solutions have been analysed \cite{doyon2017note,Bulchandani_2024,Hubner_2024}. On top of that, perhaps most notably, a bi-Hamiltonian structure was recently proposed for the special case of force-less hard rods hydrodynamics \cite{vergallo2024hamiltonian}.  {In fact, one may expect that an appropriate ``continuous-spectrum limit," from the polychromatic reduction of GHD, of the construction of \cite{vergallo2024hamiltonian}, would provide a Hamiltonian structure for the GHD equation more generally. But this appears to be a non-trivial task -- in the hard rods case, the fact that the scattering shift is constant seemed to be play an important role. As far as we are aware,} there are currently no results about Hamiltonian structure and integrability for the GHD equation in general,  {in the continuous case for $\rho(x,\theta,t)$, and} especially with external force fields.

In this paper we establish the Hamiltonian structure of the GHD equation \eqref{ghdintro}. The fluid density $\rho$ is the dynamical variable, spanning the dynamical phase space $\mathcal M$ of this two-dimensional field theory. The set of observables on $\mathcal M$ are functionals (not necessarily linear) of $\rho$. The fundamental Poisson bracket can be expressed, using the derivative of the Dirac delta-function $\delta'(x)$, the standard dressing operation ${}^{\rm dr}$, and the occupation function $n(x,\theta)$ (see e.g.~\cite{doyon2020lecture} for a review of the GHD formalism), as
\begin{equation}\label{pbintro}
\begin{aligned}
    & \{\rho(x_1,\theta_1),\rho(x_2,\theta_2)\} \\
    & \qquad= -\frc1{2\pi} \delta'(x_1-x_2)\Big(\big[\delta'(\cdot-\theta_2)\big]^{\rm dr}(x_2,\theta_1)\, n(x_2,\theta_1) +
    \big[\delta'(\cdot-\theta_1)\big]^{\rm dr}(x_1,\theta_2) \,n(x_1,\theta_2)\Big) \ .
\end{aligned}
\end{equation}
In Section \ref{sec:Ham}, Eq.~\eqref{pb}, we provide the general expression in terms of the algebra of observables on $\mathcal M$; we show it satisfies the Leibniz property and the Jacobi identity. The Hamiltonian then takes the simple form of the total fluid energy,
\begin{equation}\label{Hintro}
    H = \int \dd x\dd\theta\,E(x,\theta)\rho(x,\theta):\qquad \p_t \rho = \{\rho,H\}\mbox{ is equivalent to \eqref{ghdintro},}
\end{equation}
where $E(x,\theta)$ is the energy associated to the quasi-particle of spectral parameter $\theta$ when it is at position $x$. 

In fact, it is convenient to establish our results for an even more general form of the GHD equation, that we introduce. This form allows for non-locality in space (more precisely, it is not ``ultra-local" in space, but may still have finite interaction range), thus recovering a structural symmetry between how the spatial  and spectral  variables, $x$ and $\theta$, are treated. The dressing operation acts on the full spectral phase space, the Poisson bracket is
\begin{equation}\label{pbnonlocalintro}
    \begin{aligned}
    &\{\rho(x_1,\theta_1),\rho(x_2,\theta_2)\}\\
    &\qquad = \frc1{2\pi} \Big(\big[\delta'(\cdot-x_1)\delta'(\cdot-\theta_2)\big]^{\rm dr}(x_2,\theta_1)\, n(x_2,\theta_1) -
    \big[\delta'(\cdot-x_2)\delta'(\cdot-\theta_1)\big]^{\rm dr}(x_1,\theta_2) \,n(x_1,\theta_2)\Big) \ ,
    \end{aligned}
\end{equation}
and the Hamiltonian still takes the form \eqref{Hintro}.

 {The Poisson structure \eqref{pbnonlocalintro} and the statement in \eqref{Hintro} are the main results of this work.}  {The fact that the Hamiltonian is the total fluid energy is natural, and also appears in \cite{vergallo2024hamiltonian}\footnote{ {The definition of what we mean by energy is somewhat ambiguous, see Section 4.5 of \cite{bonnemain2022generalized} and 3.2 of \cite{bonnemain2024soliton}, and do not always coincide with that of \cite{vergallo2024hamiltonian}.}} and most works on Hamiltonian structures of fluid equations more generally. However the Poisson brackets \eqref{pbintro} and \eqref{pbnonlocalintro} are, we believe, new; in particular, it is a non-trivial problem, that we solve in this paper, to show that they indeed satisfy the appropriate requirements for a Poisson bracket (Leibniz and Jacobi).} We show that the spatially extended form of the GHD equation arises from a family of classical particle systems that we construct. This is an adaptation of the semiclassical Bethe systems introduced in \cite{ttbar1,ttbar2} to the spatially extended scaling of interactions and the presence of external force fields; it can be seen as an ``atomic reduction" of the GHD equation.

We also  {propose} that in the case without external force fields ($a^{\rm eff}=0$, $E(x,\theta) = E(\theta)$), and in the case without ``kinetic term" ($v^{\rm eff}=0$, $E(x,\theta) = E(x)$), the resulting two-dimensional field theory  {may be} integrable, by exhibiting a large family of conserved quantities in involution.

We do not require any particular form for the scattering shift or more generally the interaction kernel, besides a basic symmetry of the associated scattering phase (which is often satisfied and follows from the Kubo-Martin-Schwinger relation for the associated microscopic model \cite{Doyon_2021}). Thus our results apply to the GHD equation for the KdV soliton gas, the Lieb-Liniger quantum gas, the Toda model of classical particles, the sinh-Gordon relativistic quantum field theory, etc. We show that one of the aforementioned Hamiltonian structures in \cite{vergallo2024hamiltonian} for the hard rod gas is a special case of our construction\footnote{Preprint \cite{vergallo2024hamiltonian} appeared while this paper was in preparation.},  {when expressed in terms of moments of the field $\rho(x,t)$ and specialised to constant scattering shift.}

The paper is organised as follows. In Section \ref{sec:GHD} we introduce  elements of the GHD formalism that will be necessary for our discussion. Section \ref{sec:Ham} contains the main result of our manuscript: it is where we construct the Hamiltonian structure of the GHD equation and show that there exists an infinite family of linearly independent, conserved quantities, that are in involution with respect to the associated Poisson bracket. In Section \ref{sec:HydSys}, we go over some previously obtained results regarding the Hamiltonian structure of particular reductions of the GHD equation, and introduce the theoretical framework in which they were obtained; we also show that our construction is compatible with that of \cite{vergallo2024hamiltonian}. Finally, in Section \ref{sec:disc}, we comment on some aspects of our construction, on some restrictions that may be lifted, and on some interesting perspectives.

\section{GHD and dressing operations}\label{sec:GHD}

\subsection{Fluid density, dynamical and spectral phase spaces}\label{sec:FluDen}

The GHD equation describes the large-scale behaviour of integrable many-body systems, in terms of the propagation of local (generalised) ``equilibria" -- more precisely, local projection onto conserved quantities. At any given time, every expectation value of local observables is completely characterised by the fluid density $\rho(x,\theta)$, which is therefore the dynamical variable of interest. The fluid density can be interpreted as the density of quasi-particles in {\em spectral phase space} $\Lambda = \mathcal L \times \mathcal P$, such that $x\in \mathcal L$ and $\theta \in \mathcal P$. In this paper we concentrate on the simplest case $\mathcal L = \mathcal P = \R$, and comment on other cases in Section \ref{sec:disc}. We assume that the fluid density quickly vanishes in unbounded directions:
    \begin{equation}\label{domain}
        \Lambda = \mathbb R^2 \ , \quad \lim_{|x|+|\theta|\to \infty} \rho(x,\theta)  = 0 \ .
    \end{equation}

The {\em dynamical phase space} is the space on which the Poisson structure will be constructed. This is the space in which the dynamical variable $\rho$ lies. There are two possible setups.

In the {\em abstract setup}, $\rho$ is a {\em formal dynamical variable} out of which observables are constructed, see Subsection \ref{sec:alg}.  {In this case, v}anishing at infinity \eqref{domain} is  {to be understood} in an abstract sense.  {By this, we mean that we consider densities $\rho$ such that (i) integrals of $\rho$ over the spectral phase-space $\Lambda$ are bounded and (ii) that boundary terms, notably when performing integration by parts, cancel out. This includes, but is not limited to, the case of quickly vanishing densities as  defined in \eqref{domain}}.

In the {\em concrete setup}, $\rho$ is in a specific function space. For simplicity we assume smoothness, and thus the dynamical phase space is some space of smooth functions on $\R^2$ that asymptotically vanish,
\begin{equation}
    \mathcal M \subset \mathcal C^\infty_0(\R^2)\ .
\end{equation}
In many situations of physical interest, $\mathcal M$ is restricted to non-negative functions $\rho(x,y)\geq 0, \ \forall\,(x,y)\in \R^2$, however this is not necessary; for instance, in the KdV soliton gas, see below, it is convenient to allow for negative fluid densities. We explain in Appendix \ref{appAnalysis}, with further comments in Appendix \ref{app:alg}, how taking
\begin{equation}\label{smoothcompact}
    \mathcal M = \{|\rho|\leq \rho_*: \rho\in \mathcal C^\infty_c(\R^2)\}\ ,
\end{equation}
the set of smooth compactly supported functions that are bounded by a given function $\rho_*(x,\theta)$, or any subset thereof, gives an explicit setup for our general construction (this requires some technical conditions on our GHD data, Eqs.~\eqref{technicalOmega},  \eqref{technicalpsi} and \eqref{technicalderivative}). The use of $\mathcal C^\infty_c(\R^2)$ is convenient as it allows for a large set of GHD data and simple analysis techniques, however by adding constraints on the GHD data or using different techniques, different families of functions may be allowed. This includes Schwartz functions or other space of rapidly decaying functions, and even bounded functions not required to be smooth, with relatively weak differentiability requirement, see \cite{HubnerDoyonUniqueness}. The arguments we present in the main text and in Appendices \ref{appAnalysis} and \ref{app:alg} can be adapted to such situations.

We do not discuss under what conditions $\mathcal M$ may be invariant under the dynamics, Eq.~\eqref{ghdsum} below, although in the abstract setup this is immediate (see \cite{HubnerDoyonUniqueness} for rigorous results in a concrete setup.

\subsection{Conventional data of GHD}

In its conventional, original formulation, the GHD equation is determined by a few data: the scattering shift $\varphi(\theta,\theta')$,  the momentum function $P(\theta)$, and the energy function $E(\theta)$. This data encodes the system in terms of the scattering properties of the set of asymptotic ``objects", parametrised by $\theta$: the energy function determines the dynamics, the scattering shift fixes the two-body displacements, and the momentum sets the relation to real space. The asymptotic objects may be particles, solitons, quantum excitations, etc.; the GHD equation takes the same form in all cases. As shown in \cite{Doyon_2021}, the Kubo-Martin-Schwinger relation imposes that there exist $\phi(\theta,\theta')$ such that $\phi(\theta,\theta') = -\phi(\theta',\theta)$ and $\varphi(\theta,\theta') = \p_\theta\phi(\theta,\theta')$. But otherwise, there are no strong constraints on the data. GHD is therefore a universal framework for the hydrodynanic limit of many-body integrable systems.

As was explained in \cite{doyon2018exact}, the choice of a parametrisation $\theta$ for the asymptotic objects is largely arbitrary. In many (but not all) cases one may choose $\theta$ to correspond to the physical momentum, $P(\theta) = \theta$, or one may choose it to guarantee that the scattering shift is a function of differences of spectral parameters, $\varphi(\theta,\theta') = \varphi(\theta-\theta')$ (only in Galilean systems may it be chosen such that both hold). Moreover, as was explained in \cite{bonnemain2022generalized} and further discussed in \cite{bonnemain2024soliton}, in classical systems, even given a parametrisation $\theta$, the choice of a momentum function may be  arbitrary, being relevant in quantum models only due to the discretisation of phase space. That is, there is a general theory for the transformation of the main objects in the GHD equation under change of parametrisation $\theta\to \t\theta$, and, in classical cases, under change of momentum function $P(\theta)\to \t P(\theta)$, which keeps the GHD equation invariant. In this paper, the former transformation theory will play an important role (we will recall it), but as we consider the general form of the GHD equation, we will not discuss the latter; and we keep $P(\theta)$ in its general form.

The emergent GHD equation receives modifications when external fields, which may be varying in space and time, influence the dynamics of the many-body systems; or when the dynamics itself, say the coupling parameters of the model, is inhomogeneous in space and/or non-autonomous. These modifications can be seen as ``acceleration" terms. This was first obtained for space-varying external fields coupled to conserved densities (such as minimal coupling of electric or magnetic fields to charges), in \cite{doyon2017note}; this case corresponds to taking the energy function to be space-dependent, $E(\theta) \to E(x,\theta)$. It was then generalised to space-time varying external fields and coupling parameters in \cite{PhysRevLett.122.240606,bastianello2019generalized}, in fact admitting all data to be space-time dependent, $E(\theta) \to E(x,\theta,t)$, $P(\theta) \to P(x,\theta,t)$ and $\phi(\theta-\theta')\to \phi(x,\theta-\theta',t)$.

\subsection{Spatially extended GHD}

In this paper, we consider autonomous systems, hence all data is time independent, but we admit space-dependent energy and momentum functions.

Instead of the momentum function, the quantity that is most relevant in GHD is the spectral derivative of the momentum, $P_\theta(x,\theta) = \p_\theta P(x,\theta)$. This, divided by $2\pi$ for conventional normalisation, is physically interpreted as a bare {\em spectral phase-space state density}: $\dd x\dd \theta \,|P_\theta|/(2\pi)$ is the number of physical states that are available within the infinitesimal spectral phase space element $\dd x\dd\theta$, without accounting for interactions (the spectral phase-space state density that accounts for interaction is $\rho_{\rm s}$, defined in \eqref{defn} below). Geometrically, it is a volume density -- a discussion of the associated metric structure is left for future works. For our discussion, an object that appears to be more fundamental is the {\em phase function} $\Omega(x,\theta)$, defined such that $P(x,\theta) = \p_x \Omega(x,\theta)$.

Thus, parts of our data are phase-space dependent energy and phase functions
\begin{equation}
    E(x,\theta),\quad \Omega(x,\theta)\qquad \mbox{with}\qquad P(x,\theta) =  \Omega_x(x,\theta)
\end{equation}
where here and below indices $x,\theta,\ldots$ mean partial derivatives with respect to $x,\theta,\ldots$. For definiteness we assume\footnote{Smoothness can easily be relaxed to finite-order differentiability.} $E,\ \Omega\in \mathcal C^\infty(\R^2)$.

More importantly, we admit an extended form of the GHD equation not considered until now, which we refer to as its {\em spatially extended form}. In this form, the datum of the scattering shift $\varphi(\theta,\theta')$ is replaced by an {\em interaction kernel} $\psi(x,\theta;x',\theta')$ on $\R^2\times\R^2$, with the main requirement that it be  symmetric
\beq\label{psisymmetry}
	\psi(x,\theta;x',\theta') = \psi(x',\theta';x,\theta).
\eeq
$\psi$ is a smooth-function-valued distribution in $x',\theta'$ --  that is, it may be a generalised function and should be seen as the kernel of an integral operator. For definiteness, we require that $\psi$ be integrable in $x',\theta'$ against smooth functions supported on compact subsets of $\R^2$ and that the result be smooth functions of $x,\theta$: \begin{equation}\label{psimap}
    \int_{\R^2} \dd x'\dd\theta'\,\psi(\cdot,\cdot;x',\theta')g(x',\theta')
    \ \in \mathcal C^\infty(\R^2)
\end{equation}
for every $g\in\mathcal C_c^\infty(\R^2)$. This includes $\psi$ being any smooth function on $\R^2\times\R^2$, as well as all examples considered below. This requirement on $\psi$ is adapted to the specific context of Eq.~\eqref{smoothcompact}; for different choices of the space $\mathcal M$, one may require similar properties for a larger family of $g$.

The use of the interaction kernel $\psi$, instead of the scattering shift $\varphi$, will allow us to treat the spatial and spectral coordinates on equal footing --  restoring a structural symmetry in spectral phase space. The spatially extended GHD equation, Eq.~\eqref{ghdsum} below, is more naturally a hydrodynamic equation on two ``spatial" dimensions, with coordinates $x,\theta$. It (generically) has extended interaction range not only in spectral space $\theta$ (as is the case for the conventional GHD equation), but also in real space $x$.  This spatially extended form is convenient to consider, as it admits a Hamiltonian formulation in its full generality, which is the main object of this paper.

The spatially extended GHD equation is an equation for the dynamics of the fluid density $\rho(x,\theta,t)$. It takes the conservation form
\beq\label{ghdsum}
	\p_t \rho + \p_x (v^{\rm eff}\rho)
	+ \p_\theta(a^{\rm eff}\rho) = 0
\eeq
for effective velocity and acceleration $v^{\rm eff}(x,\theta,t)$, $a^{\rm eff}(x,\theta,t)$ which are nonlinear functionals of $\rho(\cdot,\cdot,t)$. These functionals are defined as solutions to the following linear integral equations (here taken on any time slice, keeping the time variable implicit): 
\beqa\label{veff}
    E_\theta &=& v^{\rm eff}(x,\theta) \Omega_{x\theta} + \int_{\R^2} \dd x'\dd\theta'\,\rho(x',\theta')\big[
    \psi_{x\theta}\, v^{\rm eff}(x,\theta) + \psi_{x'\theta}\, v^{\rm eff}(x',\theta')\big]\\
    -E_x &=& a^{\rm eff}(x,\theta) \Omega_{x\theta} + \int_{\R^2} \dd x'\dd\theta'\,\rho(x',\theta')\big[
    \psi_{x\theta}\, a^{\rm eff}(x,\theta) + \psi_{x\theta'}\, a^{\rm eff}(x',\theta')\big]
    \label{aeff}
\eeqa
where $E=E(x,\theta)$, $\Omega_{x\theta} = \Omega_{x\theta}(x,\theta) = P_\theta(x,\theta)$, $\psi = \psi(x,\theta;x',\theta')$ as above. Eqs.~\eqref{veff} and \eqref{aeff} are inhomogeneous Fredholm equations of the second kind; we will see below that solutions are unique in the setting of Eq.~\eqref{smoothcompact} (c.f. Appendix \ref{appAnalysis}), adapting techniques from \cite{HubnerDoyonUniqueness}.

Effectively, we are considering space-dependent external fields as in \cite{doyon2017note}, space-dependent momenta, and certain types of space-dependent, inhomogeneous and generically spatially extended scattering shifts. It is not clear if it is possible to specialise our theory to the type of local, space-dependent scattering shifts considered in \cite{bastianello2019generalized}; the form of GHD equation we obtain appears to be new.

Our construction naturally reduces to conventional GHD by imposing that the interaction kernel $\psi$ produces spatially homogeneous scattering shifts:
\beq\label{localGHD}
    \psi(x,\theta;x',\theta') = \frc12\sgn(x-x') \,\phi(\theta,\theta')
    \qquad \mbox{(conventional GHD with homogeneous scattering shift)}.
\eeq
Due to the symmetry condition \eqref{psisymmetry}, $\phi$ must be anti-symmetric, and  we recover the scattering shift of conventional GHD as $\varphi(\theta,\theta') = \p_\theta \phi(\theta,\theta')$. In this case, the spatially extended GHD equation becomes the conventional GHD equation with (in general) external inhomogeneous fields linearly coupled to the conserved densities as represented by the energy function $E(x,\theta)$; see \cite{doyon2017note}.  In our notation, it takes the form \eqref{ghdsum} with
\beqa\label{vefflocal}
    E_\theta &=& v^{\rm eff}(x,\theta) \Omega_{x\theta} + \int_{\R} \dd\theta'\,\rho(x,\theta')
    \varphi(\theta,\theta')\, \big[v^{\rm eff}(x,\theta) - v^{\rm eff}(x',\theta')\big]\\
    \label{aefflocal}
    -E_x &=& a^{\rm eff}(x,\theta) \Omega_{x\theta} + \int_{\R} \dd\theta'\,\rho(x,\theta')\big[
    \varphi(\theta,\theta')\,a^{\rm eff}(x,\theta) - \varphi(\theta',\theta) \, a^{\rm eff}(x,\theta')\big]\\
    && \qquad\qquad\qquad \mbox{(conventional GHD with homogeneous scattering shift).}\no
\eeqa
At the time of writing, this is probably the most relevant configuration. 

\begin{rema}\label{remashortrange}
We discussed above the ``interaction range" of the GHD equation, both in real and spectral space. An extended range means that the instantaneous change of $\rho(x,\theta,t)$ depends on $\rho(x',\theta',t)$ for $(x',\theta')\neq (x,\theta)$. As in many-body systems, one may also talk about ``finite-range",  ``short-range" or ``long-range" interactions. As is clear from \eqref{veff} and \eqref{aeff}, the interaction distance may be taken as the minimal value of $|x-x'|$ or $|\theta-\theta'|$ above which $\psi_{\t x\t\theta}$ vanishes for every choice of $\t x = x,\ x'$ and $\t\theta = \theta,\ \theta'$; if it is finite, we have a finite-range interaction. We may say that we have a short-range interaction if  $\psi_{\t x\t\theta}$ decays exponentially with $|x-x'|+|\theta-\theta'|$ (taking, say, the $L^1$ distance on $\R^2$), and long-range if the decay is algebraic. One may expect ``local physics'' to hold in two-dimensional spectral phase space in all cases, for the latter if the power of the algebraic decay is large enough; however a full analysis is beyond the scope of this paper.
\end{rema}

\subsection{Examples}

Important known examples of our construction occur in the case of conventional GHD \eqref{localGHD}, where typically one takes $P(x,\theta) = P(\theta)$. For instance, setting
\begin{equation}\label{LLdata}
    E(x,\theta) = \frc{\theta^2}2,\quad \Omega(x,\theta) = x\theta,\quad
    \phi(\theta,\theta') = 2\,{\rm Arctan}\frc{\theta-\theta'}c \quad \mbox{(Lieb-Liniger)}
\end{equation}
reproduces the GHD equations for the Lieb-Liniger gas with coupling $c$ \cite{castro2016emergent}; setting
\begin{equation}\label{hrdata}
    E(x,\theta) = \frc{\theta^2}2,\quad \Omega(x,\theta) = x\theta,\quad
    \phi(\theta,\theta') = a\, (\theta-\theta') \quad \mbox{(hard rods)}
\end{equation}
gives the hard rod gas for rods of length $a$; and restricting $\mathcal M$ to odd functions under $\theta\to-\theta$ on $\R^2$, and setting
\begin{equation}\label{KdV}
     E(x,\theta) = 8\theta^4, \quad \Omega(x,\theta) = 4 x \theta^2,\quad
     \phi(\theta,\theta') = 8 \Big((\theta - \theta')(\log|\theta - \theta'|-1)\Big)
     \quad \mbox{(KdV)}
\end{equation}
gives the KdV soliton gas. In the latter case, we have taken the ``convenient" choice of momentum function (see \cite[Sec 4.5]{bonnemain2022generalized}), and extended the physical domain of spectral parameters from $\R^+$ to $\R$, performing an odd continuation of $\rho$ as in \cite{congy2023dispersive}.
As $E_\theta$ is odd in $\theta$, Eq.~\eqref{veff} reproduces the usual equation of state of the KdV soliton gas. This is the way in which the KdV soliton gas fits within our general framework.

Setting $\psi(x,\theta;x',\theta') = \psi(x-x',\theta-\theta')$ along with $E(x,\theta) = E(\theta)$, $\Omega(x,\theta) = x\theta$, the spatially extended GHD equation emerges as an appropriate hydrodynamic limit of the family of classical interacting, integrable Hamiltonian particle models referred to as Bethe semiclassical systems, introduced in \cite{ttbar1,ttbar2}. In fact, we show in Appendix \ref{appparticles} that, for any $P(x,\theta)$, $E(x,\theta)$ and $\psi(x,\theta;x',\theta')$, the spatially extended GHD equation arises as an appropriate hydrodynamic limit of a microscopic Hamiltonian model that generalises the semiclassical Bethe systems. The phase function $\Omega(x,\theta)$ and the interaction kernel $\psi(x,\theta;x',\theta')$ appear naturally in the canonical transformation that defines the models.

\subsection{Spectral phase-space reparametrisation}\label{sec:reparam}

The theory of phase-space reparametrisation (generalising that introduced in \cite{doyon2018exact}) will be crucial in order to understand the dressing operations that we now discuss. For our purposes, it is sufficient to consider factorised changes of coordinates $(x,\theta)\to (\t x,\t \theta) = (f(x),g(\theta))$, where $f$ and $g$ are smooth functions with positive derivative. We note that Eqs.~\eqref{ghdsum}, \eqref{veff} and \eqref{aeff} are invariant under such transformations if the objects involved transform in the appropriate way. In order to describe the transformation properties, we say that a spectral phase space function $U(x,\theta)$ is of type $(j,k)$ if, under a change of coordinates, it transforms as
\begin{equation}\label{reparam}
    U(x,\theta) \to \t U(\t x,\t\theta) = (f_x)^{-j} (g_\theta)^{-k} U(x,\theta):\qquad \mbox{$U$ is of type $(j,k)$.}
\end{equation}
We will only consider $j,k\in\{-1,0,1\}$. Note that if $U(x,\theta)$ is of type $(0,0)$ (scalar), then $\p_x U(x,\theta)$ is of type $(1,0)$ (spatial vector field), etc. Invariance is obtained by claiming\footnote{Those transformation types are to be understood in terms of the invariance, covariance and contravariance of multilinear algebra. To avoid confusion, in definition \eqref{reparam}, it is important to note that the function $U$ is changed to the new function $\t U$ under reparametrisation; for instance, even if $\Omega = x\theta$ in Lieb-Liniger (c.f. the data \eqref{LLdata}), $\Omega$ is still of type $(0,0)$.}:
\begin{equation}\label{datatype}
    E, \Omega, \psi\quad \mbox{are of type $(0,0)$},\qquad
    \rho \quad \mbox{is of type $(1,1)$}
\end{equation}
where the statement for $\psi(x,\theta;x',\theta')$ holds independently for $x,\theta$ and $x',\theta'$. As a consequence,
\begin{equation}
    v^{\rm eff}\quad \mbox{is of type $(-1,0)$},\qquad a^{\rm eff}\quad \mbox{is of type $(0,-1)$}
\end{equation}
and this guarantees invariance of \eqref{ghdsum}. Note how the momentum function $P$ is of type $(1,0)$, and therefore $P_\theta$ is of type $(1,1)$ as it should for a density in spectral phase space (like $\rho$). The energy and momentum functions are not on the same footing, rather it is the energy and phase functions that are.

A full theory of spectral phase-space re-parametrisation would account for more general transformations $(\t x,\t\theta) = (f(x,\theta),g(x,\theta))$, including those that do not preserve orientation. We will develop this in a separate work.

\subsection{Dressing and Riemann invariants}

It is possible to rewrite the spatially extended GHD equation in terms of a ``continuum of Riemann invariants'' -- the occupation function $n(x,\theta,t)$ -- using dressing operations, as in conventional GHD. For this purpose, we need to define dressing operations. As was first realised in \cite{doyon2018exact}, the dressing operation acts differently on objects that transform differently. We generalise these concepts to spatially extended GHD as follows.
We define the occupation function as
\begin{equation}\label{defn}
    n(x,\theta) = \frc{\rho(x,\theta)}{
    \rho_{\rm s}(x,\theta)
    },\quad
    2\pi \rho_{\rm s}(x,\theta) = 
    \Omega_{x\theta} + \int_{\R^2} \dd x'\dd\theta'\,\rho(x',\theta')\psi_{x\theta}(x,\theta;x',\theta').
\end{equation}
It is clear that $n$ is a scalar:
\begin{equation}
    n\quad \mbox{is of type $(0,0)$.}
\end{equation}
A function $f\in \mathcal C^\infty(\R^2)$ is dressed, $f\mapsto f^{\rm dr}$, by solving the following linear integral equation (an inhomogeneous Fredholm equation of the second kind), which depends on the transformation type of the function $f$:
\begin{align}\label{dressing}
    f^{\rm dr}(x,\theta) = f(x,\theta) + \int_{\R^2} \frc{\dd x'\dd\theta'}{2\pi}\,
    \psi_{\t x\t\theta}(x,\theta;x',\theta')\, n(x',\theta')\,f^{\rm dr}(x',\theta')
\end{align}
where $\t x = x$ or $-x'$ and $\t \theta = \theta$ or $-\theta'$ as per the transformation type of $f$:
\begin{equation}
    \psi_{\t x\t\theta} =\lt\{
    \begin{aligned}
    & \psi_{x'\theta'}&& \mbox{($f$ is of type $(0,0)$)}\\
    & - \psi_{x'\theta}&& \mbox{($f$ is of type $(0,1)$)}\\
    & - \psi_{x\theta'}&& \mbox{($f$ is of type $(1,0)$)}\\
    & \psi_{x\theta}&& \mbox{($f$ is of type $(1,1)$).}
    \end{aligned}\rt.
\end{equation}
This dressing operation does not apply on functions with other transformation types. See Appendix \ref{appAnalysis} where we show that in the context of Eq.~\eqref{smoothcompact} the occupation function and the dressing operation are well-defined.
With this definition, we see that the dressing preserves the transformation type:
\begin{equation}
    \mbox{$f$ is of type $(j,k)$} \quad \Rightarrow\quad  \mbox{$f^{\rm dr}$ is of type $(j,k)$}.
\end{equation}
By our general requirements Eq.~\eqref{psimap} on the interaction kernel $\psi$, it also maps
\begin{equation}\label{dresmooth}
    {}^{\rm dr}:\quad \mathcal C^\infty(\R^2)\to\mathcal C^\infty(\R^2)
\end{equation}
(at least in the context of Eq.~\eqref{smoothcompact} -- see Appendix \ref{appAnalysis}). Further, if $\psi(x,\theta;x',\theta') = \psi(x-x',\theta-\theta')$, then the dressing operation does not depend on the transformation type. Combining the definition of dressing with \eqref{defn}, we find 
\begin{equation}\label{rhosOmega}
    \rho_{\rm s}(x,\theta) = \frc{\Omega_{x\theta}^{~~\rm dr}}{2\pi},
\end{equation}
that is $\rho_{\rm s}$ is the interaction-dependent spectral phase-space state density. Defining
\begin{equation}
    \Psi = \frc{\psi}{2\pi}
\end{equation}
the dressing can be written in integral-operator form as
\begin{equation}\label{dressintop}
    f^{\rm dr} = (1-\p_{\t x}\p_{\t\theta}\Psi\, n)^{-1} f.
\end{equation}
It also satisfies the symmetry relation \cite{doyon2020lecture}
\begin{equation}\label{symm}
    \int_{\R^2} \dd x\dd\theta\,n f g^{\rm dr} = \int_{\R^2} \dd x\dd\theta\,n f^{\rm dr}g
\end{equation}
for any $f,g$ such that the sum of their types is $(1,1)$ (the resulting integral is a scalar). Note how, in this symmetry relation, the dressing may act differently on $f$ and $g$, as they have different transformation types.

We observe that \eqref{veff} simply expresses the fact that $2\pi \rho_{\rm s}v^{\rm eff}$ is the dressing of $E_\theta$ (both of type $(0,1)$ as it should), and \eqref{aeff} expresses the fact that $2\pi \rho_{\rm s}a^{\rm eff}$ is the dressing of $-E_x$ (both of type $(1,0)$):
\begin{equation}\label{CRA}
    v^{\rm eff} = \frc{E_\theta^{~{\rm dr}}}{\Omega_{x\theta}^{~~{\rm dr}}},\quad
    a^{\rm eff} = -\frc{E_x^{~{\rm dr}}}{\Omega_{x\theta}^{~~{\rm dr}}}.
\end{equation}
In the context of Appendix \ref{appAnalysis} we show that, under certain technical conditions, integral equations \eqref{veff} and \eqref{aeff} indeed have unique solutions for $\rho\in\mathcal M$.

Much like in conventional GHD, we can show that $n(x,\theta)$ diagonalises the spatially extended GHD equation \eqref{ghdsum}
\begin{equation}\label{ghdn}
    \p_t n + v^{\rm eff}\p_x n + a^{\rm eff} \p_\theta n = 0 \ .
\end{equation}
 The derivation of this important fact, which generalises the usual one to the spatially extended case and to the transformation-dependent dressing operation as defined above, is provided in Appendix \ref{appnequation}. Hence, we can claim that $n(x,\theta)$ can be interpreted as a continuum of Riemann invariants (or normal modes).
With these definitions, we are now ready to establish the Hamiltonian structure of spatially extended GHD.

\begin{rema}\label{rem:dressdef}
    In many situations, such as in the GHD of quantum models, $\rho$ is non-negative, and $\rho_{\rm s}$ and $\Omega_{x\theta}$ are strictly positive. Then $n$, from Eq.~\eqref{defn}, is non-negative and upper-bounded. But in classical systems, as mentioned in the example of the KdV soliton gas descripition, $\rho$ and $\Omega_{x\theta}$ may become negative, and further $\rho_s$ might vanish and $n$ diverge. Physical quantities and the present construction still make sense; see for example discussions regarding the so-called ``condensate'' limit of the  KdV \cite{congy2023dispersive} and NLS \cite{el2020spectral} soliton gas  (with dictionary \cite{bonnemain2022generalized}). 
    Note also that condition \eqref{smoothcompact}, along with the assumptions in Appendix \ref{appAnalysis}, are sufficient for the dressing to be well-defined, but not necessary;  see for instance the analyses in \cite{kuijlaars2021minimal,HubnerDoyonUniqueness}. Regarding classical systems in particular, it is easy to see that, for any smooth function $f$, if the interaction kernel is integrable against smooth functions, $f^{{\rm dr}}$ vanishes as $n\to \infty$. Then $\tilde f := f^{{\rm dr}}/n$ remains smooth even as $n \to \infty$. For instance, when it comes to the KdV soliton gas in the condensate limit mentioned earlier, $\tilde f$ simply is the inverse Hilbert transform of $f_\theta/\pi$ \cite{congy2023dispersive}. In particular, the Poisson bracket Eq.~\eqref{pb} below remains well-defined even as $n \to \infty$, since it involves the product of $n$ with a dressed function.
\end{rema}

\begin{rema}\label{rem:dressdist}
    As hinted in the Introduction, in particular in Eqs.~\eqref{pbintro} and \eqref{pbnonlocalintro}, to define a Poisson bracket it will be convenient to introduce the dressing of distributions, which would appear to go against definition \eqref{dresmooth}. However, dressed distributions are to be interpreted in terms of their effect when integrated against a compactly supported smooth test function $g$, we will further comment on this aspect in Appendix \ref{appAnalysis}. 
\end{rema}

\section{Hamiltonian formulation and aspects of integrability}\label{sec:Ham}

In this section we establish a Hamiltonian formulation for the GHD equation; this is done for its most general (spatially extended) form, but immediately holds, by specialisation \eqref{localGHD}, to conventional GHD as well. We take a rather direct approach, inspired by Faddeev and Takhtajan's \cite{faddeev2007hamiltonian}, and define a Poisson structure on the algebra of observables on the dynamical space $\mathcal M$.

\subsection{Algebra of observables}\label{sec:alg}

For definiteness, we define the algebra of observables $\mathfrak U$ on the dynamical phase space as the algebra of real-valued functionals on $\mathcal M$ of the form
\begin{equation}\label{functionals}
    F[\rho] = \sum_{n=0}^\infty \int_{\R^{2n}}  \ c_{n}(x_1,\cdots,x_n;\theta_1, \cdots, \theta_n)\prod_{i=1}^n   \rho(x_i,\theta_i) \dd x_i\dd\theta_i \; .
\end{equation}
Here $c_0\in\R$ and, given a particular dynamical space $\mathcal M$, $c_{n},\,n\geq 1$ are elements of an appropriate space of distributions $\mathcal D_{\mathcal M}(\R^{2n})$ such that, every integral in \eqref{functionals} exists. For instance, in the context of Eq.~\eqref{smoothcompact},  $\mathcal D_{\mathcal M}(\R^{2n})$ is the space of distributions, and notably include smooth functions; while if $\mathcal M$ is a space of Schwartz functions, $\mathcal D_{\mathcal M}(\R^{2n})$ is instead the space of tempered distributions. We further require that the series \eqref{functionals} be convergent (see Appendix \ref{app:alg} for more information). Without loss of generality, we assume the $c_n$'s to be symmetric under permutations: 
$$c_{n}(x_1,\cdots,x_n;\theta_1, \cdots, \theta_n)=c_{n}(x_{\sigma(1)},\cdots,x_{\sigma(n)};\theta_{\sigma(1)}, \cdots, \theta_{\sigma(n)})$$ 
for any permutation $\sigma$ on $n$ elements. For a functional $F[\rho]$, we denote by
\beq\label{functionalderivative}
	F'[\rho](x,\theta) = \frc{\delta F[\rho]}{\delta\rho(x,\theta)} =
 \sum_{n=1}^\infty n \int_{\R^{2(n-1)}}  \ c_{n}(x,x_1\cdots,x_{n-1};\theta, \theta_1 \cdots, \theta_{n-1})\prod_{i=1}^{n-1}   \rho(x_i,\theta_i) \dd x_i\dd\theta_i
\eeq
its functional derivative, so that $F[\rho+\ep] -F[\rho]= \int \dd x\dd\theta\,F'[\rho](x,\theta) \ep(x,\theta)+ O(\ep^2)$ (higher-order functional derivatives are defined analogously). Note that $F'[\rho](x,\theta)$ is of type $(0,0)$ (it is a scalar): this can be seen by the fact that the infinitesimal perturbation $\ep(x,\theta)$ must be a phase-space density, of type $(1,1)$. The algebra $\mathfrak U$ includes polynomials in $\rho$ -- such series where only a finite number of $c_n$'s are nonzero. Given the form \eqref{functionals}, the algebra product is naturally obtained from the Cauchy product of infinite series; see Appendix \ref{app:alg} for a partial analysis of $\mathfrak U$. 

This choice of functionals $F[\rho]$, written as series that do not explicitly involve derivatives of the field $\rho$, is inspired by the theory of \emph{hydrodynamic type systems} (in the sense of Lax \cite{lax2005hyperbolic}) and comes from the notion of \emph{functionals of hydrodynamic type} that we will recall in Section \ref{sec:HydSys}. Note however that, by choosing $c_n$ to be total derivatives in some variables, integration by parts brings derivatives of $\rho$ in  \eqref{functionals}. Thus the absence of derivatives of the field is not as big as a constraint as it may initially appear, but would become stronger when considering dynamical spectral phase-space $\Lambda$.

Naturally, one may also extend the set of observables to all functionals of $\rho$ that are ``smooth'' on an appropriate space of functions (using the theory of Fr\'echet differentiability or more general frameworks, see e.g.~\cite{convenientbook}), instead of considering explicit power series \eqref{functionals}. It would be interesting to develop this in future works.

\subsection{Definition of the Poisson bracket}\label{sec:defPB}

We may now define a Poisson structure on the algebra of observables $\mathfrak U$. Let $F[\rho]$ and $G[\rho]$ be two observables, we propose the following Poisson bracket:
\beq\label{pb}
	\{F,G\} = \int_{\R^2} \frc{ \dd x\dd\theta}{2\pi}\,n\,
	\left[ F_x'\, ({G'_\theta})^{\rm dr} - G_x'\,  ({F'_\theta})^{\rm dr}\right] \; .
\eeq
Here we kept the dependences on $\rho$ and on $(x,\theta)$  implicit, and used the short-hand notation $f_x(x,\theta) = \p_x f(x,\theta)$ and $f_\theta(x,\theta) = \p_\theta f(x,\theta)$, as above. Note how the integrand is of type $(1,1)$, thus the integral is well-defined. The Poisson bracket \eqref{pb} is clearly anti-symmetric and bilinear. We show in Appendix \ref{apppoisson} that it satisfies the Leibniz rule and the Jacobi identity. Hence, this is a well-defined Poisson bracket which can be used to define Hamiltonian flows as in classical field theory.

In general the Poisson bracket maps linear functionals to nonlinear functionals. Another important aspect of the Poisson bracket \eqref{pb} is the fact that the Poisson structure it defines is degenerate, as is the case for example with the KdV Poisson structure \cite{faddeev1985poisson}.  {For example, h}ere the annihilator contains the admissible observable
\begin{equation}
    \mathcal N = \int_{\R^2} \dd x\dd\theta \ \rho(x,\theta) \; ,
\end{equation}
which Poisson-commutes with all observables since $\mathcal N'_x = \mathcal N'_\theta =0$.  {Note that $\mathcal N$ is not the only Casimir invariant of the Poisson bracket \eqref{pb}. There are, in fact, infinitely many, as we shall discuss in Section \ref{sec:consQ}.}

For free systems, where $\psi=0$, the Poisson bracket simplifies
\beq\label{pbfree}
	\{F,G\} = \int_{\R^2} \frc{\dd x\dd\theta}{\Omega_{x\theta}}\,\rho\,
	\Big( F_x'\, {G'_\theta} - G_x'\,   {F'_\theta}\Big)\qquad
	\mbox{(free).}
\eeq
This bracket, or rather its normalised version in term of the normal density \eqref{freedenocc} defined below which amounts to setting the spectral volume density $\Omega_{x\theta}$ to $1$, will be discussed in Section \ref{GHD_hydro} in relation to brackets of hydrodynamic type.

For any real smooth function $f(x,\theta)$ on $\R^2$, we denote the associated {\em linear functional} as
\begin{equation}\label{LinFun}
    Q_f = \int_{\R^2} \dd x\dd\theta\,f(x,\theta)\rho(x,\theta)\in\mathfrak U \ .
\end{equation}
This is the total charge for the quantity that takes value $f(x,\theta)$ on every quasi-particle\footnote{For instance, if the underlying system is a quantum Bethe ansatz integrable system and $f(x,\theta)=f(\theta)$ is independent of $x$, the associated operator $\hat Q_f$ has one-particle eigenvalues $\hat Q_f|\theta\ket = f(\theta)|\theta\ket$. If the underlying system is a soliton gas for a field $u(x,t)$, $\hat Q_f[u(\cdot,t)]$ represents one of the conserved quantities (mass, momentum, energy...) associated with the underlying microscopic model (KdV, NLS, Boussinesq...); then, given the single-soliton solution with parameter $\theta$, $u_\theta(x,t)$, $f(\theta)$ is obtained as $\hat Q_f[u_\theta(\cdot,t)] = f(\theta)$ (e.g. in the case of the KdV equation, the mass is defined as $\int_\R  \dd x\, u(x,t)$, such that the associated function is $f(\theta) = \int_\R  \dd x\, u_{\theta}(x,t) = \theta$).}. The Poisson bracket of total charges is
\begin{equation}\label{LinFunPB}
    \{Q_f,Q_g\}
    =\int_{\R^2} \frc{\dd x\dd\theta}{2\pi}\,
    n\,\big(
    f_xg_\theta^{~\rm dr} - g_xf_\theta^{~\rm dr}\big).
\end{equation}
Rewriting the result in terms of the dynamical variables themselves in distributional form, this gives \eqref{pbnonlocalintro}. In the case of conventional GHD, Eq.~\eqref{localGHD}, the spatial part factorises out of the dressing, and we obtain \eqref{pbintro}. 

In the free case \eqref{pbfree}, the Poisson bracket maps linear functionals to linear functionals, and we have simply
\beq\label{basicpbfree}
	\{\rho(x,\theta),\rho(x',\theta')\}
	=
	\delta'(x-x')\delta'(\theta-\theta')\,\Big(\frc{\rho(x,\theta')}{\Omega_{x\theta'}(x,\theta')}-\frc{\rho(x',\theta)}{\Omega_{x'\theta}(x',\theta)}\Big)
	\qquad \mbox{(free).}
\eeq
Note how this equation is indeed invariant under re-parametrisations \eqref{reparam} as $\rho$ is of type $(1,1)$.  In the case of flat volume element $\Omega_{x\theta}=1$, the Poisson bracket \eqref{basicpbfree} also arises from the classical mechanics of $N$ uninteracting particles, for the empirical density
\begin{equation}\label{classicalmechanics}
    \rho(x,\theta) = \sum_{i=1}^N \delta(x-x_i)\delta(\theta-\theta_i),\quad
    \{x_i,\theta_j\} = \delta_{ij}\qquad \mbox{(classical mechanics, $\Omega_{x\theta}=1$)}.
\end{equation}
In fact, the Poisson bracket \eqref{pbnonlocalintro} arises in a similar way from the classical particle models discussed in Appendix \ref{appparticles}.

\subsection{Linearisation of the Poisson bracket}\label{sec:linpb}

In general, the Poisson bracket simplifies when expressed in terms of appropriate ``normal" densities. Following and adapting \cite{DOYON2018570}, we define a new coordinate\footnote{This is slightly different from that defined in \cite{DOYON2018570}, where the factor was instead $2\pi \rho_{\rm s}/P_\theta$ in order to keep transparent transformation properties. The present transformation is however more convenient and more general.} $y = Y(x,\theta)$ as
\beq\label{metchange}
	\dd y = \Omega_{x\theta}^{~~\rm dr} \dd x = 2\pi\rho_{\rm s}\,\dd x\qquad \mbox{($\theta$ fixed, arbitrary).}
\eeq
The explicit solution can be obtained in terms of a different dressing operation, denoted $^{\rm dR}$:
\begin{equation}\label{Ysolution}
    Y(x,\theta) = \Omega_{\theta}^{~\rm dR}(x,\theta)\ ,
\end{equation}
see Appendix \ref{appDressing}. This change of coordinate transforms the $x$ direction in order to ``trivialise" the volume element $P_\theta^{~\rm dr}\dd x\dd\theta=\Omega_{x\theta}^{~~\rm dr}\dd x\dd\theta=\dd y\dd\theta$. With interaction, the volume element depends on the fluid density $\rho$, because of the dressing. Hence this is a coordinate transformation that depends on the dynamical variable itself. Without interaction, the transformation does not depend on the dynamical variable, but it still is a non-identity transformation in general: then it simply trivialises the fixed volume element $\Omega_{x\theta}\dd x\dd\theta$.

We define the \emph{normal density} as the density $\h \rho(y,\theta)$ for this coordinate. It is simply related to the occupation function (which are the normal modes for the fluid, as explained above): given the definition \eqref{defn} and the identity \eqref{rhosOmega} we have
\beq\label{freedenocc}
	\h \rho(y,\theta)\,\dd y  = \rho(x,\theta)\,\dd x\quad\Rightarrow\quad 
	\h \rho(Y(x,\theta),\theta) = \frc{n(x,\theta)}{2\pi}.
\eeq
Note that the normal density is in fact a {\em scalar} in $x,\theta$ (type $(0,0)$), while the coordinate $Y(x,\theta)$ is scalar in $x$ and a spectral vector field (type $(0,1)$). Thus, under transformations of the spatial coordinate $x$, the $y$ coordinate and normal density are unchanged -- this is because we have chosen them specifically in order to trivialise the volume element. We show in Appendix \ref{apppoisson} that in terms of the normal density, the Poisson bracket \eqref{pb} \emph{takes the free form} \eqref{basicpbfree} with unit spectral volume density,
\beq\label{pbfreeform}
	\{\h \rho(y,\theta),\h\rho(y',\theta')\}
	=
	\delta'(y-y')\delta'(\theta-\theta')\,\big(\h \rho(y,\theta')-\h \rho(y',\theta)\big) \qquad \text{(linearised)}.
\eeq
Clearly, this equation is not invariant under re-parametrisation, as, again, we have chosen a given, convenient coordinate system.

\begin{rema}\label{rem:yspace} If $\Omega_{x\theta}>0$ (as is the case for the delta-Bose and hard rods gases), then in the context of Appendix \ref{appAnalysis}, $\Omega_{x\theta}^{~~\rm dr}>0$. Then the change of coordinates \eqref{metchange} is orientation-preserving. If $\inf_{x\in\R} \Omega_{x\theta}^{~~\rm dr}>0$ then the set of all values of $y$ is $\hat{\mathcal L}_\theta = \R$ (this is the typical case); otherwise it may be an open subset $\hat{\mathcal L}_\theta \subset\R$, which may depend on $\theta$. If $\Omega_{x\theta}^{~~\rm dr}$ is not strictly positive, then the resulting space $\hat{\mathcal L}_\theta$ in which $y$ lies is not $\R$, but an open subset of a multiple cover of $\R$. In all examples of interest, this space is {\em independent of the fluid density $\rho(\cdot,\cdot)$}, and we will assume so here (if this were not so, new terms would arise, since partial derivatives with respect to $x$, $y$ or $\theta$ would also act on the bounds of integration in the $y$ space, and our derivations in Appendix \ref{apppoisson} would fail). It is by considering $Y(x,\theta)\in\hat{\mathcal L}_\theta$ that \eqref{freedenocc} holds in general; that is
\beq\label{Ymapping}
	Y(\cdot,\theta): \R\to \hat{\mathcal L}_\theta.
\eeq
\end{rema}

\begin{rema}\label{rem:metchange}
    The change of variables \eqref{metchange} is not the only one to linearise the Poisson bracket \eqref{pb}. Indeed one could keep the physical space as it is and introduce a new coordinate $p = \kappa(x,\theta)$ in order to contract the spectral space instead 
    \begin{equation}\label{specmetchange}
        \dd p = \Omega_{x\theta}^{~~{\rm dr}} \dd \theta =2\pi \rho_{\rm s}\,\dd\theta \qquad (x {\rm \ fixed, \ arbitrary).}
    \end{equation}
    In this case, an explicit solutions can be obtained in terms of yet another dressing operation (c.f. Appendix \ref{appDressing}), and the result is the ``physical momentum" commonly discussed in the Thermodynamic Bethe Ansatz  \cite{korepinbook}, the momentum the system gains upon adding a quasi-particle of spectral parameter $\theta$ at $x$,
\begin{equation}
    \kappa(x,\theta) = \Omega_x^{~{\rm Dr}}(x,\theta) \ .
\end{equation}
Defining an alternative normal density $\tilde \rho(x,p)$ as
\begin{equation}
     \tilde \rho(x,p)\dd p = \rho(x,\theta)\dd \theta \ ,
\end{equation}
the Poisson bracket \eqref{pb} again takes the free from \eqref{pbfree}
\begin{equation}
    \{\tilde \rho(x,p),\tilde\rho(x',p')\}
	=
	\delta'(x-x')\delta'(p-p')\,\big(\tilde \rho(x,p')-\tilde \rho(x',p)\big) \ .
\end{equation}
This follows immediately from the ``spectral crossing symmetry" of our construction, the structural symmetry under exchange of spatial and spectral variables, $x$ and $\theta$. We chose to highlight the change of variables \eqref{metchange} as it is more standard when it comes to GHD, since it trivialises the force-less GHD equation in which $E(x,\theta) = E(\theta)$ (thus $\,a^{\rm eff}(x,\theta)=0$) \cite{DOYON2018570}. However, the alternative transformation \eqref{specmetchange} also trivialises the GHD equations, this time when $E(x,\theta) = E(x)$ (thus $\,v^{\rm eff}(x,\theta)=0$), a situation which has only been marginally considered. A priori, there is no reason why one should restrict oneself to only contracting either physical or spectral space, and there may be a transformation involving both that trivialises the more general GHD equation. Since this would probably require a more general theory of reparametrisation than the one discussed in section \ref{sec:reparam}, we leave this for future work.
\end{rema}

%

\subsection{Hamiltonian flows}

A Hamiltonian $H$ is a functional of the fluid density $\rho$, $H \in \mathfrak U$, and, in physics, the Hamiltonian is often associated with the energy of the system of interest. Given the phase-space energy function $E(x,\theta)$, the total energy of a system in which particles are distributed according to the fluid density $\rho(x,\theta)$ is simply the linear functional $Q_E$. One of our main results is that the intuitive Hamiltonian
\begin{equation}\label{Hmain}
    H=Q_E = \int_{\R^2} \dd x\dd\theta\,
    E(x,\theta)\rho(x,\theta) \ ,
\end{equation}
is indeed the Hamiltonian whose flow with respect to the Poisson structure \eqref{pb} gives the appropriate GHD equation, given the external force field described by $E(x,\theta)$.

Let  $Q_f$ be a linear functional of type \eqref{LinFun}. From the specialised Poisson bracket \eqref{LinFunPB}, by using the symmetry relation \eqref{symm} and integration by parts, we may write
\beq
	\{Q_f,Q_E\} =\frc1{2\pi}
	\int_{\R^2} \dd x\dd\theta\,
	\big(-f \p_x (n E_\theta^{~\rm dr}) + f \p_\theta (n E_x^{~\rm dr})\big) \ ,
\eeq
and, in particular, if $Q_f = Q_\delta = \rho$, this yields
\beq\label{ghd}
	\{\rho(x,\theta),Q_E\} + \p_x (v^{\rm eff}\rho)
	+ \p_\theta (a^{\rm eff}\rho) = 0 \ ,
\eeq
where we have used the relations \eqref{CRA}, along with the definition of the occupation function \eqref{defn}, which we recall here for convenience: $v^{\rm eff} = E_\theta^{~\rm dr}/ \Omega_{x\theta}^{~~\rm dr}$, $a^{\rm eff} = -E_x^{~\rm dr}/ \Omega_{x\theta}^{~~\rm dr}$ and $n = 2\pi\rho/\Omega_{x\theta}^{~~\rm dr}$. With $\p_t \rho(x,\theta) = \{\rho(x,\theta),Q_E\}$, this indeed {\it yields the spatially extended GHD equation \eqref{ghdsum} as a Hamiltonian system}:
\beq
\p_t \rho(x,\theta) = \{\rho(x,\theta),Q_E\}=-\p_x (v^{\rm eff}\rho)
	- \p_\theta (a^{\rm eff}\rho)\,.
\eeq
As such, thanks to the Poisson structure \eqref{pb}, GHD can be seen as a classical Hamiltonian field theory in two dimensions (in spectral phase space $x,\theta$). This is the main result of this paper.

From this result, it is immediate to recover the result of \cite{DOYON2018570}, generalised here to inhomogeneous momenta and spatially extended interaction kernel: that, in the force-less case $E_x=0$, the normal density $\hat\rho(y,\theta)$ satisfies the Liouville equation. Indeed, in the force-less case, using \eqref{freedenocc}, the Hamiltonian is written as
\beq
	Q_E = \int_{\R^2} \dd x\dd\theta\,E(\theta)\rho(x,\theta)
	=\int_\R \dd \theta \int_{\h{\mathcal L}_\theta}\dd y\, E(\theta)\h\rho(y,\theta)
\eeq
and hence, the linearised Poisson bracket \eqref{pbfreeform} with $\p_t \h\rho(y,\theta) = \{\h\rho(y,\theta),Q_E\}$ gives
\beq\label{hrhoGHD}
	\p_t \h\rho(y,\theta)+ E_\theta \p_y\h\rho(y,\theta)=0 \qquad (E_x=0).
\eeq
Likewise, according to Remark \ref{rem:metchange}, in the case $E_\theta=0$ (the ``kinetic-less" case), the density $\t\rho(x,p)$ satisfies
\begin{equation}
	\p_t \t\rho(x,p) - E_x \p_p\t\rho(x,p)=0\qquad (E_\theta=0).
\end{equation}
 {This last case is mostly formal and, a priori, not (yet) physically relevant. Rather, we mention it to highlight the symmetry between $x$ and $\theta$ in our construction, which naturally appears as we introduce spatial inhomogeneity in the interaction kernel $\psi$.}

Clearly, any linear functional $Q_h$ of type \eqref{LinFun} generates its own GHD-like flow. If $E_x=0$, then any linear functional $Q_h$ such that $h_x=0$ are conserved quantities of the GHD equations and are in involution with respect to the (specialised) Poisson bracket \eqref{LinFunPB}
\beq\label{case1}
	\{Q_h,Q_E\} = 0\qquad (E_x=0,\ h_x=0).
\eeq
This corresponds to the case of conventional GHD describing the evolution of an integrable many body system without external potential. From the point of view of the microscopic model underlying the GHD equation, the functionals $Q_h$ are the higher conserved charges of the hierarchy, i.e.~these are the commuting flows of the hierarchy. Similarly, if $E_\theta=0$, linear functionals $Q_g$ such that $g_\theta=0$ are, again, conserved and in involution
\beq\label{case2}
	\{Q_g,Q_E\} = 0 \qquad (E_\theta=0,\ g_\theta=0).
\eeq
From the point of view of the microscopic model, these flows correspond, this time, to dynamics generated by Hamiltonians that are inhomogeneous in space, but that couple to a single conserved quantity of the hierarchy, which is the total particle density.

In both cases above, we see that the GHD equation admits a large number of conserved quantities in involution, which are extensive in phase-space. However, this is not sufficient to claim integrability: as GHD is a two-dimensional field theory, we would expect a family of conserved quantities specified by functions of two variables, in effect fixing ``every point" of phase-space. We will now discuss this.

\begin{rema}
    Note that the Liouville equation \eqref{hrhoGHD} we obtain is slightly different from the result of \cite{DOYON2018570} (assuming the condition \eqref{smoothcompact} is met)
    \begin{equation}
        \p_t\hat\rho + v^{\rm gr}(\theta)\p_y\hat\rho=0 \; ,
    \end{equation}
    where $v^{\rm gr}=E_\theta/\Omega_{x\theta}$. This is because the change of metric introduced in \cite{DOYON2018570} to trivialise the conventional GHD equation is not the one defined in Eq.~\eqref{metchange}, rather it is
    \begin{equation}\label{metchangeGeoViewpoint}
        \dd y = \frac{\Omega_{x\theta}^{\rm dr}}{\Omega_{x\theta}}\dd x \ .
    \end{equation}
    As such, in the context of \cite{DOYON2018570}, in the absence of interaction we have $\dd y = \dd x$, which is not the case in this manuscript, as remarked upon in Section \ref{sec:linpb}, following Eq.~\eqref{Ysolution}.
\end{rema}

\subsection{Extensive conserved quantities in involution}\label{sec:consQ}

It is known (see e.g.~\cite{10.21468/SciPostPhys.6.6.070}  {Appendix C}\footnote{ {To make upcoming computations simpler, we use a definition \eqref{tildeQf} of $\tilde Q_f$  which is slightly different from that of \cite{10.21468/SciPostPhys.6.6.070} Eq. (17), where our $f$ corresponds to their $f$ divided by $n$.}}) that, for general force field $E(x,\theta)$, given any  {continuously differentiable} function of the normal modes $f(n)$, functionals of the type
\beq\label{tildeQf}
	\tilde Q_f = \int_{\R^2} \dd x\dd\theta\,f(n(x,\theta))\rho_{\rm s}(x,\theta) \ ,
\eeq
are invariant under the GHD evolution
\beq\label{pbQtilde}
	\{\tilde Q_f,Q_E\}=0.
\eeq
Indeed, this follows immediately from the GHD equation \eqref{ghdsum}, its diagonalisation or normal mode decomposition \eqref{ghdn}, and the definition of the occupation function \eqref{defn}. This includes, in particular, the entropy function of the underlying microscopic model, conserved under the GHD equation \cite{doyon2017note} (for instance for classical particle systems, one chooses $f(n) = -n {\left[\log n -1\right]}$), and the total number of particles ($f(n) = n$, which is $\tilde Q_f = Q_1$).

Given the result \eqref{pbQtilde}, it is now clear that, in both cases considered previously, \eqref{case1} and \eqref{case2}, there are more conserved quantities besides those written there. Indeed, if $E_x=0$, given any function of two variables $f(n,\theta)$ the quantity
\beq\label{Qftheta}
	\tilde Q_f = \int_{\R^2} \dd x\dd\theta\,f(n(x,\theta),\theta)\rho_{\rm s}(x,\theta)\qquad \mbox{(case $E_x=0$)}
\eeq
is invariant under the GHD evolution. This includes all the quantities previously highlighted by \eqref{case1}, with $\tilde Q_f = Q_h$ for $f(n,\theta) = n h(\theta)$, and notably the Hamiltonian with $h(\theta) = E(\theta)$. Moreover, if instead $E_\theta=0$, given any function $f(x,n)$ the quantity
\beq\label{Qfx}
	\tilde Q_f = \int_{\R^2} \dd x\dd\theta\,f(x,n(x,\theta))\rho_{\rm s}(x,\theta)\qquad \mbox{(case $E_\theta=0$)}
\eeq
is invariant under the GHD evolution. This includes those higlighted by \eqref{case2}, with $\tilde Q_f = Q_g$ for $f(x,n) = n g(x)$, and, once again, the Hamiltonian with $g(x) = E(x)$.

In general, quantities $\tilde Q_f$ of type \eqref{tildeQf} are nonlinear functionals of $\rho$. In fact, if the interaction kernel has finite range, short range, or long range with a quick enough algebraic decay (see Remark \ref{remashortrange}), then they are ``extensive", with a density $f(n(x,\theta))\rho_{\rm s}(x,\theta)$ ``supported in a neighbourhood of $(x,\theta)$": it depends only on the dynamical variable $\rho(x',\theta')$ at spectral phase-space points $(x',\theta')$ near to $(x,\theta)$. Extensivity of conserved quantities is known to be an important concept in many-body physics. However, it is not clear to us how much it plays a role in the present construction, and in some examples (hard rods, the KdV  {equation}) the interaction appears to be of infinite range in spectral space. Notwithstanding this, in the general case, the GHD equation possesses an infinite number of (extensive) conserved quantities parametrised by a function of a single variable, while in the special cases $E_x=0$ and $E_\theta=0$, they are parametrised by a function of two variables.

 {The above discussion suggests that GHD, seen as a classical field theory, may be integrable. However, t}o claim the GHD equation is integrable (in some accepted field theoretic version of Liouville theorem that has been established for several $(1+1)-$dimensional systems, see \cite{faddeev2007hamiltonian}), another crucial point is that the {\em conserved quantities must be  {non trivially} in involution,  {that is not Casimirs}}.  {This last point is particularly important. It highlights the fact that conservation of functionals of type \eqref{tildeQf} cannot be used as an argument to suggest integrability of the GHD equation since they are, in fact, Casimir elements of our Poisson bracket \eqref{pb}. This can be easily checked by using the change of coordinates \eqref{metchange} which linearises the Poisson bracket, then Eq.~\eqref{tildeQf} becomes
\beq
   \tilde  Q_f = \frc1{2\pi} \int_{\R}\dd\theta\int_{\hat{\mathcal L}_\theta} \dd y\,f(2\pi \h\rho(y,\theta))\; ,
\eeq
where the space $\hat{\mathcal L}_\theta$ has been introduced in Remark \ref{rem:yspace}. Note that in terms of $\hat\rho(y,\theta)$, these are now truly extensive, with an ultra-local density. Hence, we have
\beq
	\frc{\delta \tilde Q_f}{\delta \hat \rho(y,\theta)}
	= f_n(2\pi \hat \rho(y,\theta)) \; ,
\eeq
where as usual we use the index notation for derivatives, with in particular $f_n(n,\theta) = \p f(n,\theta)/\p n$. Using the normal form of the Poisson bracket \eqref{pbfreeform} -- or the equivalent form \eqref{pby} from Appendix \ref{ssectequiv} --, we can therefore evaluate for any $G \in \mathfrak U$
\begin{equation}\label{invol1}
\begin{aligned}
\{\tilde Q_f,G\} &= \int_{\R}\dd\theta\int_{\hat{\mathcal L}_\theta} \dd y\,
    \h\rho\, \big[(f_n)_y\left(\frac{\delta G}{\delta \hat \rho}\right)_\theta - \left(\frac{\delta G}{\delta \hat \rho}\right)_y (f_n)_\theta\big] \\
    &= 2\pi \int_{\R}\dd\theta\int_{\hat{\mathcal L}_\theta} \dd y\, \frac{\delta G}{\delta \hat \rho}\left[\left(\hat \rho f_{nn} \hat \rho_\theta\right)_y - \left(\hat \rho f_{nn} \hat \rho_y\right)_\theta\right]
    \\
    &= 2\pi \int_{\R}\dd\theta\int_{\hat{\mathcal L}_\theta} \dd y\, \frac{\delta G}{\delta \hat \rho} (f_{nn} + f_{nnn}\hat\rho)\left[\hat \rho_y\hat\rho_\theta - \hat \rho_\theta\hat\rho_y\right] = 0 
\end{aligned}   \; .
\end{equation}
When the energy function depends both on $x$ and $\theta$, we were not able to find non-Casimir conserved quantities in involution and we do not expect GHD to be integrable. This is consistent with the atomic reduction of GHD (discussed in Appendix \ref{appparticles}) which corresponds to Hamiltonian systems of particles that can not be expected to be integrable if the energy is both $x-$ and $\theta-$dependent, while they are integrable if it only depends on $x$ or $\theta$.} 

 {Functionals of type \eqref{Qftheta} (respectively of type \eqref{Qfx}) are not Casimirs and, when the energy function is such that $E_x = 0$ (respectively $E_\theta=0$), they are indeed in involution. In the following, we only discuss the family associated with $E_x=0$, that is functionals of type \eqref{Qftheta}; involution of functionals of type \eqref{Qfx} follows from a similar derivation. Again, t}o make things simpler, we  {make use of the change of coordinates \eqref{metchange}
\beq
   \tilde  Q_f = \frc1{2\pi} \int_{\R}\dd\theta\int_{\hat{\mathcal L}_\theta} \dd y\,f(2\pi \h\rho(y,\theta),\theta)\; , \qquad \frc{\delta \tilde Q_f}{\delta \hat \rho(y,\theta)}
	= f_n(2\pi \hat \rho(y,\theta),\theta) \; .
\eeq}
Using the normal form of the Poisson bracket \eqref{pbfreeform}, we therefore obtain
\begin{equation}\label{invol1}
\begin{aligned}
\{\tilde Q_f,\tilde Q_g\} &= \int_{\R}\dd\theta\int_{\hat{\mathcal L}_\theta} \dd y\,
    \h\rho\, \big[(f_n)_y(g_n)_\theta - (g_n)_y (f_n)_\theta\big] \\
    &= 2\pi \int_{\R}\dd\theta\int_{\hat{\mathcal L}_\theta} \dd y\,
    \h\rho\, \big[f_{nn} \h\rho_y(2\pi g_{nn}\h\rho_\theta +g_{n\theta}) - g_{nn} \h\rho_y (2\pi f_{nn}\h\rho_\theta + f_{n\theta})\big]\\
    &= 2\pi \int_{\R}\dd\theta\int_{\hat{\mathcal L}_\theta} \dd y\,
    \h\rho\h\rho_y \, \big[f_{nn} g_{n\theta} - g_{nn} f_{n\theta}\big]\\
\end{aligned}   \; ,
\end{equation}
where $(f_n)_\theta = \p_\theta \big(f_n(2\pi\h\rho(y,\theta),\theta)\big)$ while $f_{n\theta} = \p_\theta f_{n}(n,\theta)\big|_{n=2\pi \h\rho(y,\theta)}$, etc. Note that $f_{nn} g_{n\theta} - g_{nn} f_{n\theta}$ is a function of $\h\rho(y,\theta)$ and $\theta$ only (that is, its $y$ dependence comes solely from $\h\rho(y,\theta)$); we denote it as $h(\h\rho(y,\theta),\theta)$. Then we obtain
\beq\label{invol2}
    \{\tilde Q_f,\tilde Q_g\} = 2\pi\int_{\R}\dd\theta\int_{\hat{\mathcal L}_\theta} \dd y\,
    \h\rho\h\rho_y \, h(\h\rho,\theta)
    = 2\pi \int_{\R}\dd\theta\int_{\hat{\mathcal L}_\theta} \dd y\,
    \p_y j(\h\rho,\theta) \; ,
\eeq
where $j(\h\rho,\theta) = \int_{0}^{\h\rho} \dd\h\rho'\,\h\rho' h(\h\rho',\theta)$ is a primitive of $\h\rho h(\h\rho,\theta)$ with respect to $\h\rho$ (at $\theta$ fixed). Hence the result is, using the fact that $\hat\rho(y)$ at the asymptotic boundaries of $\hat{\mathcal L}_\theta$ takes the values $n(\pm\infty,\theta)/2\pi = 0$ thanks to the mapping \eqref{Ymapping},
\beq\label{invol}
    \{\tilde Q_f,\tilde Q_g\} = 2\pi\int_{\R} \dd\theta\,( j(0,\theta) - j(0,\theta)) = 0.
\eeq
This shows that {, under appropriate conditions on the energy function,} quantities of either type \eqref{Qftheta} or \eqref{Qfx} are all invariant under the GHD evolution \emph{and} in involution with respect to the Poisson bracket \eqref{pb}. 

 {One last requirement for integrability is that the family of conserved quantities should be ``large enough". Although this is not a fully accurate concept, for field theories lying in two dimensions of space -- here spectral phase space $(x,\theta)$ --, intuitively one would need a family of conserved quantities that explores the full two dimensions. In the case $E_x=0$, for instance, one may see $\theta$ as parametrising an infinite set of coupled equations (the GHD equations at fixed $\theta$) for fields lying in one dimension of space $x$. Then, one may use $f(n,\theta) = f(n)\delta(\theta-\theta_0)$ to construct functionals $\tilde Q_f$ of type \eqref{Qftheta}, which provide a full family of extensive conserved quantities for each parameter $\theta_0$; indeed the family appears to be large enough. Such an approach has in fact been considered before, we will discuss it in Section \ref{sec:HydSys}, and we shall see we indeed have additional reasons to presume the GHD equation is integrable in either cases $E_x=0$ or $E_\theta=0$.}

\section{Overview of the previously known structures of the GHD equation}\label{sec:HydSys}

Our work is not the first one to deal with the Hamiltonian structure and/or integrability of the GHD equations. However, in most cases, previous works have not dealt with the GHD equations directly, focusing instead on particular reductions. Moreover, their language was often not that of GHD, but that of differential geometry. In this section we provide a brief overview of the state of the art, showcasing previous known results regarding reductions of the GHD equations and introducing the theoretical framework in which they were derived, in order to better highlight what our construction brings to the table.

\subsection{Reduced GHD equations as systems of hydrodynamic type}

Investigation into the integrability of the GHD equations \eqref{ghdsum}-\eqref{veff}-\eqref{aeff} can be traced back to the work of El, Kamchatnov, Pavlov, and Zykov in \cite{el2011kinetic}. In that work, the authors studied the so-called kinetic equations of the KdV soliton gas, i.e. a special case of the conventional and homogeneous GHD equation: $E(x,\theta) = E(\theta),\,P(x,\theta) = P(\theta),\,a^{\rm eff}(x,\theta)=0$, with the choice of data given by \eqref{localGHD} and \eqref{KdV}. They established integrability of the kinetic equations via generalised hodograph transform \cite{tsarev1991geometry} under the  ``cold gas'' (a.k.a ``polychromatic gas'' \cite{congy2024riemann}) reduction 
\begin{equation}\label{coldgas}
    \rho(x,\theta,t) = \sum_{i=1}^m  {\rho^i(x,t)\delta(\theta-\theta^i)} \ , 
\end{equation}
which generically transforms the GHD equations into an $m$-component, linearly degenerate, hyperbolic system of hydrodynamic type for the propagation of the weights $\{ {\rho^i}\}_{i=1}^m$ according to their effective velocities $ {v^i(x,t) \equiv v^{\rm eff}(x,\theta^i,t)}$. In the particular case they investigated, the resulting system takes the form
\begin{equation}\label{PolyKinEq}
    \partial_t  {\rho^i + \partial_x\left(v^i\rho^i\right)} = 0 \ , 
\end{equation}
\begin{equation}\label{PolyVeff}
    { v^i = 4(\theta^i)^2 + \frac{1}{\theta^i} \sum_{k\neq i}  \log\left|\frac{\theta^i - \theta^k}{\theta^i + \theta^k} \right|(v^i-v^k)} \ .
\end{equation}
It is important to note that the resulting system admits a Riemann invariants representation
\begin{equation}\label{PolyDiag}
     {\partial_t r^i + v^i \partial_x r^i} = 0 \ ,
\end{equation}
\begin{equation}\label{PolyRinv}
     {r^i  = \frac{1}{\rho^i} \left(1 +\sum_{k\neq i} \frac{1}{\theta^i} \log\left|\frac{\theta^i - \theta^k}{\theta^i + \theta^k}\right|\rho^k \right)} =  {\frac{\pi}{4 \theta^i n(x,\theta^i,t)}} \ ,
\end{equation}
the GHD analogue of which is the ``diagonalised'' equation \eqref{ghdn}\footnote{Note that, given the transport equation \eqref{PolyDiag}, any function of a Riemann invariant $f( {r^i})$ is also a Riemann invariant. This is also the main reason why relation \eqref{pbQtilde} holds.}. This aspect is crucial in establishing integrability: to show that a diagonal system of hydrodynamic type is integrable it is sufficient to show that it is either Hamiltonian, or semi-Hamiltonian \cite{tsarev1985poisson,pavlov1988hamiltonian}. We will now discuss those two notions and introduce the framework through which reduced GHD equations have been studied.

\begin{rema}
    We highlight the fact that the effective velocities $v_i$ in the system of reduced GHD equations \eqref{PolyKinEq} are also the characteristic velocities that appear in the Riemann invariants representation \eqref{PolyDiag}. This is generically not the case when it comes to systems of hydrodynamic type but is a specificity of reduced GHD equations. The equivalent situation occurs with the effective velocity and acceleration in the full GHD equation, see \eqref{ghdsum}, \eqref{ghdn}. In the force-less case of conventional GHD, this plays an important role in \cite{HubnerDoyonQuadrature}.
\end{rema}

\subsection{Integrable systems of hydrodynamic type}

The theory of Hamiltonian and semi-Hamiltonian systems of hydrodynamic type is well-developed, we provide here some basic elements and, for more information, we refer the interested reader to the reviews \cite{dubrovin1989hydrodynamics,tsarev1991geometry}.
In this section we consider an $m$-component system of hydrodynamic type 
\begin{equation}\label{HydTypeSys}
     {\partial_t \rho^i + \partial_x\left[v^i(\rho^1,\cdots\rho^m) \rho^i\right] = 0} \quad i = 1,\cdots m \ , 
\end{equation}
similar to the reduced GHD equations \eqref{PolyDiag} but with arbitrary velocities. Given this, an integral is said to be of hydrodynamic type if it is of form
\begin{equation}
 I = \int_{\mathbb R} \dd x \ P( {\rho^1, \cdots \rho^m}) \ ,
\end{equation}
where the densities $P$ are functions of the components $\{ {\rho^i}\}_{i=1}^m$ only, and do not depend on their derivatives. This notion is what informed our choice regarding the algebra of observables $\mathfrak U$ in the previous section. A system of hydrodynamic type is then said to be Hamiltonian if there exists an integral of hydrodynamic type $H$ such that
\begin{equation}
    \partial_t  {\rho^i = \{\rho^i,H\}} \ ,
\end{equation}
where the Poisson bracket  (summing over repeated indices) 
\begin{equation}\label{PBHHS}
    \{I, J\} = \int_{\R} \dd x \frac{\delta I}{\delta  {\rho^i}(x)}  {A^{ij}} \frac{\delta J}{\delta  {\rho^j}(x)} \ ,
\end{equation}
is of Dubrovin-Novikov type \cite{dubrovin1983hamiltonian}
\begin{equation}\label{PBDN}
    \quad  {A^{ij} = g^{ij}(\rho^1,\cdots\rho^n)\partial_x - g^{is}(\rho^1,\cdots\rho^n)\Gamma^j_{sk}(\rho_1,\cdots\rho_n)\partial_x \rho^k} \ .
\end{equation}
Here the variational derivatives are defined from
\begin{equation}
     {I[\rho^1,\cdots\rho^i + \delta\rho^i,\cdots\rho^m] - I[\rho^1,\cdots \rho^m] = \int_{\R}\dd x \frac{\delta I}{\delta \rho^i(x)} \delta\rho^i(x) + o(\delta\rho^i)} \ .
\end{equation}
Moreover, we require that the coefficients $g_{ij}
$ define a contravariant, flat, pseudo-Riemannian metric (notably $\det g_{ij}\neq 0$ and $g_{ij}=g_{ji}$), and that $ {\Gamma_{jk}^i}$ are Christoffel symbols of the associated Levi-Civita connection. These conditions ensure that the Poisson bracket is skew-symmetric and satisfies both the Leibniz rule and the Jacobi identity.

If a Hamiltonian system of hydrodynamic type admits a Riemann invariants representation\footnote{Again, in general, the characteristic velocities $v_i$ are not necessarily the same as in Eq.~\eqref{HydTypeSys}. However, we limit ourselves to this situation since it is generically the case when it comes to polychromatic reductions of the GHD equation.}
\begin{equation}
    \partial_t  {r^i + v^i \partial_x r^i} = 0 \ ,
\end{equation}
then, under this change of variable, the metric becomes diagonal $ {g^{ij}(r^1,\cdots r^m) = g^{ij}(r^1,\cdots r^m)\delta^{ij}}$  and the Christoffel symbols simplify \cite{tsarev1985poisson}
\begin{equation}
     {\Gamma_{ij}^k = 0 \ , \quad \Gamma_{ij}^i = \partial_{r^j}\log\sqrt{g_{ii}} = \frac{\partial_{r^j}v^i}{v^j-v^i}} \ , \quad i\neq j \neq k \ .
\end{equation}
In particular, this last expression implies the relation
\begin{equation}\label{semiHam}
     {\partial_{r^j}\frac{\partial_{r^k} v^i}{v^k-v^i} = \partial_{r^k}\frac{\partial_{r^j} v^i}{v^j-v^i}} \ , \quad i,j = 1,\cdots m \ , \quad i\neq j \neq k \ ,
\end{equation}
known as the \emph{semi-Hamiltonian} property \cite{pavlov1988hamiltonian}. Note that a diagonalisable system of hydrodynamic type may possess this property without being Hamiltonian, examples of this include the equations of isotachophoresis \cite{babskii2012mathematical}, ideal chromatography \cite{rozdestvenskii1983systems,tsarev2000integrability}, or an extended Born-Infeld equation written in hydrodynamic form \cite{pavlov1988hamiltonian,brenier2004hydrodynamic}.

A semi-Hamiltonian system of hydrodynamic type is also integrable: it possesses infinitely many linearly independent integrals of hydrodynamic type
that are invariant under the dynamics \eqref{PolyDiag}, are in involution and generate commuting Hamiltonian flows including the system Hamiltonian, and form a complete set in the sense of \cite{tsarev1991geometry}. Moreover such a system is (implicitly) solvable via the generalised hodograph transform developed by Tsarev \cite{tsarev1985poisson}
\begin{equation}\label{hodo}
    x +  {v^i} t = w_i^{(\tau)} \ ,
\end{equation}
where the functions $\{w_i^{(\tau)}\}_{i=1}^m$ solve the overdetermined linear system
\begin{equation}\label{HydFlows}
     {\frac{\partial_{r^j}w_i^{(\tau)}}{w_j^{(\tau)}-w_i^{(\tau)}} = \frac{\partial_{r^j}v^i}{v^j-v^i}} \ , \quad i,j = 1,\cdots m \ , \quad i\neq j \ ,
\end{equation}
and specify the commuting flows
\begin{equation}
     {\partial_\tau r^i + w_i^{(\tau)} \partial_x r^i} = 0 \ ,
\end{equation}
for the time (group parameter) $\tau$, such that $ {\partial_\tau\left(\partial_t r^i\right) = \partial_t\left(\partial_\tau r^i\right)}$.

Admittedly, the generalised hodograph method may seem rather unwieldy: assuming one manages to solve the over-determined system \eqref{HydFlows}, one would then need to invert the transform \eqref{hodo}. However, there exist (relevant) circumstances, that we highlight here briefly for reference, under which this procedure simplifies. In particular, this is the case if the system is linearly degenerate
\begin{equation}\label{lindeg}
     {\partial_{r^i} v^i} = 0 \ , \quad \text{for any } i=1,\cdots m\ ,
\end{equation}
a condition that also implies the system does not support classical shocks \cite{rozdestvenskii1983systems}. Ferapontov showed in \cite{ferapontov1991integration} that, of the infinitely many linearly independent commuting flows possessed by a linearly degenerate semi-Hamiltonian system, only $m$ of them are linearly degenerate as well (including the trivial ones $w_i^{(\tau_0)} = 1$ and $w_i^{(\tau_1)} =  {v^i}$); he then provided a method to solve \eqref{HydFlows} by quadrature. Further simplification can also be obtained if the system belongs to the Egorov class \cite{pavlov2003tri}, i.e. if it features a unique pair of conservation laws such that
\begin{equation}
     {\partial_t A(r^1,\cdots r^m) + \partial_x B(r^1,\cdots r^m) =0 \ , \quad \partial_t B(r^1,\cdots r^m) + \partial_x C(r^1,\cdots r^m)} = 0 \ ,
\end{equation}
which essentially expresses the fact that the system is Galilean invariant. The Egorov property implies that the metric is \emph{potential}: there exists a function $\mathcal G {(r^1,\cdots r^m)}$ such that $g_{ii}=\partial_{ {r^i}} \mathcal G$, Pavlov then proved  that the potential is in fact the first conserved density $\mathcal G {(r^1,\cdots r^m) = A(r^1,\cdots r^m)}$ \cite{pavlov1994exact,pavlov1994multi}. Furthermore, Egorov systems lie in the isomorphism class derived in \cite{el2011kinetic} in which Riemann invariants can be computed explicitly.

\subsection{Structure of GHD equations: state of the art}

In \cite{el2011kinetic} the authors showed that the cold gas reduction of the kinetic equations of the KdV soliton gas is linearly degenerate, Egorov and indeed possesses the semi-Hamiltonian property for any $m$. Moreover, they provided explicit solutions (in terms of the Riemann invariants $\{ {r^i}\}_{i=1}^m$ and of the weights $\{ {\rho^i}\}_{i=1}^m$) for $m=3$, and they found the following solution by quadrature for arbitrary $m$
\begin{equation}
    x +  {4(\theta^i)^2 t = \int^{ {r^i}} \dd\xi\,\frac{\xi\phi^i(\xi)}{\zeta(\xi)} + \sum_{j\neq i}\frac{1}{\theta^i\theta^j}\log\left|\frac{\theta^i-\theta^j}{\theta^i+\theta^j}\right|\int^{ {r^j}}\dd\xi\,\frac{\phi^j(\xi)}{\zeta(\xi)}} \; ,
\end{equation}
where the $ {\phi^i}$'s and $\zeta$ play the role of functional degrees of freedom\footnote{For instance, quasi-periodic (finite-gap like) solutions are obtained by imposing that the $ {\phi^i}$'s are polynomials of degree less than $m$ and that $\zeta = \prod_{k=1}^K(\xi-\lambda_k)$, where the $\lambda_k$'s are real constants and where $K=2m+1$ if $m$ is odd or $2m+2$ if it is even.}. The Hamiltonian structure of the cold gas reduction was then explored in \cite{dubrovin2011linearly} and \cite{Bulchandani_2017}. The cold gas reduction was extended in \cite{pavlov2012generalized} to account for non-constant $\theta_i$'s, in which case the diagonal system \eqref{PolyDiag} now takes the form
\begin{equation}\label{PolyDiagExt1}
    \partial_t  {r^i + v^i \partial_x r^i + p^i \partial_x \theta^i}= 0 \ ,
\end{equation}
where the $ {p^i}$'s are functions of both the Riemann invariants and the $ {\theta^i}$'s, and is supplemented by $m$ more equations describing the propagations of the $ {\theta^i}$'s
\begin{equation}\label{PolyDiagExt2}
    \partial_t  {\theta^i + v^i \partial_x \theta^i} = 0 \ .
\end{equation}
More recently it was shown in \cite{ferapontov2022kinetic} that the extended system \eqref{PolyDiagExt1}-\eqref{PolyDiagExt2}, whose matrix is composed of $m$ $2\times 2$ Jordan blocks, is linearly degenerate  {(using the same definition \eqref{lindeg} as before, which does not involve the $p^i$'s or $\theta^i$'s)} and integrable via an extension of the generalised hodograph method\footnote{Note that  {the Jordan-block reducible} equations \eqref{PolyDiagExt1}-\eqref{PolyDiagExt2} do not correspond to reduced versions of the spatially extended GHD equations. They rather are reduced versions of conventional GHD equations  {on a} dynamical  {spectral phase space} $\Lambda {(x,t)}$, a case we allude to in Section \ref{sec:MoreGenDomain} {, but do not directly address in this paper}.}. Hamiltonian structure of this extended system (along with other ones corresponding to different models like Lieb-Liniger, sinh-Gordon, DNLS...) was established in \cite{vergallo2023hamiltonian,vergallo2024hamiltonian}. In addition, a Hamiltonian formulation was proposed for the (non-reduced) conventional GHD of the hard rods model, represented as an infinite integrable hydrodynamic chain (see \cite{pavlov2003integrable} for a review of that topic) associated with the spectral moments of the fluid density $\rho$. We will comment on this point further in the next section. 

At least when it comes to standard (conventional and homogeneous) GHD, continuum generalisations of some of the aforementioned aspects have been discussed in \cite{Bulchandani_2017,DOYON2018570,bulchandani2017solvable,doyon2020lecture,bonnemain2022generalized}. In particular the GHD equations are diagonalisable  {(for constant domain $\Lambda$)}, linearly degenerate 
\begin{equation}
    \frac{\delta v^{{\rm eff}}(\theta)}{\delta n(\theta)} = 0 \quad \forall \theta \ ,
\end{equation}
and the existence of a self-conserved current (i.e. the Egorov property) was used in \cite{spohn2020collision,yoshimura2020collision} to justify the collision rate ansatz yielding the effective velocity \eqref{CRA}. Moreover, GHD equations satisfy the semi-Hamiltonian property
\begin{equation}
    \int \dd\nu \left[\frac{\delta}{\delta n(\nu)}\left(\frac{\delta v^{{\rm eff}}(\eta)/\delta n(\mu)}{v^{{\rm eff}}(\mu)-v^{{\rm eff}}(\eta)}\right)\right]=\int \dd\mu \left[\frac{\delta}{\delta n(\mu)}\left(\frac{\delta v^{{\rm eff}}(\eta)/\delta n(\nu)}{v^{{\rm eff}}(\nu)-v^{{\rm eff}}(\eta)}\right)\right], \quad \mu \ne \nu \ne \eta \, .
\end{equation}
This means we expect integrability to hold, at least when it comes to standard GHD, beyond the previously considered reductions. In fact, as we have shown in Section \ref{sec:Ham}, even in the present case of spatially extended GHD, the constitutive system of equations \eqref{ghdsum}-\eqref{veff}-\eqref{aeff} is diagonalisable \emph{and} Hamiltonian, and in the homogeneous (force-less) case possesses a seemingly sufficiently large family of extensive conserved quantities in involution, suggesting GHD may be integrable even in its more general, spatially extended form. 

Lastly, a geometric approach to conventional GHD was developed in \cite{DOYON2018570}, based on the change of metric \eqref{metchange}. As mentioned in this change of metric trivialises the conventional GHD equation, Eq.~\eqref{hrhoGHD}, viz.
\begin{equation}\label{TrivlocGHD}
    \partial_t \h \rho(y,\theta,t) + v^{{\rm gr}}(\theta) \partial_y \h \rho(y,\theta,t) = 0 \ ,  
\end{equation}
where we introduced the group velocity $v^{{\rm gr}}:=E_\theta$, which is the (bare) velocity at which a quasi-particle of parameter $\theta$ moves in the absence of interactions. Importantly, the group velocity is only a function of $\theta$ and does not involve the dressing operation\footnote{Recall that in force-less GHD $E(x,\theta)=E(\theta)$.}, contrary to the effective velocity. This allows for an implicit solution by the method of characteristics, indeed
\begin{equation}
    \h \rho(y,\theta,t) = \h \rho(y-v^{{\rm gr}}(\theta)t,\theta,0) \ ,
\end{equation}
is solution of \eqref{TrivlocGHD}. As such, recalling the identity \eqref{freedenocc}, we have
\begin{equation}\label{solchar}
    n(x,\theta,t) = n(U(x,\theta,t),\theta,0) \ ,
\end{equation}
where the function $U$ is determined by
\begin{equation}
    Y(U(x,\theta,t),\theta,0) = Y(x,\theta,t) - v^{{\rm gr}}(\theta)t \ ,
\end{equation}
or, equivalently, according to the identity \eqref{metchange}
\begin{equation}\label{intchar}
    2\pi\left(\int_{-\infty}^x \dd z \rho_s(z,\theta,t) -\int_{-\infty}^{U(x,\theta,t)} \dd z \rho_s(z,\theta,0)\right)= v^{{\rm gr}}(\theta)t \ .
\end{equation}
Note that time only appears as a parameter in equations \eqref{solchar}-\eqref{intchar}, and a solution at arbitrary time can be obtained directly by solving these equations iteratively \cite{DOYON2018570}; there is no need to solve the GHD equations using, for instance, a finite element method.

\subsection{Hamiltonian structure of the hard rod gas: comparison with Vergallo-Ferapontov} \label{appcomparison}

While this manuscript was in preparation, Vergallo and Ferapontov (VF) posted the preprint \cite{vergallo2024hamiltonian}, where two Hamiltonian structures for the polychromatic reductions of the GHD equation (or soliton gas kinetic equation) were constructed. Most interestingly, they took the continuous (non-reduced) limit of their construction for the special case of the hard rod gas, obtaining Hamiltonian structures for the corresponding conventional GHD equation. We now verify that one of the VF Hamiltonian structures for the hard rod gas (the one that is ``local"
\footnote{In this instance, ``local'' is to be interpreted as referring to an operator of Dubrovin-Novikov type \eqref{PBDN}, while ``non-local'' refers to a generalisation of this type of operators, as the one introduced in \cite{mokhov1990non,ferapontov1991differential}, in which, notably, the coefficients $g_{ij}$ define a metric of constant curvature.}) is indeed a special case of our construction.  {We note that it is a non-trivial matter to take the continuous limit in the general setup discussed in \cite{vergallo2024hamiltonian}, although we expect such a limit to reproduce our more general construction, in its force-less and spatially local specialisation.}

The VF construction holds for the extended polychromatic reductions of the conventional GHD equation \eqref{PolyDiagExt1}-\eqref{PolyDiagExt2}, under the constraint that one may choose a parametrisation $\theta$ such that the scattering shift takes the form $\varphi(\theta,\theta') = a(\theta)a(\theta') g (b(\theta)-b(\theta'))$ for arbitrary functions $a,b$ and even function $g$. By contrast, our construction works directly for the non-reduced GHD equation, and under the constraint that $\varphi(\theta,\theta') = \p_\theta \phi(\theta,\theta')$ where $\phi(\theta,\theta') = -\phi(\theta',\theta)$ is odd. Notwithstanding the reduction, the constraints of the GHD data appear to be non-related; however it is a simple matter to check that all special cases considered in \cite{vergallo2024hamiltonian} are also covered by our construction. 

The hard rod gas corresponds to the special choice of GHD data \eqref{hrdata}. In section 3.2 of their paper, VF present the Hamiltonian structure of the infinite hydrodynamic chain generated by the spectral moments of the fluid density (keeping their notation)
\begin{equation}
    A^m(x) = \int \dd \theta\,\theta^m\rho(x,\theta).
\end{equation}
Note that they construct their Hamiltonian structure with respect to the Hamiltonian density $h_{\rm VF} = -A^2/2$. Traditionally, in physics, the Hamiltonian of a system is associated with its energy, the density of which is $h = A^2/2$ in the hard rod case. As we went with the conventions of physics, in order for our and VF's construction to agree, the two proposed Poisson brackets must differ by a sign. 

We will now show, by direct computation, that our Poisson bracket specialises to the one proposed by VF in \cite{vergallo2024hamiltonian} in the case of conventional GHD with choice of data \eqref{hrdata}. To that end, we compute the Poisson bracket between two arbitrary spectral moments: using the fundamental bracket \eqref{pbintro} of conventional GHD, it is a simple matter to write
\begin{equation}
    \{A^m(x),A^n(x')\} = B^{mn}(x)\delta(x-x') \ ,
\end{equation}
where $B^{mn}(x)$ is the following differential operator
\begin{equation}
    B^{mn}(x) = -\frc1{2\pi} \Big[
    \p_x\, \int \dd\theta\,h_m^{\rm dr}(x,\theta) \p_\theta h_n(\theta)\,n(x,\theta)
    + \int \dd\theta\,\p_\theta h_m(\theta)  h_n^{\rm dr}(\theta)\,n(x,\theta)\,\p_x
    \Big] \ ,
\end{equation}
and $h_m(\theta) = \theta^m$. In the case of the hard rod gas, it is well known that the quantities of interest simplify as follows:
\begin{equation}
    2\pi\rho_{\rm s}(x,\theta) = 1-A^0(x),\qquad f^{\rm dr}(x,\theta)
    = f(x,\theta) - a \int_\R \dd\theta f(x,\theta)\rho(x,\theta) \ ,
\end{equation}
and, therefore, $h_m^{\rm dr}(x,\theta) = \theta^m - a A^m(x)$. Putting all of these together, we end up with an explicit expression for the Hamiltonian operator in terms of the moments
\begin{equation}\label{Bnous}
    B^{mn}(x) = -\Big[
    \p_x \frc{n}{1-aA^0}\big(A^{m+n-1} - a A^m A^{n-1}\big)
    + \frc{m}{1-aA^0} \big(A^{m+n-1} - a A^n A^{m-1}\big)\p_x
    \Big] \ .
\end{equation}

Conversely, in their paper, VF obtain the Poisson bracket
\begin{equation}
    \{A^m(x),A^n(x')\}_{\rm VF} = B_{\rm VF}^{mn}(x)\delta(x-x') \ ,
\end{equation}
with the Hamiltonian operator (indices being encoded in matrix form)
\begin{equation}\label{BVF}
    B_{\rm VF}(x) = J \Big( \frc1{1-aA^0} U\p_x + \p_x U^T \frc1{1-aA^0}\Big)J^T
    + \frc{a}{1-aA^0}\big(PA_x^T - A_xP^T\big) \ ,
\end{equation}
where
\begin{equation}\begin{aligned}
    U^{mn} &= m A^{m+n-1} \ ,\\
    J^{mn} &= \delta_{mn} - aA^m \delta_{n,0} \ ,\\
    P^m &= mA^{m-1}\ .
\end{aligned}
\end{equation}
In order to compare the operators $B$ and $B_{VF}$, we bring both in the standard form \eqref{Bnous}, and find that, indeed,
\begin{equation}\begin{aligned}
    B^{mn}_{\rm VF} &= \frc1{1-aA^0}\big(
    (m+n)A^{m+n-1} - amA^{m-1}A^n-anA^{n-1}A^m\big)\p_x
    + \\ &\quad +\,
    \frc{n}{1-aA^0}(A^{m+n-1}_x-aA^mA^{n-1}_x - a A^m_x A^{n-1})
    + \\ &\quad +\,
    \frc{anA^0_x}{(1-aA^0)^2}\big(
    aA^mA^{n-1} - A^{m+n-1}\big) \\
    & = -B^{mn}\ .
\end{aligned}
\end{equation}
Hence, the Hamiltonian structure associated with the hard rod gas VF constructed corresponds to a special (conventional) case of our general result on the Hamiltonian structure of the GHD equation. That is arguable, of course, but we believe our formulation to be slightly simpler than VF's, provided one is comfortable with the dressing formalism. Our field theoretic formulation also seems to be more direct, as it deals explicitly with the GHD equation, rather than with the infinite hydrodynamic chain characterised by the spectral moments of the fluid density. It would be interesting to understand the relation between the two constructions more generally in conventional GHD, beyond the hard rod gas, and to construct the analogue of the second Hamiltonian structure they presented.

\section{Discussion and perspectives}\label{sec:disc}

In this paper, we introduced a new, more general (spatially extended) form of the GHD equations \eqref{ghdsum}-\eqref{veff}-\eqref{aeff}, which accounts for space-dependent external potentials, space-dependent momenta and spatially extended scattering kernels. We showed that, even in its spatially extended form, the GHD equations constitute a Hamiltonian system (technically, under the sufficient but not necessary condition \eqref{smoothcompact} regarding the functional space of fluid densities). We constructed the associated Poisson structure on an appropriate algebra of observables $\mathfrak U$, for which we took inspiration from the notion of functional of hydrodynamic type \cite{dubrovin1989hydrodynamics}. We further showed that the GHD equations admit an infinite set of linearly independent (extensive) conserved quantities that are in mutual involution in the sense of Liouville. Although this is no proof {, as we currently do not know if the Hamiltonian flows generated by the conserved quantities we highlight are complete}, this strongly suggests that the GHD equations might be integrable. And, in fact, integrability of the conventional GHD equations had been previously investigated under the so-called ``polychromatic reduction'' {, for which a bi-Hamiltonian structure has been evinced \cite{vergallo2023hamiltonian}}. As such, we provided a brief overview of the previous results and of the framework in which they were obtained: the theory of systems of hydrodynamic type. Clarifying the precise connection between systems of hydrodynamic type and GHD, beyond its polychromatic reductions, would be useful; especially since the formalism used by those two theories is fairly different. We believe that we have taken a first step towards this goal by showing that our results agree with those presented in the recent preprint \cite{vergallo2024hamiltonian}.

In this Section, to conclude our discussion, we go over the potential extensions of our construction, and further reflect on the connection between GHD and systems of hydrodynamic type. For clarity, we will examine those two aspects in different subsections.

\subsection{More general dynamical spaces}\label{sec:MoreGenDomain}

In section \ref{sec:FluDen}, we introduced the spectral phase space $\Lambda = \mathcal L \times \mathcal P$ but, in the rest of the text we restricted ourselves to the case $\Lambda =\R^2$. Moreover we have assumed that the fluid density quickly vanishes in unbounded directions, i.e. $\mathcal M \subset \mathcal C^\infty_0(\R^2)$. However, even discarding boundary conditions in space, there are many physically relevant situations in which  $\mathcal P$ may be a more general manifold, that may include many disconnected components (representing, for instance, many particle types; see the GHD description of this situation in \cite{doyon2017note}), be of higher dimension \cite{el2020spectral}, or have the topology of the circle as in quantum spin chains (see e.g.~\cite{takahashibook}). A natural extension of our work would be to consider more general dynamical spaces $\mathcal M$. For instance, one may consider the additional cases
\begin{itemize}
    \item[i)] the fluid density is periodic in physical space but quickly vanishes in spectral space:
    \begin{equation}
     \mathcal L = [x_0,x_0+ \mathsf L] \ , \quad \mathcal P = \mathbb R \ ,  \quad \rho(x,\theta) = \rho(x+\mathsf L,\theta) \ , \quad \lim_{|\theta|\to \infty} \rho(x,\theta) = 0 \ ;
\end{equation}
    \item[ii)] the fluid density quickly vanishes  in physical space but is periodic in spectral space:
    \begin{equation}
     \mathcal L = \mathbb R \ , \quad \mathcal P = [\theta_0, \theta_0+ \mathsf P] \ , \quad  \rho(x,\theta) = \rho(x,\theta + \mathsf P) \ , \quad \lim_{|x|\to \infty} \rho(x,\theta) = 0  \ ; 
     \end{equation}
     \item[iii)] the fluid density is periodic in both physical and spectral space:
    \begin{equation}
     \Lambda = [x_0,x_0+ \mathsf L]\times[\theta_0, \theta_0+ \mathsf P] \ ,  \quad \rho(x,\theta) = \rho(x+\mathsf L,\theta) \ , \quad  \rho(x,\theta) = \rho(x,\theta + \mathsf P) \ ;
     \end{equation}
\end{itemize}
where we give periodic directions the topology of the circle and where the condition \eqref{smoothcompact} becomes
\begin{equation}\label{smoothcompactLambda}
    \mathcal M = \{|\rho|\leq \rho_*: \rho\in \mathcal C^\infty_c(\Lambda)\}\ .
\end{equation}
One may even consider that the fluid density lies on a domain  $\Lambda \subset \mathbb R^2$ (which may not be fully connected), taking arbitrary values on its boundary. This last case is useful in GHD, when considering zero-entropy states \cite{doyon2017large}, and in soliton gases \cite{el2021soliton}, notably when dealing with the so-called soliton condensates \cite{congy2023dispersive}. This may be seen as a singular limiting case of the situation discussed in the main text of this paper, where the fluid density on $\R^2$ has less and less regularity at the boundary of $\Lambda$. Importantly, in this case, the domain $\Lambda$ is in general dynamical (in the sense that it depends on time).

Note that, even in the cases (i), (ii) and (iii) we do not require the GHD data to be periodic. In fact, we may rather assume $E\in \mathcal C^\infty(\Lambda)$ and $\Omega$ smooth on the universal cover of $\Lambda$ such that its mixed derivative $\Omega_{x\theta}$ is continuous on $\Lambda$ (hence periodic in compact directions in cases (i)-(iii)). Using the notation $\p_x^{-1}\mathcal C^\infty(\Lambda),\, \p_\theta^{-1}\mathcal C^\infty(\Lambda)$, etc., in order to represent pre-images of derivatives on the universal cover of $\Lambda$, we require
\begin{equation}
    \Omega\in \p_x^{-1}\p_\theta^{-1}\mathcal C^\infty(\Lambda) = \{f\in \mathcal C^\infty(\R^2):f_{x\theta}\big|_{\Lambda} \in\mathcal C^\infty(\Lambda)\}.
\end{equation}
Similarly, the condition \eqref{psimap} on the interaction kernel should be changed into, respectively for the different cases $u =\psi , \ \psi_{\tilde x}, \ \psi_{\tilde \theta}$ or $\psi_{\tilde x\tilde\theta}$,
\begin{equation}\label{psimapextended}
    \int_\Lambda \dd x'\dd\theta'\,u(\cdot,\cdot;x',\theta')f(x',\theta')
    \ \in \ \p_x^{-1}\p_\theta^{-1}\mathcal C^\infty(\Lambda),\; \p_\theta^{-1}\mathcal C^\infty(\Lambda),\; \p_x^{-1}\mathcal C^\infty(\Lambda),\;\mbox{or}\
    \;\mathcal C^\infty(\Lambda)\; ,
\end{equation}
and for every $f\in\mathcal C_c^\infty(\Lambda)$. 

Our construction still holds in the cases (i) and (ii): our derivation is based on the change of metric giving the normal fluid density, either in real space Eq.~\eqref{metchange} or  in spectral space Eq.~\eqref{specmetchange}. In the case (i), we may use Eq.~\eqref{specmetchange}, while in the case (ii), we may use Eq.~\eqref{metchange}, and the derivation is unchanged. Thus \eqref{pb} still is a valid Poisson bracket and \eqref{Hmain} still generates the GHD equation, with integrations over $\Lambda$ instead of $\R^2$.

However, in the case (iii) our derivation appears to have technical problems. The change of metric may still be done in periodic directions. In particular, the property \eqref{psimapextended} of the interaction kernel guarantees that the change of coordinates \eqref{metchange} introduced in Section \ref{sec:Ham} remains well-defined. Indeed, $Y\in\p_x^{-1}\mathcal C^\infty(\Lambda)$ -- at least in the setup of Eq.~\eqref{smoothcompactLambda}, see Appendix \ref{appDressing} -- which is the right space for a change of coordinates. As before, under this change of coordinates, the space $\mathcal L$ is mapped onto $\hat{\mathcal L}_\theta$ which depends on $\theta$ (and also, in general, on the fluid density), and thus $(y,\theta)$ lies on $\hat \Lambda = \bigsqcup_{\theta \in\mathcal P} \hat{\mathcal L}_\theta$. If $\Omega_{x\theta}>0$, $\rho_{\rm s}$ is strictly positive and $Y(x,\theta)$ is monotonic in $x$, so it maps $\mathcal L$ onto $\h{\mathcal L_\theta}$ as a subset of $\R$ (or all of $\R$), which is then given the induced topology; on this subset its inverse $X(\cdot,\theta)$ exists
\beq\label{Xdefinition}
	X:\hat\Lambda \to \Lambda,\ (y,\theta)\mapsto X(y,\theta)\quad :\quad Y(X(y,\theta),\theta) = y.
\eeq
Otherwise, $Y$ is smooth but not necessarily monotonic, and $\h{\mathcal L_\theta}$ is, as before, taken as a multiple cover of a subset of $\R$; then $Y(\cdot,\theta)$ is a smooth diffeomorphism $\mathcal L\to \h{\mathcal L_\theta}$, with $X(\cdot,\theta):\h{\mathcal L_\theta}\to \mathcal L$ its inverse. As before, one must require $\hat{\mathcal L}_\theta$ not to depend on the fluid density (see Remark \ref{rem:yspace}).

However, the lack of periodicity of the interaction kernel affects the adjoint operator $\mathcal O^\dag$ in Appendix \ref{app:varfree}, as this  requires integration by part. Our derivation fails at that point. Similar technical issues would occur when trying to show the conserved quantities of the GHD are in involution in Section \ref{sec:consQ}, Eqs.~\eqref{invol2}-\eqref{invol} in particular. We believe \eqref{pb} still is a valid Poisson bracket, but we leave a more in depth analysis for future work.

\subsection{GHD as a system of hydrodynamic type}\label{GHD_hydro}

Our Hamiltonian formulation of the spatially extended GHD equation led us to consider the linearised Poisson bracket \eqref{pbfreeform}. 
On functionals, it is the specialisation to $\Omega_{x\theta}=1$ and $\Psi=0$ of \eqref{pb}, and thus can be written\footnote{Here we assume $\hat{\mathcal L}_\theta = \R$ for simplicity; see Remark \ref{rem:yspace}.} as 
\begin{eqnarray}
    \{F,G\} &=& \int_{\R^2} \dd y\dd\theta\dd y'\dd \theta '\,\frc{\delta F}{\delta\h\rho(y,\theta)}\{\h \rho(y,\theta),\h\rho(y',\theta')\}
	\frc{\delta G}{\delta\h\rho(y',\theta')}\nonumber\\
 &=&\int_{\R^2} \dd y\dd\theta\,\h\rho(y,\theta)\,\left(\partial_y\frc{\delta F}{\delta\h\rho(y,\theta)}
	\partial_{\theta}\frc{\delta G}{\delta\h\rho(y,\theta)}-\partial_\theta\frc{\delta F}{\delta\h\rho(y,\theta)}
	\partial_{y}\frc{\delta G}{\delta\h\rho(y,\theta)}\right)\,.\label{eq:pbyfunctionals}
 \end{eqnarray}
This is a type of non-canonical Poisson bracket known as a Lie-Poisson bracket \cite{thiffeault1998invariants}, which is rather well-known \cite{ZakharovHamil2}. It is a special case of a Poisson bracket first proposed in \cite{morrison1980maxwell} for the Maxwell-Vlasov equations. The Jacobi identity for the general bracket of \cite{morrison1980maxwell} did not hold (this was fixed in \cite{marsden1982hamiltonian}, see also the comment \cite{weinstein1981comments}) but it did anyway for the special case which is of interest for us here. In any case, we prove the Jacobi identity directly in Appendix \ref{ssectJacobi}. Other contexts in which  the same Poisson bracket also appears include the Benney equation \cite{zakharov1981benney,kupershmidt1987hydrodynamical} (see also \cite{chesnokov2017stability} for a study of its reductions via the theory of hydrodynamic type systems), the two-dimensional ideal fluid \cite{piterbarg1995poisson} or the Kida vortex \cite{jayawardana2022clebsch}. Even some stochastic extensions have been considered, see for instance the recent paper \cite{luesink2024casimir}.

Note that we can also write
\beq
    \{F,G\}=\int_{\R^2} \dd y\dd\theta\,{\cal K}(y,\theta)\,\frc{\delta F}{\delta\h\rho(y,\theta)}\otimes
	\frc{\delta G}{\delta\h\rho(y,\theta)} \; ,
 \eeq
in terms of the Hamiltonian operator 
\beq
{\cal K}(y,\theta)=\h\rho(y,\theta)\,\partial_y\wedge\partial_{\theta}\,.
\eeq
In this last form, it is of course reminiscent of the Dubrovin-Novikov hydrodynamic bracket \cite{dubrovin1983hamiltonian}, Eq.~\eqref{PBDN},  and its various generalisations \cite{mokhov1988dubrovin,mokhov2008classification,mokhov1998symplectic}. Despite the fact recalled above that this Poisson bracket has been known for a long time, to the best of our knowledge, it seems that its Hamiltonian operator falls outside the many investigations of Hamiltonian operators for PDEs of hydrodynamic type. The various generalisations of the original Dubrovin-Novikov bracket can be labelled by three integers: $N$ for the number of fields, $n$ for the number of independent variables and $d$ for the highest order of the differential operators involved. The case $n=d=1$, $N$ arbitrary corresponds to the original works by Dubrovin and Novikov. The cases $N$ and $n$ arbitrary have only been considered and (partially) classified when $d=1$, and the cases $N$ arbitrary and $d>1$ have only been considered when $n=1$, see e.g. \cite{mokhov2008classification,vergallo2021homogeneous,ferapontov2014projective,ferapontov2016towards,vergallo2023projective} and references therein. The case at hand here corresponds to $N=1$, $n=2$ and $d=2$. We introduced it with a concrete motivation in mind, and it appears in a rather special form, but we believe it could stimulate research on the case $d\ge 2$ (with $n, \ N$ arbitrary). In fact, there is an alternative, tantalising way of visualising \eqref{pbfreeform} as a large $N$ limit of a $n=1$, $d=1$ Jordan block type Hamiltonian operator as described in \cite{pavlov2012generalized,ferapontov2022kinetic,vergallo2023hamiltonian,vergallo2024hamiltonian}: first because of the equivalence highlighted in Section \ref{appcomparison}, second because the $\delta'(\theta-\theta')$ terms in the Poisson brackets could be interpreted as coupling nearest neighbours in spectral space, in which case we would expect $g_{ij}$ in Eq.~\eqref{PBDN} to be block diagonal. At this point, this is admittedly rather formal, but we believe properly investigating the connection between those interpretations may prove fruitful.

\subsection{Lagrangian (multiform) formulation}

Systems of hydrodynamic type are known to have profound connections with differential geometry. It is therefore not surprising that they are amenable to Lagrangian formulations which offer variational counterparts of their Hamiltonian formulation (when they have one). For the linearised Poisson bracket \eqref{pbfreeform} and a linear Hamiltonian as in \eqref{Hmain}, a Lagrangian formulation is obtainable from results of \cite{dharetal1993}; this can in principle be rewritten in terms of the (interacting) fluid density $\rho$. However, such traditional variational formulations know nothing of possible integrable structures. There have been attempts at capturing integrability in a Lagrangian formulation by emulating the bi-Hamiltonian picture and proposing the notion of bi-Lagrangian systems, see \cite{mokhov1998symplectic}. More recently, the notion of Lagrangian multiforms, originally introduced in \cite{lobb2009lagrangian}, has emerged as a variational framework for integrability. It has been developed for almost all incarnations of integrable systems -- discrete and continuous, finite and infinite dimensional -- but not yet for systems of hydrodynamic type. Lagrangian multiform theory for continuous infinite dimensional systems in $1+1$d \cite{suris2016lagrangian,sleigh2019variational,sleigh2020variational,caudrelier2021multiform,caudrelier2024classical} and $2+1$d \cite{sleigh2023lagrangian} is relatively well understood so it is natural and desirable to try to cast the GHD equation treated in this paper (which lives in $2+1$d), and some of its reductions (which live in $1+1$d), into this framework. We believe this alternative approach could shed additional light on the integrability features of GHD.

The Hamiltonian formulation we have provided, and also potential Lagrangian formulations (conventional or multiform), should help shed light on a number of other problems related to GHD, such as its quantisation (e.g.~is there a dressed Moyal bracket?), and the characterisation of random GHD configurations as may occur in possible turbulent regimes \cite{biagettin2024therma}.

\section*{Acknowledgments}

BD is grateful to Alexander Abanov for discussions and for pointing out important references, to Olalla Castro-Alvaredo for discussions, and to the Nor-Amberd School in Theoretical Physics (21-30 May 2024, Tsaghkadzor, Armenia) where interesting lectures were given and useful exchanges occurred. TB and VC wish to acknowledge useful discussions with Gennady El, Evgeny Ferapontov, Pierandrea Vergallo and Antonio Moro who also pointed out useful references. BD, TB and VC would like to thank the CIRM for their support and hospitality during the workshop “Emergent Hydrodynamics of Integrable Systems and Soliton Gases” when preliminary work for this manuscript was undertaken. TB and BD were supported by the Engineering and Physical Sciences Research Council (EPSRC) under grant EP/W010194/1. 

\section*{Compliance with Ethical Standards}
The authors have no conflicts of interest to declare. All coauthors have seen and agree with the contents of the manuscript and there is no financial interest to report.

\appendix

\section{Hamiltonian system of particles and its hydrodynamic limit}\label{appparticles}
This Appendix generalises the construction made in \cite{ttbar1,ttbar2} to space-dependent energy functions (that is, to the addition of force terms due to external space-varying fields coupled to conserved densities), and to more general interaction kernels. The derivation of the spatially extended GHD equation is also based on a different scaling of the interaction kernel that we present below.

Although we omit the explicit calculation, we mention that the models we present in fact correspond to the {\em atomic reduction} of the (spatially extended) GHD equation, making the assumption \eqref{classicalmechanics} with $x_i=x_i(t)$ and $\theta_i=\theta_i(t)$ in the GHD equation \eqref{ghdsum}.

\subsection{Definition of the system}
We define a Hamiltonian system on phase space $\R^{2N}$ as follows. Consider the function
\begin{equation}
    \Phi(\bs x,\bs \theta) = \sum_{i=1}^N \Omega(x_i,\theta_i)
    + \frc12 \sum_{i\neq j} \psi(x_i,\theta_i;x_j,\theta_j)
\end{equation}
as a generating function for a canonical transformation between two sets of canonical coordinates on $\R^{2N}$
\begin{equation}
    (\bs x,\bs p) \leftrightarrow (\bs y,\bs \theta) \ ,
\end{equation}
where we used a bold font to denote vectors, e.g. $\bs x = (x_i)_{i=1}^N$. Given the generating function $\Phi$ our new set of canonical coordinates takes the form
\begin{align}\label{yieq}
    y_i &= \p_{\theta_i} \Phi = \Omega_\theta(x_i,\theta_i) +\sum_{j\neq i}
    \psi_\theta(x_i,\theta_i;x_j,\theta_j) \\
    p_i &= \p_{x_i} \Phi = \Omega_x(x_i,\theta_i) + \sum_{j\neq i}
    \psi_x(x_i,\theta_i;x_j,\theta_j),
    \label{pieq}
\end{align}
where we used the symmetry of $\psi$, Eq.~\eqref{psisymmetry}. Here and below we use indices $x$ and $\theta$ to denote derivatives with respect to the spatial and spectral variable, respectively. The two sets of coordinates $(\bs x,\bs p)$ and $(\bs y,\bs \theta)$ being canonical, we naturally have the fundamental relations
\begin{equation}
    \{x_i,x_j\} = \{p_i,p_j\} = \{y_i,y_j\}=\{\theta_i,\theta_j\}=0 \ ,
\end{equation}
and 
\begin{equation}
    \{x_i,p_j\} = \{y_i,\theta_j\}=\delta_{ij} \ .
\end{equation}
Additionally, we define our Hamiltonian, in terms of the energy function $E\in \mathcal C^{\infty}(\R^2)$, as
\begin{equation}\label{particleH}
    H = \sum_{i=1}^N E(x_i,\theta_i).
\end{equation}

The coordinates $(\bs x,\bs p)$ represent the real particle coordinates, while, following the nomenclature of \cite{ttbar1,ttbar2}, $(\bs y,\bs \theta)$ can be referred to as ``asymptotic coordinates". However, in general, the asymptotic analysis of \cite{ttbar1,ttbar2} does not hold here, as the Hamiltonian is not solely a function of $\bs \theta$. Further, the model \eqref{particleH} is not necessarily integrable. However, if $E(x,\theta) = E(\theta)$ is independent of $x$, then we see that each $\theta_i$ is conserved and $y_i$ evolves linearly in time. In this case, with appropriate asymptotic properties on $\psi$, $\bs\theta$ indeed represent the asymptotic momenta and $\bs y$ are simply related to impact parameters (in the sense of \cite{doyon2020lecture}). Then, the model is integrable, by an extension of the analysis made in \cite{ttbar1,ttbar2} (which we omit here).

\subsection{Properties of the generating function $\Phi$}
In order for Eq.~\eqref{particleH} to be a well-defined function on the real phase space $(\bs x,\bs p)$, and for \eqref{yieq} and \eqref{pieq} to give rise to a well-defined canonical transformation, Eqs.~\eqref{yieq} and \eqref{pieq} need to be intertible for $\bs x$ and $\bs\theta$, respectively. This holds if the following matrix is positive definite for all $\bs x,\bs\theta$:
\begin{equation}\label{defGaudin}
    \p_{x_i}\p_{\theta_j} \Phi = \Gamma_{ij}(\bs x,\bs\theta) = \Big(\Omega_{x\theta}(x_i,\theta_i) + \sum_{k\neq i}\psi_{x\theta}(x_i,\theta_i;x_k,\theta_k)\Big)\delta_{ij}
    + \psi_{x\theta'}(x_i,\theta_i;x_j,\theta_j)(1-\delta_{ij})
\end{equation}
where, following the notations of the main text, here and below we use 
\begin{equation}
    \psi_{x\theta'}(a,d;c,d) = \p_{x}\p_{\theta'}\psi(x,\theta;x',\theta')|_{(a,b;c,d)} \ ,\mbox{ etc.}
\end{equation}
Note that the matrix $\Gamma$ can be interpreted as an analogue of the Gaudin matrix in the theory of quantum integrable systems, the determinant of which is the Jacobian of the transformation that maps the quasi-momenta to the quantum numbers (c.f. Sections 4.3.2 and 4.3.3 of \cite{gaudin2014bethe}). The positive definite condition $\sum_{ij} \Gamma_{ij}v_i v_j>0$ (for all $\bs v\in \R^N$ with $|\bs v|=1$) amounts to
\begin{equation}\label{condition}
    \sum_{i=1}^N v_i^2 \Omega_{x\theta}(x_i,\theta_i) + \frc12\sum_{j\neq i}\left[
    (v_i^2 + v_j^2) \psi_{x\theta}(x_i,\theta_i;x_j,\theta_j)
    + 2v_iv_j \psi_{x\theta'}(x_i,\theta_i;x_j,\theta_j)\right]>0 \ ,
\end{equation}
thanks to the symmetry \eqref{psisymmetry}. For instance, this holds if $\Omega_{x\theta}, \ \psi_{x\theta}>0$ and $|\psi_{x\theta'}|\leq \psi_{x\theta}$, which includes the cases considered in \cite{ttbar1,ttbar2}. The result is a smooth diffeomorphism of $\R^2$ if $\Omega$ and $\psi$ are smooth. Alternatively, a more compact way to write this condition is
\begin{equation}
    \sum_{i=1}^N \hat v_i^x\hat v_i^\theta \Omega(x_i,\theta_i) + \frc12\sum_{j\neq i}
    (\hat v_i^x + \hat v_j^x)
    (\hat v_i^\theta + \hat v_j^\theta)
    \psi(x_i,\theta_i;x_j,\theta_j) >0 \ ,
\end{equation}
where $\h v_i^x = v_i \p_{x_i}$ and $\h v_i^\theta = v_i \p_{\theta_i}$.
A full analysis of the set of $\Omega$ and $\psi$ giving invertibility of \eqref{yieq} and \eqref{pieq} would be interesting, but is beyond the scope of this paper. Below we assume that the transformation \eqref{yieq}, \eqref{pieq} is a smooth diffeomorphism of $\R^2$.

\subsection{Equations of motion}
Denote $\dot y_i|_{\bs\theta}=\sum_j\p_{x_j}\p_{\theta_i}\Phi\,\dot x_j$ (that is, the time derivative of the right-hand side of \eqref{yieq} keeping $\bs\theta$ fixed), and $\dot p_i|_{\bs x}=\sum_j\p_{x_i}\p_{\theta_j}\Phi\,\dot \theta_j$. Then we show below that the equations of motion generated by $H$, that is $\dot x_i = \{x_i,H\},\,\dot p_i = \{p_i,H\}$, and $\dot y_i = \{y_i,H\},\,\dot\theta_i = \{\theta_i,H\}$, are equivalent to
\begin{equation}\label{ypdoteq}
    \dot y_i|_{\bs\theta} = E_\theta(x_i,\theta_i),\quad
    \dot p_i|_{\bs x} = -E_x(x_i,\theta_i).
\end{equation}
Indeed, by definition of $H$ \eqref{particleH}
\begin{equation}
    \dot y_i = \{y_i,H\} = E_\theta(x_i,\theta_i) + \sum_{j=1}^N \{y_i,x_j\}E_x(x_j,\theta_j) \ .
\end{equation}
Then, recalling the definition \eqref{yieq}, $y_i = \p_{\theta_i}\Phi(\bs x,\bs\theta)$, we have
\begin{equation}
    \{y_i,x_j\} = \sum_{l=1}^N \p_{\theta_i}\p_{\theta_l}\Phi(\bs x,\bs\theta) \{\theta_l,x_j\} \ ,
\end{equation}
and therefore
\begin{equation}
    \sum_{j=1}^N \{y_i,x_j\}E_x(x_j,\theta_j)
    =
    \sum_{l=1}^N \p_{\theta_i}\p_{\theta_l}\Phi(\bs x,\bs\theta) \{\theta_l,H\}
    =
    \sum_{l=1}^N \p_{\theta_i}\p_{\theta_l}\Phi(\bs x,\bs\theta) \dot\theta_l \ .
\end{equation}
As $\dot y_i=\sum_j\p_{x_j}\p_{\theta_i}\Phi\,\dot x_j + \sum_j\p_{\theta_j}\p_{\theta_i}\Phi\,\dot \theta_j$, we find the first equation of \eqref{ypdoteq}. A similar derivation starting with
\begin{equation}
    \dot p_i = \{p_i,H\} = -E_x(x_i,\theta_i) + \sum_{j=1}^N\{p_i,\theta_j\} E_\theta(x_j,\theta_j)
\end{equation}
gives the second equation of \eqref{ypdoteq}. As we assumed that the transformation of coordinates $(\bs x,\bs p)\leftrightarrow (\bs y,\bs\theta)$ is smooth  and invertible, then Eqs.~\eqref{veffparticles} (consequence of the first equation of \eqref{ypdoteq}) can be inverted for $\dot x_i$, and hence that equation along with the second equation of \eqref{ypdoteq} fully fix the dynamics. Hence, the inverse statement immediately holds. 

Given \eqref{ypdoteq}, we may eventually write the relations
\begin{align}\label{veffparticles}
    E_\theta(x_i,\theta_i) &=
    \Omega_{x\theta}(x_i,\theta_i)\dot x_i
    +\sum_{j\neq i}\big(\psi_{x\theta}(x_i,\theta_i;x_j,\theta_j) \dot x_i
    +
    \psi_{x'\theta}(x_i,\theta_i;x_j,\theta_j) \dot x_j\big) \ , \\
    -E_x(x_i,\theta_i) &=
    \Omega_{x\theta}(x_i,\theta_i)\dot \theta_i
    +\sum_{j\neq i}\big(\psi_{x\theta}(x_i,\theta_i;x_j,\theta_j) \dot \theta_i
    +
    \psi_{x\theta'}(x_i,\theta_i;x_j,\theta_j) \dot \theta_j\big) \ ,
    \label{aeffparticles}
\end{align}
which are reminiscent of Eqs.~\eqref{veff}-\eqref{aeff}.

\subsection{Hydrodynamic limit}
We now argue that, in an appropriate hydrodynamic limit, we recover the spatially extended GHD equation \eqref{ghdsum}. For this purpose, we introduce a large parameter $L$, such that in the limit $L\to\infty$ one recovers the hydrodynamic equation. The data $E, \ \Omega, \ \psi$ depend on $L$ as follows:
\begin{equation}
    E(x,\theta) = \b E(x/L,\theta) \ ,\quad
    \Omega(x,\theta) = L\b \Omega(x/L,\theta) \ ,\quad
    \psi(x,\theta;x',\theta')
    = \b\psi(x/L,\theta;x'/L,\theta') \ ,
\end{equation}
where $\b E,\b\Omega,\b\psi$ are independent of $L$. Consider the empirical density
\begin{equation}\label{rhoempirical}
    \rho(\b x,\theta) = L^{-1}\sum_{i=1}^N\delta(\b x-x_iL^{-1})\delta(\theta-\theta_i) \ .
\end{equation}
With $\b t = tL^{-1}$, this satisfies the continuity equation
\begin{equation}\label{continuityparticles}
    \p_{\b t} \rho(\b x,\theta) + \p_{\b x}\left(
    L^{-1} \sum_{i=1}^N \dot x_i \delta(\b x-x_iL^{-1})\delta(\theta-\theta_i)\right)
    + \p_\theta \left(
    L^{-1} \sum_{i=1}^N \dot \theta_i L\, \delta(\b x-x_iL^{-1})\delta(\theta-\theta_i)\right) = 0 \ .
\end{equation}
Note that under this scaling, the two terms in the generating function $\Phi$ scale uniformly,
\begin{equation}
    \Phi \to L^2\b \Phi[\rho] = L^2\left(\int_{\R^2} \dd \b x\dd\theta\,
    \rho(\b x,\theta)\b\Omega(\b x,\theta)
    +\int_{\R^{2}} \dd \b x\dd\theta\dd \b x'\dd\theta'\,
    \rho(\b x,\theta)\rho(\b x',\theta')\b\psi(\b x,\theta;\b x',\b\theta')
    \right)
\end{equation}
(the term with $i=j$ in the sum on the second term is  a subleading correction, scaling as $O(L)$).

In the limit $L\to\infty$, we assume that: (i) the number of particles increases proportionally, $N\propto L$; (ii) the positions are spread on a region of length $\propto L$; and (iii) the asymptotic momenta condense on a region of order $O(1)$. Thus, the interparticle distance scales as $O(1)$, while the separation between nearby momenta scales as $O(L^{-1})$. Then the empirical density \eqref{rhoempirical} tends, weakly, to a finite, nonzero function. Now let $\b x_i = x_i/L$ and assume that $\dot x_i$ can be written as a smooth function $v^{\rm eff}(\b x_i,\theta_i)$ of $\b x_i,\theta_i$ with a uniformly smooth limit as $L\to\infty$, and likewise $\dot \theta_iL$ can be written as a smooth function $a^{\rm eff}(\b x_i,\theta_i)$ (both functionals of $\rho$). This should be the case for good enough particle distributions under the above scaling (see for example \cite{spohn2012large}). Then Eq.~\eqref{veffparticles} becomes
\begin{align}
    &\b E_\theta(\b x_i,\theta_i)
    =\\
    &\b\Omega_{x\theta}(\b x_i,\theta_i)
    v^{\rm eff}(\b x_i,\theta_i)
    +
    \int_{\R^{2}} \dd \b x'\dd\theta'\,
    \rho(\b x',\theta')\left[
    \b\psi_{x\theta}(\b x_i,\theta_i;\b x',\theta') v^{\rm eff}(\b x_i,\theta_i)
    +
    \b\psi_{x'\theta}(\b x_i,\theta_i;\b x',\theta') v^{\rm eff}(\b x',\theta')\right] \ .
    \no
\end{align}
As the set of $(\b x_i,\theta_i)$'s becomes dense, this should hold for all $(\b x_i,\theta_i) = (\b x,\theta)\in\R^2$. This reproduces \eqref{veff} (omitting the over-bars). Similarly, \eqref{aeffparticles} becomes
\begin{align}
    &\b E_x(\b x_i,\theta_i)
    =\\
    &\b\Omega_{x\theta}(\b x_i,\theta_i)
    a^{\rm eff}(\b x_i,\theta_i)
    +
    \int_{\R^{2}} \dd \b x'\dd\theta'\,
    \rho(\b x',\theta')\left[
    \b\psi_{x\theta}(\b x_i,\theta_i;\b x',\theta') a^{\rm eff}(\b x_i,\theta_i)
    +
    \b\psi_{x'\theta}(\b x_i,\theta_i;\b x',\theta') a^{\rm eff}(\b x',\theta')\right]
    \no
\end{align}
and we recover \eqref{aeff}. Together with \eqref{continuityparticles}, we find that the hydrodynamic equation \eqref{ghdsum} indeed emerges under this scaling.

\section{Analysis of the dressing operation}\label{appAnalysis}

The dressing operation is defined in \eqref{dressing}. Here we show that there exists a non-negative function $\rho_*$ on $\R^2$, such that for any smooth function $f\in \mathcal C^{\infty}(\R^2)$, the dressed quantity $f^{\rm dr}$ (with all possible transformation types) is unique, well-defined and smooth on $\R^2$, for all $\rho\in\mathcal C^{\infty}_c(\R^2)$ with $|\rho|\leq \rho_*$ (i.e. for every $\rho\in\mathcal M$ with the dynamical space $\mathcal M$ defined as in Eq.~\eqref{smoothcompact}). This holds under certain technical conditions on $\Omega$ and on the interaction kernel $\psi$: Eqs.~\eqref{technicalOmega}, \eqref{technicalpsi}, \eqref{Psiabs} and \eqref{technicalderivative}. The techniques we use are based on (and slightly extend) those of \cite{HubnerDoyonUniqueness}.

As already mentioned in Remark \ref{rem:dressdef}, the conditions above for the dressed function to exist and be smooth are in no way necessary. They are fulfilled in the Lieb-Liniger gas \eqref{LLdata} and the hard-rod gas \eqref{hrdata}, as well as in the GHD of the family of models of Appendix \ref{appparticles}, but not in the KdV soliton gas \eqref{KdV}. Notably, different sets of sufficient conditions in the Lieb-Liniger model have been fully worked out in \cite{HubnerDoyonUniqueness}. Further, in the KdV soliton gas, even though the conditions are not met, the existence and uniqueness of $\Omega_{x\theta}^{\rm dr}$ and of $E_\theta^{\rm dr}$ have been proven in \cite{kuijlaars2021minimal}.

\subsection{Existence of dressed functions}

In order for dressed functions to exist, one needs to ensure the occupation function $n$, defined by Eq.~\eqref{defn}, exists and remains bounded. To that end, we first require that
\beq\label{technicalOmega}
	\Omega_{x\theta}(x,\theta) \neq 0,\quad (x,\theta) \in \R^2 \; ,
\eeq
and we also require that there exists a non-negative, locally integrable function $\mu$ on $\R^2$, such that the following bounds hold for all $(x,\theta)\in\R^2$:
\begin{equation}\label{technicalpsi}
    \int_{\R^2} \dd x'\dd\theta' \,|\psi_{x\theta}(x,\theta;x',\theta')|\,\mu(x',\theta')<|\Omega_{x\theta}(x,\theta)|\ .
\end{equation}
In this equation, the absolute value $|\psi_{x\theta}|$ of the mixed-derivative of the interaction kernel $\psi$ is involved; recall that we had only required \eqref{psimap} (along with \eqref{psisymmetry}). Thus, in particular, we require here that $|\psi_{x\theta}|$ is also an integral kernel. In fact, we also ask that this be the case for derivatives with respect to both $\t x = x,\,x'$ and both $\t \theta = \theta,\,\theta'$: for every positive continuous, compactly supported $\nu\in \mathcal C_c(\R^2)_+ = \{\nu\in\mathcal C_c(\R^2):\nu(x,\theta)>0\,\forall\,(x,\theta)\in\R^2\}$, there is $V\in\mathcal C(\R^2)$ with
\begin{equation}\label{boundV}
    \inf_{(x,\theta)\in\R^2} V(x,\theta)>0
\end{equation}
such that
\begin{equation}\label{Psiabs}
    \int_{\R^2} \dd x'\dd\theta' \,|\psi_{\t x\t\theta}(x,\theta;x',\theta')|\,\nu(x',\theta')<V(x,\theta) \ .
\end{equation}

The function $\mu$, and the pairs $(\nu,V)$, are not part of the data for the GHD equation itself, but influence the space $\mathcal M$ in which the fluid density $\rho$ must lie. We note that the bound \eqref{technicalpsi} is invariant under re-parametrisation, with $\mu$ of type $(1,1)$.

As we work under the assumption \eqref{psimap}, for all $\rho\in\mathcal C^{\infty}_c(\R^2)$ with $|\rho|\leq \mu$, we have smooth $\rho_{\rm s}(x,\theta)\neq 0\;\forall (x,\theta)\in\R^2$, hence the occupation function $n$, defined in \eqref{defn}, is also in $\mathcal C^\infty_c(\R^2)$. Moreover, we can show that, for any non-negative function $h$ on $\R^2$, there exists a non-negative function $\rho_*$ such that $|n|\leq h$ for all $|\rho|\leq \rho_*$. Indeed, let
\begin{equation}\label{defm}
    m(x,\theta) = \frc1{|\Omega_{x\theta}(x,\theta)|} \int \dd x'\dd\theta' \, |\psi_{x\theta}(x,\theta;x',\theta')|\,\mu(x',\theta')\ ,
\end{equation}
where $\mu$ is the function defined in the first condition of \eqref{technicalpsi}. The function $m$ exists and satisfies $0\leq m(x,\theta)< 1 , \;\forall (x,\theta)\in\R^2$ by definition. Then, take for instance
\begin{equation}\label{rhostarfunction}
    \rho_* = \rho_*[h] := \frc{1-m}{2\pi} \frc{|\Omega_{x\theta}|}{\sqrt{1+\Omega_{x\theta}^2}}\,
    \frc{\mu}{\sqrt{1+\mu^2}} \frc{h}{\sqrt{1+h^2}} < \mu \ ,
\end{equation}
which implies $2\pi|\rho_{\rm s}| \geq |\Omega_{x\theta}|(1-m)$, and therefore $|n|\leq h$ for any $|\rho|\leq \rho_*$.

Now, consider the dressing of a smooth function $f\in\mathcal C^{\infty}(\R^2)$, and suppose $n$ is supported on a compact $C\in\R^2$. One can then multiply the dressing equation satisfied by $f$, Eq.~\eqref{dressing}, by any continuous, positive function $w\in \mathcal C(\R^2)_+:=\{w\in\mathcal C(\R^2):w(x,\theta)> 0\,\forall\,(x,\theta)\in\R^2\}$, and look for a solution for $wf^{\rm dr}$ as a Liouville-Neumann series
\begin{equation}
    wf^{\rm dr} = \sum_{k=0}^\infty
    \Big(w\,\Psi_{\t x\t\theta}\,\frc{n}{w}\Big)^k wf\ ,
\end{equation}
where we recall that $\Psi$ is an integral operator defined in Eq.~\eqref{dressintop}. One can see that, if
\begin{equation}\label{conditiondressing}
    q:=\sup_{(x,\theta)\in C} w(x,\theta) \int_{\R^2} \frc{\dd x'\dd\theta'}{2\pi w(x',\theta')}\,|\psi_{\t x\t\theta}(x,\theta;x',\theta')|\,n(x',\theta') < 1 \; ,
\end{equation}
then the Liouville-Neumann series is absolutely convergent on $C$ provided
\begin{equation}
    ||f||_{w,C} = \sup_{(x,\theta)\in C} w(x,\theta) |f(x,\theta)|<\infty \; ,
\end{equation} 
since each term is then bounded by $q^k ||f||_{w,C}$. Thus Eq.~\eqref{dressing} has, on $C$, a unique solution on the Banach space $\mathcal B_w(C)$ of functions $f$ (on $\R^2$) such that $wf$ is bounded on $C$, with norm $||f||_{w,C}$. Then the dressing operation is indeed well-defined, and gives the same function (on $C$) for any $0<w' \leq w$ such that \eqref{conditiondressing} holds (with $w'$ in place of $w$). As $C$ is compact, $||f||_{w,C}<\infty$ for every $f\in \mathcal C^\infty(\R^2)$, and thus under condition \eqref{conditiondressing}, the dressing equation \eqref{dressing} has a unique solution for every $f\in \mathcal C^\infty(\R^2)$. By the bound \eqref{Psiabs} and the result expressed around \eqref{rhostarfunction}, we see that with
\beq\label{hw}
	h(x,\theta) = \frc{2\pi \chi_C(x,\theta)\nu(x,\theta)}{V(x,\theta)}\ ,\qquad
	w(x,\theta) = \frc1{V(x,\theta)},
\eeq
where $\chi_C$ is the characteristic function of the set $C$, the condition \eqref{conditiondressing} is satisfied for every $|n|\leq h$. Therefore, the dressing of any smooth $f$ exists on $C$ for every $|\rho|\leq\rho_*[2\pi\chi_C\nu/V]$. As $C$ is arbitrary, the dressing exists and gives a unique function on $\R^2$ for every $\rho\in \mathcal C^\infty_c(\R^2),\,|\rho|\leq\rho_*$ with
\beq\label{rhostar}
	\rho_* = (1-m)
	\frc{|\Omega_{x\theta}|}{\sqrt{1+\Omega_{x\theta}^2}}\,
    \frc{\mu}{\sqrt{1+\mu^2}} \frc{\nu}{\sqrt{V^2+(2\pi\nu)^2}}
\eeq
where we recall that $m$ is defined in \eqref{defm} and $\mu,\,\nu,\,V$ in \eqref{technicalpsi}. This gives the explicit space $\mathcal M$, Eq.~\eqref{smoothcompact}.

\subsection{Smoothness of dressed functions}

The result of the dressing $f^{\rm dr}(x,\theta)$ of a smooth function $f\in\mathcal C^\infty(\R^2)$ is, in general, a locally integrable function. For smoothness, we note that each term of the Liouville-Neumann series is smooth, hence we must simply show convergence of the series of derivatives. For simplicity, we analyse only the first derivatives in $x$ and $\theta$ (a similar analysis hold for higher derivatives), and we consider four situations:
\begin{equation}\label{technicalderivative}
\begin{aligned}
    \mbox{either } &\mbox{(i) $|\p_x \psi_{\t x\t\theta}|$ is locally integrable,}\\ 
    \mbox{or } &\mbox{(ii) $\p_x\psi_{\t x\t\theta} = -\p_{x'}\psi_{\t x\t\theta}$ ;}\\  
    \mbox{either } &\mbox{(iii) $|\p_\theta \psi_{\t x\t\theta}|$ is locally integrable,}\\ 
    \mbox{or } &\mbox{(iv) $\p_\theta\psi_{\t x\t\theta} = -\p_{\theta'}\psi_{\t x\t\theta}$ .}
\end{aligned}
\end{equation}
If $\psi_{\t x\t\theta}$ is a smooth function, then the first and third alternatives hold; if it is a distribution then, in general, they do not. However, in many examples, at least alternatives (ii) and/or (iv) hold. For instance, for local GHD, Eq.~\eqref{localGHD}, in the examples of the Lieb-Liniger gas \eqref{LLdata} and hard rod gas \eqref{hrdata}, alternatives (i) and (iv) hold. For the spatially extended GHD arising from the models of Appendix \ref{appparticles}, alternatives (i) and (iii) hold. The conditions \eqref{technicalderivative} are sufficient, but as mentioned not necessary, in order to have differentiability.

Consider the $x$ derivative for definiteness. On a generic term of the Liouville-Neumann series for $f^{\rm dr}$ this gives
\begin{equation}
    a_k = \p_x \Psi_{\t x\t\theta} n
    (\Psi_{\t x\t\theta} n)^k f\ ,
\end{equation}
where we assume that $n$ is supported on a compact $C\subset \R^2$. In the case (i) this is bounded as
\begin{equation}
\begin{aligned}
    |a_k(x,\theta)|&\leq\int_{\R^2} \frc{\dd x'\dd\theta'}{2\pi}
    |\p_x \psi_{\t x\t\theta}(x,\theta;x',\theta')|
    \frc{n(x',\theta')}{w(x',\theta')}
    ||(\psi_{\t x\t\theta} n)^k f ||_{w,C} \\
    &\leq 
    \int_{\R^2} \frc{\dd x'\dd\theta'}{2\pi}
    |\p_x \psi_{\t x\t\theta}(x,\theta;x',\theta')|
    \frc{n(x',\theta')}{w(x',\theta')}
    q^k
    ||f ||_{w,C}
\end{aligned}    \ ,
\end{equation}
using \eqref{conditiondressing}. Since $n/w\in\mathcal C_c(\R^2)$, the result of the integral is finite by \eqref{Psiabs} and, because $0\leq q<1$ under condition \eqref{conditiondressing}, the series converges.

In the case (ii), we have
\begin{equation}
    a_k = \sum_{j=0}^k
    (\Psi_{\t x\t\theta}n)^j \Psi_{\t x\t\theta} \p_{x'}n
    (\Psi_{\t x\t\theta}n)^{k-j} f
    + (\Psi_{\t x\t\theta}n)^k \p_{x'} f \ .
\end{equation}
The last term gives rise to a converegent series by the results already established, because $\p_{x'} f(x',\theta')$ is a smooth function of $x',\theta'$. We bound the first term for $(x,\theta)\in C$ as
\begin{equation}
\begin{aligned}
    |w (x,\theta)a_k(x,\theta)| \leq \sum_{j=0}^k \ q^j\ 
    \Bigg(\sup_{(x,\theta)\in C}
    w(x,\theta)\int_{\R^2} \frc{\dd x'\dd\theta'}{2\pi}\,
    |\psi_{\t x\t\theta}(x,\theta;x',\theta')|
    \frc{|\p_{x'} n(x',\theta')|}{w(x',\theta')}
    \Bigg)
    \;q^{k-j}\;||f||_w \ .
\end{aligned}
\end{equation}
By \eqref{boundV} and \eqref{hw}, the function $w$ is upper bounded, as is, in fact, $\p_x n/w\in\mathcal C_c(\R^2)$ by \eqref{hw}. Then the condition \eqref{Psiabs} implies that the result inside the parentheses is a finite quantity, say $b$. Therefore $|a_k(x,\theta)| \leq kq^k b||f||_w/|w (x,\theta)|$, and again the series converge.

\subsection{Dressing of distributions}

As per Remark \ref{rem:dressdist}, it is also convenient to define the dressing of distributions. In fact, dressed distributions are to be interpreted in terms of their effect when integrated against a compactly supported smooth test function $g$. For instance, when it comes to the fundamental Poisson bracket \eqref{pbnonlocalintro}, we can then make use of the symmetry relation \eqref{symm}, 
    \begin{equation}
    \int_{\R^2}\dd x'\dd\theta'\, \left[\delta'(\cdot-x)\delta'(\cdot-\theta)\right]^{{\rm dr}}(x',\theta')n(x',\theta')f(x',\theta') = - \partial_x\partial_\theta\left[n(x,\theta)f^{{\rm dr}}(x,\theta)\right] \ ,
\end{equation}
and, in this case, the right hand side is indeed well-defined. But, more generally, looking at the definition of the dressing \eqref{dressing}, let $f$ be a distribution and $g$, again, a compactly supported smooth function, we may write
\begin{equation}\label{dressdistrib}
\begin{aligned}
    \int_{\R^2}\dd x\dd\theta\, g(x,\theta)f^{{\rm dr}}(x,\theta) &= \int_{\R^2}\dd x\dd\theta\, g(x,\theta)f(x,\theta)\\ 
    & \ + \int_{\R^2}\dd x'\dd\theta'\,\left(\int_{\R^2}\dd x\dd\theta\, \psi_{\t x\t\theta}(x,\theta;x',\theta')g(x,\theta)\right)n(x',\theta')\,f^{\rm dr}(x',\theta')\\
    &= \int_{\R^2}\dd x\dd\theta \,g(x,\theta)f(x,\theta)\\ 
    & \ + \int_{\R^2}\dd x'\dd\theta'\,\left(\int_{\R^2}\dd x\dd\theta \psi_{\t x\t\theta}(x,\theta;\cdot,\cdot)g(x,\theta)\right)^{{\rm dr}}(x',\theta')\,n(x',\theta')\,f(x',\theta') \ .
\end{aligned}
\end{equation}
Because of the symmetry \eqref{psisymmetry} and the condition \eqref{psimap} on the integral kernel $\psi$, everything in \eqref{dressdistrib} is well-defined.

\section{Diagonalisation of the spatially extended GHD equation} \label{appnequation}

In this Section we show that the occupation function $n$ diagonalises the GHD equation \eqref{ghdsum}. To that end, we recall the definition of the occupation function \eqref{defn}, along with the relations \eqref{rhosOmega} and \eqref{CRA}, from which we may write the following identities
\begin{equation}\label{iddiag}
    \rho = \frc1{2\pi} n \Omega_{x\theta}^{~~\rm dr},\qquad v^{\rm eff}\rho = \frc1{2\pi} nE_\theta^{~{\rm dr}},\qquad
    a^{\rm eff}\rho = -\frc1{2\pi} nE_x^{~{\rm dr}}.
\end{equation}
The derivation works for $\Lambda=\R^2$, Eq.~\eqref{domain}, as well as for $\Lambda$ as in the cases (i), (ii) and (iii) of Subsection \ref{sec:MoreGenDomain}.

As per \eqref{datatype}, both $\Omega$ and $E$ are of type $(0,0)$, meaning $\Omega_{x\theta}$ is of type $(1,1)$, $E_\theta$ of type $(0,1)$ and $E_x$ of type $(1,0)$. From the dressing operation \eqref{dressing}, if $f\in \mathcal C^{\infty}(\R^2)$ is of type $(0,1)$, we have
\begin{eqnarray}
    \p_x f^{\rm dr} &=& f_x + \int_\Lambda \frc{\dd x'\dd\theta'}{2\pi} 
    (-)\psi_{x x'\theta}\,n(x',\theta')\, f^{\rm dr}(x',\theta')\n
    &=& f_x + \int_\Lambda \frc{\dd x'\dd\theta'}{2\pi}\,\psi_{x\theta}\, \Big(
    n(x',\theta')\, \p_{x'}f^{\rm dr}(x',\theta')
    +
    \p_{x'}n(x',\theta')\,f^{\rm dr}(x',\theta')\Big) \ ,
\end{eqnarray}
and therefore, in operatorial form,
\begin{equation}
    \p_x f^{\rm dr} = (1-\Psi_{x\theta}\,n)^{-1}\Big(f_x + \Psi_{x\theta}\, n_x\,f^{\rm dr}\Big)
    = f_x^{~\rm dr} + (1-\Psi_{x\theta}\,n)^{-1}\Psi_{x\theta}\,n_x\,f^{\rm dr}.
\end{equation}
If $f$ is instead of type $(1,0)$, then similarly
\begin{equation}
    \p_\theta f^{\rm dr}
    = f_\theta^{~\rm dr} + (1-\Psi_{x\theta}\,n)^{-1}\Psi_{x\theta}\,n_\theta\,f^{\rm dr}.
\end{equation}
Finally, if $f$ is of type $(1,1)$, then
\begin{equation}
    \p_t f^{\rm dr} = \p_t f + \int_\Lambda \frc{\dd x'\dd\theta'}{2\pi}
    \,\psi_{x\theta}\,\Big(n(x',\theta')\, \p_{t}f^{\rm dr}(x',\theta')
    +
    \p_{t}n(x',\theta')\,f^{\rm dr}(x',\theta')\Big) \ ,
\end{equation}
thus
\begin{equation}
    \p_t f^{\rm dr} = (\p_t f)^{\rm dr} + (1-\Psi_{x\theta}\,n)^{-1}\Psi_{x\theta}\,\p_t n\,f^{\rm dr}.
\end{equation}
Hence, substituting the identities \eqref{iddiag} in the GHD equation \eqref{ghdsum}, and noting that $\p_t \Omega_{x\theta}=0$, we obtain
\begin{equation}
    0 + nE_{x\theta}^{~~\rm dr} - nE_{x\theta}^{~~\rm dr} +
    \big((1-\Psi_{x\theta}\,n)^{-1}\Psi_{x\theta}+1\big)\,\Big(
    \Omega_{x\theta}^{\rm dr}\,\p_t n
    +
    E_\theta^{~\rm dr}\, \p_x n
    -
    E_x^{~\rm dr}\,\p_\theta n
    \Big) = 0 \ ,
\end{equation}
which leads to the diagonalised equation \eqref{ghdn} by inversion of the operator $(1-\Psi_{x\theta}\,n)^{-1}\Psi_{x\theta}+1$.

\section{Alternative dressing formulations} \label{appDressing}

Following and extending the general idea and notation introduced in \cite[Suppl Mat II.A]{ruggiero2020quantum}, it is possible to define different types of dressing operations that will prove useful in Appendix \ref{apppoisson}. First recall the usual ``{\bf dr}essing'' operation \eqref{dressing}
\begin{align}
    f^{\rm dr}(x,\theta) = f(x,\theta) + \int_\Lambda \frc{\dd x'\dd\theta'}{2\pi}\,
    \psi_{\t x\t\theta}(x,\theta;x',\theta')\, n(x',\theta')\,f^{\rm dr}(x',\theta') \ ,
\end{align}
where $\t x = x$ or $-x'$ and $\t \theta = \theta$ or $-\theta'$ as per the transformation type of $f$:
\begin{equation}
    \psi_{\t x\t\theta} =\lt\{
    \begin{aligned}
    & \psi_{x'\theta'}&& \mbox{($f$ is of type $(0,0)$)}\\
    & - \psi_{x'\theta}&& \mbox{($f$ is of type $(0,1)$)}\\
    & - \psi_{x\theta'}&& \mbox{($f$ is of type $(1,0)$)}\\
    & \psi_{x\theta}&& \mbox{($f$ is of type $(1,1)$).}
    \end{aligned}\rt.
\end{equation}
Similarly, we define the following alternative operations.
\begin{itemize}
    \item ``{\bf Dr}essing'' is
\begin{align}\label{Dressing}
    f^{\rm Dr}(x,\theta) = f(x,\theta) + \int_\Lambda \frc{\dd x'\dd\theta'}{2\pi}\,
    \psi_{\t x}(x,\theta;x',\theta')\,n(x',\theta') \, \p_{\theta'} f^{\rm Dr}(x',\theta') \ ,
\end{align}
for $f$ of type either $(0,0)$ or $(1,0)$ (with $\t x = -x'$ or $\t x = x$, resp.), that is, scalar in the spectral variable;

\item ``{\bf dR}essing'' is
\begin{align}\label{dRessing}
    f^{\rm dR}(x,\theta) = f(x,\theta) + \int_\Lambda \frc{\dd x'\dd\theta'}{2\pi}\,
    \psi_{\t \theta}(x,\theta;x',\theta')\,n(x',\theta') \, \p_{x'} f^{\rm dR}(x',\theta') \ ,
\end{align}
for $f$ of type either $(0,0)$ or $(0,1)$ (with $\t \theta = -\theta'$ or $\t \theta = \theta$, resp.), that is, scalar in the spatial variable;

\item ``{\bf DR}essing'' is
\begin{align}\label{DRessing}
    f^{\rm DR}(x,\theta) = f(x,\theta) + \int_\Lambda \frc{\dd x'\dd\theta'}{2\pi}\,
    \psi(x,\theta;x',\theta') \,n(x',\theta') \,\p_{x'}\p_{\theta'} f^{\rm DR}(x',\theta') \ ,
\end{align}
for $f$ of type $(0,0)$, that is, scalar in both variables. 
\end{itemize}
Alternatively, in terms of integral operators, we may write
\begin{equation}\label{intDR}
    f^{\rm Dr} = (1-\p_{\t x}\Psi\, n\,\p_{\theta})^{-1} f,\quad
    f^{\rm dR} = (1-\p_{\t \theta}\Psi\, n\,\p_{x})^{-1} f,\quad
    f^{\rm DR} = (1-\Psi\, n\,\p_{x}\p_\theta)^{-1} f.
\end{equation}
All dressing operations preserve the transformation type $(i,j)$ of the function. Further, as per the requirements on $\psi$ (see Eq.\eqref{psimap}), they act as follows:
\begin{equation}\begin{aligned}
    {}^{\rm dr}:&\quad \mathcal C^\infty(\Lambda)\to \mathcal C^\infty(\Lambda)\ , \\
    {}^{\rm Dr}:&\quad \mathcal \p_\theta^{-1}C^\infty(\Lambda)\to \p_\theta^{-1}\mathcal C^\infty(\Lambda) \ , \\
    {}^{\rm dR}:&\quad \mathcal \p_x^{-1}C^\infty(\Lambda)\to \p_x^{-1}\mathcal C^\infty(\Lambda) \ , \\
    {}^{\rm DR}:&\quad \mathcal \p_x^{-1}\p_\theta^{-1}C^\infty(\Lambda)\to \p_x^{-1}\p_\theta^{-1}\mathcal C^\infty(\Lambda) \ ,
    \end{aligned}
\end{equation}
or rather, they do so at least in the context of $ \mathcal M = \{|\rho|\leq \rho_*: \rho\in \mathcal C^\infty_c(\Lambda)\}$, with a domain $\Lambda$ either as in Eq.~\eqref{domain}, or which falls under one of the categories discussed in Section \ref{sec:MoreGenDomain}. Moreover, the following basic properties are immediate from the definitions \eqref{intDR}
\begin{equation}\label{dDRall}
    f_\theta^{~{\rm dr}} = (f^{\rm Dr})_\theta,\quad
    f_x^{~{\rm dr}} = (f^{\rm dR})_x,\quad
    f_\theta^{~{\rm dR}} = (f^{\rm DR})_{\theta},\quad
    f_x^{~{\rm Dr}} = (f^{\rm DR})_{x},\quad
    f_{x\theta}^{~~{\rm dr}} = (f^{\rm DR})_{x\theta}.
\end{equation}
These, along with the symmetry property \eqref{symm}, allow us to write the Poisson bracket \eqref{pb} in various ways, which are nonetheless equivalent, involving different types of dressing operations
\beqa\label{pb2}
	\{F,G\} &=& \int_\Lambda\frc{ \dd x\dd\theta}{2\pi}\,n\,
	\left[ F'_x\,  \big({G'}^{\rm Dr}\big)_\theta -  G'_x\, \big({F'}^{\rm Dr}\big)_\theta\right]\n
	&=& \int_\Lambda \frc{ \dd x\dd\theta}{2\pi}\,n\,
	\left[({F_x'})^{\rm dr}\, G'_\theta - ({G_x'})^{\rm dr}\, F'_\theta\right]\n
	&=& \int_\Lambda \frc{ \dd x\dd\theta}{2\pi}\,n\,
	\left[\big( {F'}^{\rm dR}\big)_x\, G'_\theta - \big( {G'}^{\rm dR}\big)_x\,  F'_\theta\right].
\eeqa

Note that in the case of conventional GHD, in which the interaction kernel takes the form \eqref{localGHD}, both {\bf dr}essing and {\bf Dr}essing specialise to integral operators on spectral space $\theta$ only\footnote{The other dressing operations still make sense, but act on the full spectral phase space.}
\begin{equation}
    f^{\rm dr} = (1-\p_{\t\theta} \Theta\, n)^{-1} f,\quad
    f^{\rm Dr} = (1-\Theta\, n\,\p_{\theta})^{-1} f
    \qquad \mbox{(conventional GHD with homogeneous coupling)}
\end{equation}
where $\Theta = \phi/(2\pi)$. In this case, the {\bf Dr}essing operation $f\mapsto f^{\rm Dr}$ is in fact a standard operation in the context of the Thermodynamic Bethe ansatz, and one  writes it as
\begin{equation}
    f^{\rm Dr}(\theta) = f(\theta) -
    \int_{\mathcal P} \dd\theta' n(\theta') F(\theta'|\theta)\p_{\theta'}f(\theta')
\end{equation}
in terms of the ``shift function" or ``backflow function" \cite[Chap 1]{korepinbook}, defined as
\begin{equation}
    F(\theta|\alpha) =
    \frc{\phi(\cdot,\alpha)^{\rm dr}(\theta)}{2\pi}.
\end{equation}
Since the {\bf Dr}essing operation defined as such is known to satisfy the first equation of \eqref{dDRall}, we show the equivalence with our formulation \eqref{Dressing} as follows
\begin{align}
    f^{\rm Dr}(\theta) &=
    f(\theta) - \int_{{\mathcal P}} \frc{\dd\theta'}{2\pi}
    n(\theta')\phi(\cdot,\theta)^{\rm dr}(\theta')\p_{\theta'}f(\theta')\n
    &=f(\theta) - \int_{{\mathcal P}} \frc{\dd\theta'}{2\pi}
    n(\theta')\phi(\theta',\theta)f_\theta^{~{\rm dr}}(\theta')\n
    &=f(\theta) - \int_{{\mathcal P}} \frc{\dd\theta'}{2\pi}
    n(\theta')\phi(\theta',\theta)\p_{\theta'}f^{{\rm Dr}}(\theta')\n
    &=f(\theta) - \int_{{\mathcal P}} \frc{\dd x'\dd\theta'}{2\pi}
    \p_{\t x} \frc12\sgn(x-x')\phi(\theta',\theta) n(\theta')\p_{\theta'}f^{{\rm Dr}}(\theta')
\end{align}
where in the second line we used the symmetry \eqref{symm}, and in the third we used the first equation of \eqref{dDRall}. Note that the same result is obtained for $\t x=x$ or $\t x = -x'$. This indeed reproduces \eqref{Dressing}. 

\section{Properties of the Poisson bracket}\label{apppoisson}

For operation \eqref{pb} to define a valid Poisson bracket it must be bilinear, skew-symmetric, and satisfy both Leibniz's rule and the Jacobi identity. While the first two of these properties are trivially verified, the latter two are not. This appendix provides the proof that operation \eqref{pb} indeed defines a valid Poisson bracket. On top of that, we show the linearised form \eqref{pbfreeform} under the change of metric \eqref{metchange}, which proves useful in establishing the involution \eqref{invol}.

The strategy is as follows. We prove the Leibniz property of the general Poisson bracket \eqref{pb} directly in Subsection \ref{appleibniz}. For the Jacobi identity, we proceed in two steps which take advantage of the fact that the validity of the Jacobi identity is independent of the choice of coordinates. In Subsection \ref{ssectequiv}, we show that the linearised Poisson bracket \eqref{pby} in terms of $\h\rho$ and the general Poisson bracket \eqref{pb} in terms of $\rho$ are equivalent under a change of fluid density coordinate. This also relies on a technical result proved in Appendix \ref{app:varfree}. Then the Jacobi identity for \eqref{pby} is proved directly in Subsection \ref{ssectJacobi}.

We start with some clarifications on the algebra of observables in Subsection \ref{app:alg}.

\subsection{Algebra of observables}\label{app:alg}

Observables are appropriate functionals of $\rho\in\mathcal M$. One may wish to specify a general algebra of functionals on $\mathcal M$ and use, say, Fr\'echet differentiability theory for their functional derivatives. In particular, the product rule and chain rule hold for Fr\'echet derivatives. Instead, we take the more straightforward series form of observables described in Subsection \ref{sec:alg}, which we repeat for convenience:
\begin{equation}\label{functionalsapp}
    F[\rho] = \sum_{n=0}^\infty \int_{\R^{2n}}  \ c_{n}(x_1,\cdots,x_n;\theta_1, \cdots, \theta_n)\prod_{i=1}^n   \rho(x_i,\theta_i) \dd x_i\dd\theta_i \; ,
\end{equation}
$c_0\in\R$, $c_n\in\mathcal D_{\mathcal M}(\R^{2n})$, with the functional derivatives defined by \eqref{functionalderivative}. The product rule is shown to hold in the following but, to give a complete picture, we would have to show that functional derivatives are elements of $\mathcal D_{\mathcal{M}}(\R^2)$, as well as their dressing, and that the Poisson bracket \eqref{pb}  gives a functional within the required space. This is beyond the scope of this paper.

 We will now make the space $\mathfrak U$ slightly more precise and show that it forms an algebra, both in the {\em abstract setup}, and in the {\em concrete setup} of Eq.~\eqref{smoothcompact} (see Subsection \ref{sec:FluDen}). We also show that functional derivatives, as defined in \eqref{functionalderivative}, satisfy the product rule, which is important for the Leibniz property of the Poisson bracket (Subsection \ref{appleibniz}). We do not show the chain rule (used in Subsection \ref{ssectequiv}), which should nevertheless hold (as it does for instance for Fréchet derivatives). In the abstract setup, it is then relatively simple to show that the Poisson bracket \eqref{pb} maps $\mathfrak U\otimes \mathfrak U \to \mathfrak U$. In the concrete setup, this should hold for $\rho_*$ in \eqref{smoothcompact} small enough (possibly smaller than \eqref{rhostar}).

It is simpler for the discussion not to assume $c_n$ in \eqref{functionalsapp} to be symmetric. Instead, we consider equivalence classes $[c_n]=\{\tilde c_n: \tilde c_n\sim c_n\}$ of the functions $c_{n}$ in \eqref{functionalsapp} whereby two functions are related if they differ by a function which is antisymmetric under at least one element of the symmetric group $S_n$ acting on the variables: $c_{n}\sim \tilde{c}_{n}$ if and only if there exists $f_n\in\mathcal D_{\mathcal M}(\R^{2n})$ and $\tau\in S_n$ such that
\beq
c_{n}=\tilde{c}_{n}+f_n\,,~~f_n(x_1,\cdots,x_n;\theta_1, \cdots, \theta_n)=-f_n(x_{\tau(1)},\cdots,x_{\tau(n)};\theta_{\tau(1)}, \cdots, \theta_{\tau(n)})\,.
\eeq
It is easy to see that if the functionals $F[\rho]$, $\widetilde{F}[\rho]$ associated to $c_n$, $\tilde{c}_n$, $n\ge 0$, are such that $c_n\sim\tilde{c}_n$ for all $n\ge 0$, then $F[\rho]=\widetilde{F}[\rho]$ (in both setups).  We denote $[\mathcal D_{\mathcal M}(\R^{2n})]=\{ [c_n]: c_n\in\mathcal D_{\mathcal M}(\R^{2n})\}$
the space of equivalence classes of $c_n$. The concatenation of functions gives rise to a well-defined product, which we denote $\cdot$, and which acts as $\cdot : [\mathcal D_{\mathcal M}(\R^{2n})]\times [\mathcal D_{\mathcal M}(\R^{2m})] \to [\mathcal D_{\mathcal M}(\R^{2(n+m)})]$. On representatives it is
\begin{equation}
\begin{aligned}
        &(c_n\cdot c_m)(x_1,\ldots,x_{n+m};\theta_1,\ldots,\theta_{n+m}) \\
    &\qquad\qquad =
    c_n(x_1,\ldots,x_n;\theta_1,\ldots,\theta_n) c_m(x_{n+1},\ldots,x_{n+m};
    \theta_{n+1},\ldots,\theta_{n+m}),
\end{aligned}
\end{equation}
and we define\footnote{The fact the product $\cdot$ is well-defined follows naturally: for every $f_n,\, f_m$ anti-symmetric, under some $\tau_n\in S_n,\,\tau_m\in S_m$ respectively, $(c_n+f_n)\cdot (c_m+f_m) = c_n\cdot c_m + f_{n+m}$ for some $f_{n+m}$ which is  anti-symmetric under some $\tau_{n+m}\in S_{n+m}$.} $[c_n]\cdot [c_m]= [c_n\cdot c_m]$.

In the {\em abstract setup}, $\mathfrak U$ is the algebra of formal functional series in $\rho$, that is, an observable is a sequence of equivalence classes of functions $F[\rho] = ([c_0],[c_1],[c_2],\ldots)$ and $\mathfrak U = \R\times [\mathcal D_{\mathcal M}(\R^2)]\times [\mathcal D_{\mathcal M}(\R^4)]\times\cdots$. The algebra product is obtained from the Cauchy product of series, which gives, for every $F[\rho]= (c_n)_n$, $G[\rho]=(d_n)_n \in\mathfrak U$,
\begin{equation}\label{algebraproduct}
    (F G)[\rho]
    = \Big( \sum_{k=0}^n [c_k] \cdot [d_{n-k}]\Big)_n \in \mathfrak U\ .
\end{equation}
The Cauchy product and concatenation of functions are commutative and associative operations, thus
\begin{equation}
    F G =
    G F ,\quad
    (F (G H)) =
    ((F G) H)\ . 
\end{equation}

For the {\em concrete setup of Eq.~\eqref{smoothcompact}}, an element of the algebra $\mathfrak U$ is an infinite series \eqref{functionalsapp}, with the algebra product taken as point-wise multiplication
\begin{equation}\label{pointwise}
    (FG)[\rho]=F[\rho]G[\rho].
\end{equation}
We note that for every $\rho\in\mathcal M$, each integral in \eqref{functionalsapp} exists and, in addition, we require that: (i) the series for $F[\rho]$ be absolutely convergent; (ii) the series for all functional derivatives of $F[\rho]$ (see \eqref{functionalderivative} for the first derivative) and all their $x,\theta$ derivatives, be, as distributions (e.g.~integrated against compactly supported smooth functions in the context of Eq.~\eqref{smoothcompact}), absolutely convergent. 
Absolute convergence guarantees that point-wise multiplication \eqref{pointwise} is equivalent to the Cauchy product, hence to the algebra product defined as \eqref{algebraproduct}; absolute convergence is preserved under the algebra product.

Concerning functional derivatives, we may be more precise as follows. Recall that $F'[\rho](x,\theta)$ are given by Eq.~\eqref{functionalderivative} for the symmetric choice of representative $c_n$. We may define the operation $|_{(x,\theta)}:\mathcal D_{\mathcal M}(\R^{2n})\to \mathcal D_{\mathcal M}(\R^{2(n-1)})$ as
\begin{equation}
    c_n|_{(x,\theta)} = \sum_{k=1}^n c_n(x_1,\ldots,\underbrace{x}_{k^{\rm th}\;\text{position}},\ldots,x_{n-1};\theta_1,\ldots,\underbrace{\theta}_{k^{\rm th}\;\text{position}},\ldots,x_{n-1})\ .
\end{equation}
One can show that this acts well on equivalence classes, so we can define $[c_n]|_{(x,\theta)} = [c_n|_{(x,\theta)}]$. We note that under concatenation, this operation acts as a {\em differentiation},
\begin{equation}\label{differentiationconc}
    (c_n\cdot c_m)|_{(x,\theta)}
    = c_n|_{(x,\theta)}\cdot c_m + c_n \cdot c_m|_{(x,\theta)}.
\end{equation}
Then Eq.~\eqref{functionalderivative} is equivalent to
\begin{equation}
    F'[\rho](x,\theta)
    =\sum_{n=1}^\infty
    \int_{\R^{2(n-1)}}
    c_n|_{(x,\theta)}\prod_{i=1}^{n-1}\rho(x_i,\theta_i)\dd x_i\dd\theta_i.
\end{equation}

In the concrete setup, for every $F[\rho]\in\mathfrak U$ and every $\rho\in\mathcal M$, the series defining $F'[\rho](x,\theta)$ results in a function of $(x,\theta)$ that lies in $\mathcal D_{\mathcal M}(\R^2)$, as per our conditions. In the abstract setup, each coefficient in the series is, as a function of $(x,\theta)$, an element of $\mathcal D_{\mathcal M}(\R^2)$. This viewpoint is taken, for instance, when dressing the functional derivative, $(F'[\rho])^{\rm dr}(x,\theta)$ (dressing smooth functions or more generally distributions, see Appendix \ref{appAnalysis}), using linearity of the dressing operation in the abstract setup in order to apply it on each term of the series.

The functional derivative is also a map
\begin{equation}
    \R^2\to\mathfrak U : (x,\theta)\mapsto F'[\rho](x,\theta).
\end{equation}
That is, for every $(x,\theta)\in\R^2$, the series for $F'[\rho](x,\theta)$ gives rise, as distribution, to an element of $\mathfrak U$. This is immediate in the abstract setup, and follows from the conditions stated in the concrete setup.
Therefore, the products $F'[\rho](x,\theta)G[\rho]$, etc., are well-defined.

The product rule then follows immediately from the Cauchy product and the differentiation-like property \eqref{differentiationconc}:
\begin{align}
    (F[\rho] G[\rho])'(x,\theta)
    &= \Bigg(\Big( \sum_{k=0}^n [c_k] \cdot [d_{n-k}]\Big)_n\Bigg)'(x,\theta)\n
    &= \Big( \sum_{k=0}^n \Big([c_k] \cdot [d_{n-k}]\Big)|_{(x,\theta)}\Big)_n\n
    &= \Big( \sum_{k=0}^n [c_k]|_{(x,\theta)} \cdot [d_{n-k}] +
    [c_k] \cdot [d_{n-k}]|_{(x,\theta)}\Big)_n\n
    &= F'[\rho](x,\theta)G[\rho]
    + F[\rho]G'[\rho](x,\theta)\ .
    \label{productrule}
\end{align}

For $f\in \mathcal D_{\mathcal M}(\R^2)$, its dressing $f^{\rm dr}(x,\theta)$ gives rise to an element of $\mathfrak U$ (again as a distribution). This follows from the Liouville-Neumann series; it is immediate in the abstract setup, but would necessitate a more accurate analysis in the concrete setup, which we omit here. A similar analysis can be done for $(F'[\rho])^{\rm dr}(x,\theta)$.

\subsection{Leibniz rule}\label{appleibniz}

We now show that the Poisson bracket \eqref{pb} satisfies the Leibniz's rule 
\begin{equation}\label{LebRule}
    \{F_1F_2,G\} = F_1\{F_2,G\} + F_2 \{F_1,G\} \ ,
\end{equation} 
with the algebra product on $\mathfrak U$ defined as in Subsection \ref{app:alg}.
Indeed, by writing \eqref{pb}, via integration by part and the symmetry property \eqref{symm}, as
\beq
	\{F,G\} = \int_{\R^2} \frc{\dd x\dd\theta}{2\pi}\,
	\Big(F' \p_x \big(n {G'_\theta}^{\rm dr}\big) - \p_\theta \big({G'_x}^{\rm dr} n\big) F'\Big),
\eeq
the product rule \eqref{productrule} immediately gives \eqref{LebRule}.

\subsection{Equivalence of the normal-density and fluid-density formulations}\label{ssectequiv}

The Poisson bracket for functionals of the normal density is determined by the fundamental linearised bracket \eqref{pbfreeform} along with Leibniz's rule, viz.~Eq.~\eqref{eq:pbyfunctionals} which we write here for generic $\hat{\mathcal L}_\theta$ (Remark \ref{rem:yspace}):
\beq\label{pby}
	\{F,G\} = \int_{\R}\dd\theta\int_{\hat{\mathcal L}_\theta} \dd y\,\h\rho(y,\theta)\,\Bigg(
	\p_y \frc{\delta F}{\delta\h\rho(y,\theta)}
	\p_\theta \frc{\delta G}{\delta\h\rho(y,\theta)}
	-
	\p_y \frc{\delta G}{\delta\h\rho(y,\theta)}
	\p_\theta \frc{\delta F}{\delta\h\rho(y,\theta)}\Bigg)\,.
\eeq
We now show that \eqref{pby} implies \eqref{pb}; the steps can be retraced backwards to show the implication in the opposite direction. For our purposes, we will need to make use of the following identity 
\beq\label{variationfree}
	\frc{\delta F}{\delta\h\rho(y,\theta)}
	= {F'}^{\rm dR}\left(X(y,\theta),\theta\right) \ ,
\eeq
that expresses the variational derivative of a functional of type \eqref{functionals} in terms of the {\bf dR}essing \eqref{dRessing}, and that we derive in Appendix \ref{app:varfree}.
From this we may evaluate the first half of the Poisson bracket \eqref{pby} 
\beqa
	\lefteqn{\int_{\R}\dd\theta\int_{\hat{\mathcal L}_\theta} \dd y\,\h\rho(y,\theta)
	\p_y \frc{\delta F}{\delta\h\rho(y,\theta)}\,
	\p_\theta \frc{\delta G}{\delta\h\rho(y,\theta)}}&&\n
	&=&
	\int_{\R^2} \dd x\dd\theta\,\rho(x,\theta)\,
	\Big(\p_y \frc{\delta F}{\delta\h\rho(y,\theta)}\Big)\Big|_{y=Y(x,\theta)}\Bigg(
	\p_\theta \frc{\delta G}{\delta\h\rho(Y(x,\theta),\theta)}
	-
	2\pi\rho_{\rm s}(x,\theta)\,
	\Big(\p_y \frc{\delta G}{\delta\h\rho(y,\theta)}\Big)\Big|_{y= Y(x,\theta)}\Bigg) \n
	&=&
	\int_{\R^2} \frc{\dd x\dd\theta}{2\pi}\,n(x,\theta)\,
	\p_x \frc{\delta F}{\delta\h\rho(Y(x,\theta),\theta)}\,
	\p_\theta \frc{\delta G}{\delta\h\rho(Y(x,\theta),\theta)} \n && -\;
	2\pi\int_{\R^2} \dd x\dd\theta\,\rho(x,\theta)\rho_{\rm s}(x,\theta)\,
	\Big(\p_y \frc{\delta F}{\delta\h\rho(y,\theta)}\,
	\p_y \frc{\delta G}{\delta\h\rho(y,\theta)}\Big)\Big|_{y=Y(x,\theta)}
\eeqa
Similarly, we may also evaluate the second half of the bracket \eqref{pby}, obtained by exchanging $F\leftrightarrow G$, with the opposite sign. But we note that the second term on the right-hand side of the last equality is symmetric under $F\leftrightarrow G$, hence cancels out in \eqref{pby}. Thus, omitting it and using \eqref{variationfree}, we find
\beq
	\int_{\R^2} \frc{\dd x\dd\theta}{2\pi}\,
	n(x,\theta)\,
	\p_x {F'}^{\rm dR}(x,\theta)
	\p_\theta {G'}^{\rm dR}(x,\theta) 
	=
	\int_{\R^2} \frc{\dd x\dd\theta}{2\pi}\,
	n(x,\theta)\,
	{F'_x}^{\rm dr}(x,\theta)
	\p_\theta {G'}^{\rm dR}(x,\theta) \ .
\eeq
We use the integral operator formulation $g^{\rm dR} = g + \Psi_{-\theta'}n \p_x g^{\rm dR}$ in order to write the general identity
\beq
	\p_\theta g^{\rm dR} = 
	g_\theta - \Psi_{\theta\theta'} g_x^{~\rm dr}\ .
\eeq
Therefore we obtain
\beqa
	\lefteqn{
	=
	\int_{\R^2} \frc{\dd x\dd\theta}{2\pi}\,
	n(x,\theta)\,
	{F'_x}^{\rm dr}(x,\theta)
	G'_\theta(x,\theta)\; +
	}&&\n
	&&
	\int_{{\R^4}} \frc{\dd x\dd\theta\dd x'\dd\theta'}{(2\pi)^2}\,
	n(x,\theta)
	n(x',\theta')\,
	\psi_{\theta\theta'}(x,\theta;x',\theta')
	\,{F'_x}^{\rm dr}(x,\theta)
	{G'_x}^{\rm dr}(x',\theta').\label{eqta}
\eeqa
As $\psi_{\theta\theta'}(x,\theta;x',\theta')$ is symmetric under exchange $(x,\theta)\leftrightarrow (x',\theta')$, in \eqref{pby}, where again we anti-symmetrise under $F\leftrightarrow G$, the double integral terms (i.e. the second term on the right-hand side of \eqref{eqta} and its companion from the other half of \eqref{pby}) cancel out. The result is \eqref{pb}.

\subsection{Jacobi identity}\label{ssectJacobi}

Thanks to the equivalence shown in Subsection \ref{ssectequiv}, it is sufficient to show the Jacobi identity in the normal density formulation. Moreover, thanks to Leibniz's rule and to the fact that, in the normal density formulation, the space of linear functionals is preserved, it is sufficient to show that \eqref{pbfreeform} satisfies
\beq\label{jacobihrho}
	\sum_{\mbox{cyclic permutations}\ 1\to2\to3}\{\{\h\rho(x_1,\theta_1),\h\rho(x_2,\theta_2)\},\h\rho(x_3,\theta_3)\}=0.
\eeq
That this implies the Jacobi identity for arbitrary functionals of $\h\rho$ is then a standard result (see e.g.~\cite{abraham2012manifolds} Section 8.1).

We evaluate (in a short-hand notation)
\beqa
	\lefteqn{\{\{\h\rho(x_1,\theta_1),\h\rho(x_2,\theta_2)\},\h\rho(x_3,\theta_3)\}}&& \n
	&=&
	\delta'_{x_{12}}\delta'_{\theta_{12}}\,(
	\delta'_{x_{13}}\delta'_{\theta_{23}}\,(
	\h\rho(x_1,\theta_3)-\h\rho(x_3,\theta_2))
	-
	\delta'_{x_{23}}\delta'_{\theta_{13}}\,(
	\h\rho(x_2,\theta_3)-\h\rho(x_3,\theta_1))
	).
\eeqa
For convenience, we will refer to each term in the right-hand side as, from left to right, (i),( ii), (iii) and (iv). In every such term, we extract the derivatives with respect to the four variables that are not repeated\footnote{For instance, for terms (i) and (ii), with the factors $\delta'_{x_{12}}\delta'_{\theta_{12}}\delta'_{x_{13}}\delta'_{\theta_{23}}$, these are $\p_{x_2}$, $\p_{\theta_1}$, $\p_{x_3}$, $\p_{\theta_3}$.}. For each term (i)-(iv), two new terms arise: a term of type I  where the derivatives are applied to the full expression, and a term of type II where one of the derivatives is applied to the normal density $\h\rho(\cdot,\cdot)$ -- the derivative that acts non-trivially on it. More explicitly, in the case of term (i), we have
\begin{equation}
\begin{aligned}
    &\text{Type I term:}  \qquad -\p_{x_2}\p_{\theta_1}\p_{x_3}\p_{\theta_3}\left[(\prod\delta)\h\rho(x_1,\theta_3)\right] \ ,\\
    &\text{Type II term:} \qquad \p_{x_2}\p_{\theta_1}\p_{x_3} \left[(\prod\delta) \p_{\theta_3}\h\rho(x_1,\theta_3)\right] \ ,
\end{aligned}
\end{equation}
where we introduced the notation
\begin{equation}
    (\prod\delta)= \delta_{x_{12}}\delta_{\theta_{12}}\delta_{x_{13}}\delta_{\theta_{23}} \ .
\end{equation}
In the terms of type I, there is a product of delta functions that imposes $x_1=x_2=x_3$ and $\theta_1=\theta_2=\theta_3$ (for instance $(\prod\delta)=\delta_{x_{12}}\delta_{\theta_{12}}\delta_{x_{13}}\delta_{\theta_{23}}$). Each such product is equivalent and, thus, in each term of type I, for (i)-(iv), we can set the normal density to $\h\rho(x_1,\theta_1)$. Since terms (i) and (ii) have the same set of extracted derivatives and opposite sign, they cancel; similarly for terms (iii) and (iv). In the end, only terms of type II remain; by using the delta functions, in such terms we have either $\p_{x_1}\h\rho(x_1,\theta_1)$ or $\p_{\theta_1}\h\rho(x_1,\theta_1)$. Hence, we are left with
\beqa
	\lefteqn{\{\{\h\rho(x_1,\theta_1),\h\rho(x_2,\theta_2)\},\h\rho(x_3,\theta_3)\}}&&\n
	&=&
	(\p_{x_2}\p_{\theta_1}\p_{x_3}-\p_{x_1}\p_{\theta_2}\p_{x_3})\left[(\prod\delta)\,\p_{\theta_1}\rho(x_1,\theta_1)\right]\n
	&& +\;
	(\p_{x_1}\p_{\theta_2}\p_{\theta_3}-\p_{x_2}\p_{\theta_1}\p_{\theta_3})\left[(\prod\delta)\,\p_{x_1}\rho(x_1,\theta_1)\right].
\eeqa
The sum of cyclic permutations of the first line and of the second line both vanish, which shows the Jacobi identity \eqref{jacobihrho}.

\section{Variational derivative with respect to the normal density $\hat\rho$}\label{app:varfree}

In this appendix, we derive the identity \eqref{variationfree}, expressing the variational derivative of a functional of type \eqref{functionals} in terms of the {\bf dR}ressing Eq.~\eqref{dRessing}. We recall the identity for the sake of convenience:
\begin{equation}
    \frc{\delta F}{\delta\h\rho(y,\theta)}
	= {F'}^{\rm dR}(X(y,\theta),\theta)
\end{equation}
where $X(y,\theta)$ is defined in \eqref{Xdefinition}.

For our purposes, we must first recall the relation between the spectral fluid density and its normal counterpart \eqref{freedenocc}, viz.
\beq\label{rhorhohat1}
	\rho(x,\theta) = 2\pi \rho_{\rm s}(x,\theta)\h\rho(Y(x,\theta),\theta).
\eeq
We may now evaluate the functional derivative ${\delta \rho(x,\theta)}/{\delta\h\rho(y,\alpha)}$, first by computing
\beq
	\frc{\delta (2\pi \rho_{\rm s}(x,\theta))}{\delta\h\rho(y,\alpha)}
	= \int_{{\R^2}} \dd x'\dd\theta'\,\partial_x\partial_\theta \psi(x,\theta;x',\theta')
	\frc{\delta \rho(x',\theta')}{\delta\h\rho(y,\alpha)} \ ,
\eeq
and then, using the definition of $\rho_s$ \eqref{rhosOmega}, and that of $Y$ \eqref{Ysolution} in terms of the {\bf dR}essing \eqref{dRessing}, along the dressing relations \eqref{dDRall}, yields
\beqa\label{appderY1}
	\frc{\delta Y(x,\theta)}{\delta\h\rho(y,\alpha)}
 &=& \int_{{\R^2}} \dd x' \dd\theta'\,\partial_{\theta} \psi(x,\theta;x',\theta')
	\frc{\delta \rho(x',\theta')}{\delta\h\rho(y,\alpha)}\ .
\eeqa
Putting all of this together, we eventually obtain
\beqa
	\frc{\delta \rho(x,\theta)}{\delta\h\rho(y,\alpha)}
	&=&
	2\pi\rho_{\rm s}(x,\theta)\,\delta(y-Y(x,\theta))\delta(\alpha-\theta)\n
	&&
	+\;
	\Bigg(\h\rho(Y(x,\theta),\theta)
	\int_{{\R^2}} \dd x'\,\dd\theta'\,\partial_x\partial_\theta \psi(x,\theta;x',\theta')\n
	&&
	\quad +\;
	2\pi\rho_{\rm s}(x,\theta)\p_y \h\rho(Y(x,\theta),\theta)
	\int_{{\R^2}} \dd x' \dd\theta'\,\partial_{\theta} \psi(x,\theta;x',\theta')
	\Bigg)\;
	\frc{\delta \rho(x',\theta')}{\delta\h\rho(y,\alpha)}\n
	&=&
	\delta(x-X(y,\alpha))\delta(\theta-\alpha)\n
	&&
	+\;
	\Bigg(n(x,\theta)
	\int_{{\R^2}} \,\frc{\dd x'\dd\theta'}{2\pi}\,\partial_x\partial_\theta \psi(x,\theta;x',\theta')\n
	&&
	\qquad +\;
	\p_x n(x,\theta)
	\int_{{\R^2}} \,\frc{\dd x'\dd\theta'}{2\pi}\,\partial_\theta \psi(x,\theta;x',\theta')
	\Bigg)\;
	\frc{\delta \rho(x',\theta')}{\delta\h\rho(y,\alpha)}\n
 &\equiv& \delta(x-X(y,\alpha))\delta(\theta-\alpha)+ \left({\cal O}\frc{\delta \rho(\cdot_1,\cdot_2)}{\delta\h\rho(y,\alpha)}\right)(x,\theta) \ ,
\eeqa
where $X$ is defined by \eqref{Xdefinition}, and where ${\cal O}$ is the operator
$$({\cal O}f)(x,\theta)=\partial_x\left(n(x,\theta)
	\int_{{\R^2}} \,\frc{\dd x'\dd\theta'}{2\pi}\,\partial_\theta \psi(x,\theta;x',\theta')f(x',\theta')  \right)\,.$$
 Its adjoint is given by 
 $$({\cal O}^\dagger f)(x,\theta)=	\int_{{\R^2}} \,\frc{\dd x'\dd\theta'}{2\pi}\,n(x',\theta')\partial_{-\theta'} \psi(x,\theta;x',\theta')\partial_{x'}f(x',\theta')\,,$$
 which is the operator appearing in the {\bf dR}ressing (for a function of type $(0,0)$).
Therefore
\beqa
	\frc{\delta F}{\delta\h\rho(y,\alpha)}
	&=&
	\int_{{\R^2}} \dd x\dd\theta\,
	F'[\rho](x,\theta)
	\Big[(1-{\cal O})^{-1}\delta(\cdot_1 - X(y,\alpha))\delta(\cdot_2-\alpha)\Big](x,\theta)\n
	&=&
\int_{{\R^2}} \dd x\dd\theta\,
	((1-{\cal O^\dagger})^{-1}F'[\rho])(x,\theta)
		\delta(x - X(y,\alpha))\delta(\theta-\alpha)\n	
  &=&
	\int_{{\R^2}} \dd x\dd\theta\,
	{F'[\rho]}^{\rm dR}(x,\theta)
	\,
	\delta(x - X(y,\alpha))\delta(\theta-\alpha)\n
	&=& {F'[\rho]}^{\rm dR}(X(y,\alpha),\alpha)
\eeqa
which indeed yields the identity \eqref{variationfree}.

\bibliographystyle{ieeetr}
\bibliography{Biblior,Biblio-2019-01-03r,Bibliographyr}

\begin{thebibliography}{10}

\bibitem{spohn2012large}
H.~Spohn, {\em Large scale dynamics of interacting particles}.
\newblock Springer Science \& Business Media, 2012.

\bibitem{calogero2012integrability}
F.~Calogero, N.~Ercolani, H.~Flaschka, V.~Marchenko, A.~Mikhailov, A.~Newell,
  E.~Schulman, A.~Shabat, E.~Siggia, V.~Sokolov, {\em et~al.}, {\em What is
  integrability?}
\newblock Springer Science \& Business Media, 2012.

\bibitem{faddeev2007hamiltonian}
L.~Faddeev and L.~Takhtajan, {\em Hamiltonian methods in the theory of
  solitons}.
\newblock Springer Science \& Business Media, 2007.

\bibitem{doyon2020lecture}
B.~Doyon, ``Lecture notes on generalised hydrodynamics,'' {\em SciPost Physics
  Lecture Notes}, p.~018, 2020.

\bibitem{spohn2023hydrodynamic}
H.~Spohn, ``Hydrodynamic scales of integrable many-particle systems,'' {\em
  preprint arXiv:2301.08504}, 2023.

\bibitem{Doyon_2017}
B.~Doyon and H.~Spohn, ``Dynamics of hard rods with initial domain wall
  state,'' {\em Journal of Statistical Mechanics: Theory and Experiment},
  vol.~2017, p.~073210, jul 2017.

\bibitem{el2021soliton}
G.~A. El, ``Soliton gas in integrable dispersive hydrodynamics,'' {\em Journal
  of Statistical Mechanics: Theory and Experiment}, vol.~2021, no.~11,
  p.~114001, 2021.

\bibitem{bonnemain2022generalized}
T.~Bonnemain, B.~Doyon, and G.~El, ``Generalized hydrodynamics of the kdv
  soliton gas,'' {\em Journal of Physics A: Mathematical and Theoretical},
  vol.~55, no.~37, p.~374004, 2022.

\bibitem{suret2023solitonRev}
P.~Suret, S.~Randoux, A.~Gelash, D.~Agafontsev, B.~Doyon, and G.~El, ``Soliton
  gas: Theory, numerics and experiments,'' 2023.

\bibitem{Bastianello_2022}
A.~Bastianello, B.~Bertini, B.~Doyon, and R.~Vasseur, ``Introduction to the
  special issue on emergent hydrodynamics in integrable many-body systems,''
  {\em Journal of Statistical Mechanics: Theory and Experiment}, vol.~2022,
  p.~014001, jan 2022.

\bibitem{doyon2017note}
B.~Doyon and T.~Yoshimura, ``{A note on generalized hydrodynamics:
  inhomogeneous fields and other concepts},'' {\em SciPost Phys.}, vol.~2,
  p.~014, 2017.

\bibitem{PhysRevLett.122.240606}
A.~Bastianello and A.~De~Luca, ``Integrability-protected adiabatic
  reversibility in quantum spin chains,'' {\em Phys. Rev. Lett.}, vol.~122,
  p.~240606, Jun 2019.

\bibitem{bastianello2019generalized}
A.~Bastianello, V.~Alba, and J.-S. Caux, ``Generalized hydrodynamics with
  space-time inhomogeneous interactions,'' {\em Physical Review Letters},
  vol.~123, no.~13, p.~130602, 2019.

\bibitem{ZakharovHamil1}
V.~Zakharov, ``Hamiltonian formalism for hydrodynamic plasma models,'' {\em
  Sov. Phys. JETP}, vol.~33, p.~927, 1971.

\bibitem{ZakharovHamil2}
V.~E. Zakharov and E.~Kuznetsov, ``Hamiltonian formalism for nonlinear waves,''
  {\em Phys.-Usp}, vol.~40, p.~1087, 1997.

\bibitem{OLVER1982233}
P.~J. Olver, ``A nonlinear {H}amiltonian structure for the euler equations,''
  {\em Journal of Mathematical Analysis and Applications}, vol.~89, no.~1,
  pp.~233--250, 1982.

\bibitem{MorrisonHamil}
J.~P. Morrison, ``Hamiltonian description of the ideal fluid,'' {\em Rev. Mod.
  Phys.}, vol.~70, p.~467, 1998.

\bibitem{10.21468/SciPostPhys.14.5.103}
G.~M. Monteiro, A.~G. Abanov, and S.~Ganeshan, ``{Hamiltonian structure of 2D
  fluid dynamics with broken parity},'' {\em SciPost Phys.}, vol.~14, p.~103,
  2023.

\bibitem{el2011kinetic}
G.~El, A.~Kamchatnov, M.~V. Pavlov, and S.~Zykov, ``Kinetic equation for a
  soliton gas and its hydrodynamic reductions,'' {\em Journal of Nonlinear
  Science}, vol.~21, no.~2, pp.~151--191, 2011.

\bibitem{Bulchandani_2017}
V.~B. Bulchandani, ``On classical integrability of the hydrodynamics of quantum
  integrable systems,'' {\em Journal of Physics A: Mathematical and
  Theoretical}, vol.~50, p.~435203, oct 2017.

\bibitem{vergallo2023hamiltonian}
P.~Vergallo and E.~V. Ferapontov, ``Hamiltonian systems of {J}ordan block type:
  delta-functional reductions of the kinetic equation for soliton gas,'' {\em
  Journal of Mathematical Physics}, vol.~64, no.~10, 2023.

\bibitem{vergallo2024hamiltonian}
P.~Vergallo and E.~V. Ferapontov, ``Hamiltonian aspects of the kinetic equation
  for soliton gas,'' 2024.

\bibitem{bulchandani2017solvable}
V.~B. Bulchandani, R.~Vasseur, C.~Karrasch, and J.~E. Moore, ``Solvable
  hydrodynamics of quantum integrable systems,'' {\em Physical review letters},
  vol.~119, no.~22, p.~220604, 2017.

\bibitem{DOYON2018570}
B.~Doyon, H.~Spohn, and T.~Yoshimura, ``A geometric viewpoint on generalized
  hydrodynamics,'' {\em Nuclear Physics B}, vol.~926, pp.~570--583, 2018.

\bibitem{Bulchandani_2024}
V.~B. Bulchandani, ``Revised {E}nskog equation for hard rods,'' {\em Journal of
  Statistical Mechanics: Theory and Experiment}, vol.~2024, p.~043205, apr
  2024.

\bibitem{Hubner_2024}
F.~H\"ubner, ``Mesoscopic impurities in generalized hydrodynamics,'' {\em
  Journal of Statistical Mechanics: Theory and Experiment}, vol.~2024,
  p.~033102, mar 2024.

\bibitem{bonnemain2024soliton}
T.~Bonnemain and B.~Doyon, ``Soliton gas of the integrable {B}oussinesq
  equation and its generalised hydrodynamics,'' {\em arXiv preprint
  arXiv:2402.08669}, 2024.

\bibitem{ttbar1}
B.~Doyon, F.~H\"ubner, and T.~Yoshimura, ``Generalised {$T\bar T$}-deformations
  of classical free particles,'' {\em arXiv preprint}, p.~arXiv:2312.14855,
  2023.

\bibitem{ttbar2}
B.~Doyon, F.~H\"ubner, and T.~Yoshimura, ``New classical integrable systems
  from generalized {$T\bar T$}-deformations,'' {\em arXiv preprint},
  p.~arXiv:2311.06369, 2023.

\bibitem{Doyon_2021}
B.~Doyon and J.~Durnin, ``Free energy fluxes and the
  {K}ubo–{M}artin–{S}chwinger relation,'' {\em Journal of Statistical
  Mechanics: Theory and Experiment}, vol.~2021, p.~043206, apr 2021.

\bibitem{HubnerDoyonUniqueness}
F.~H\"ubner and B.~Doyon, ``Existence and uniqueness of solutions to the
  generalized hydrodynamics equation,'' 2024.

\bibitem{doyon2018exact}
B.~Doyon, ``Exact large-scale correlations in integrable systems out of
  equilibrium,'' {\em SciPost Phys}, vol.~5, no.~5, p.~054, 2018.

\bibitem{castro2016emergent}
O.~A. Castro-Alvaredo, B.~Doyon, and T.~Yoshimura, ``Emergent hydrodynamics in
  integrable quantum systems out of equilibrium,'' {\em Physical Review X},
  vol.~6, no.~4, p.~041065, 2016.

\bibitem{congy2023dispersive}
T.~Congy, G.~El, G.~Roberti, and A.~Tovbis, ``Dispersive hydrodynamics of
  soliton condensates for the {K}orteweg--de {V}ries equation,'' {\em Journal
  of Nonlinear Science}, vol.~33, no.~6, p.~104, 2023.

\bibitem{el2020spectral}
G.~El and A.~Tovbis, ``Spectral theory of soliton and breather gases for the
  focusing nonlinear {S}chr{\"o}dinger equation,'' {\em Physical Review E},
  vol.~101, no.~5, p.~052207, 2020.

\bibitem{kuijlaars2021minimal}
A.~Kuijlaars and A.~Tovbis, ``On minimal energy solutions to certain classes of
  integral equations related to soliton gases for integrable systems,'' {\em
  Nonlinearity}, vol.~34, no.~10, p.~7227, 2021.

\bibitem{lax2005hyperbolic}
P.~D. Lax, ``Hyperbolic systems of conservation laws ii,'' in {\em Selected
  Papers Volume I}, pp.~233--262, Springer, 2005.

\bibitem{convenientbook}
A.~Kriegl and P.~W. Michor, {\em The Convenient Setting of Global Analysis}.
\newblock American Mathematical Society, 1997.

\bibitem{faddeev1985poisson}
L.~Faddeev and L.~Takhtajan, ``Poisson structure for the kdv equation,'' {\em
  letters in mathematical physics}, vol.~10, pp.~183--188, 1985.

\bibitem{korepinbook}
V.~E. Korepin, N.~M. Bogoliubov, and A.~G. Izergin, {\em Quantum inverse
  scattering method and correlation functions}, vol.~3.
\newblock Cambridge university press, 1997.

\bibitem{10.21468/SciPostPhys.6.6.070}
J.-S. Caux, B.~Doyon, J.~Dubail, R.~Konik, and T.~Yoshimura, ``{Hydrodynamics
  of the interacting Bose gas in the Quantum Newton Cradle setup},'' {\em
  SciPost Phys.}, vol.~6, p.~070, 2019.

\bibitem{tsarev1991geometry}
S.~P. Tsarev, ``The geometry of hamiltonian systems of hydrodynamic type. the
  generalized hodograph method,'' {\em Mathematics of the USSR-Izvestiya},
  vol.~37, no.~2, p.~397, 1991.

\bibitem{congy2024riemann}
T.~Congy, H.~T. Carr, G.~Roberti, and G.~A. El, ``Riemann problem for
  polychromatic soliton gases: a testbed for the spectral kinetic theory,''
  2024.

\bibitem{tsarev1985poisson}
S.~Tsarev, ``On {P}oisson brackets and one-dimensional systems of hydrodynamic
  type,'' in {\em Sov. Math. Doklady}, vol.~31, p.~488, 1985.

\bibitem{pavlov1988hamiltonian}
M.~V. Pavlov, ``Hamiltonian formalism of weakly nonlinear hydrodynamic
  systems,'' {\em Theor. Math. Phys.;(United States)}, vol.~73, no.~2, 1988.

\bibitem{HubnerDoyonQuadrature}
F.~H\"ubner and B.~Doyon, ``A new quadrature for the generalized hydrodynamics
  equation and absence of shocks in the lieb-liniger model,'' 2024.

\bibitem{dubrovin1989hydrodynamics}
B.~A. Dubrovin and S.~P. Novikov, ``Hydrodynamics of weakly deformed soliton
  lattices. differential geometry and {H}amiltonian theory,'' {\em Russian
  Mathematical Surveys}, vol.~44, no.~6, p.~35, 1989.

\bibitem{dubrovin1983hamiltonian}
B.~Dubrovin, S.~Novikov, {\em et~al.}, ``The hamiltonian formalism of
  one-dimensional systems of hydrodynamic type and the {B}ogolyubov-{W}hitham
  averaging method,'' {\em Doklady Akademii Nauk SSSR}, vol.~270, no.~4,
  pp.~781--785, 1983.

\bibitem{babskii2012mathematical}
V.~G. Babskii, M.~Y. Zhukov, and V.~I. Yudovich, {\em Mathematical theory of
  electrophoresis}.
\newblock Springer Science \& Business Media, 2012.

\bibitem{rozdestvenskii1983systems}
B.~Rozdestvenskii and N.~Yanenko, ``Systems of quasilinear equations and their
  applications to gas dynamics, translated from the second russian edition by
  {JR} {S}chulenberger,'' {\em Translations of Mathematical Monographs},
  vol.~55, 1983.

\bibitem{tsarev2000integrability}
S.~P. Tsarev, ``Integrability of equations of hydrodynamic type from the end of
  the 19th to the end of the 20th century,''

\bibitem{brenier2004hydrodynamic}
Y.~Brenier, ``Hydrodynamic structure of the augmented {Born-Infeld}
  equations,'' {\em Archive for rational mechanics and analysis}, vol.~172,
  pp.~65--91, 2004.

\bibitem{ferapontov1991integration}
E.~Ferapontov, ``Integration of weakly nonlinear hydrodynamic systems in
  {R}iemann invariats,'' {\em Physics Letters A}, vol.~158, no.~3-4,
  pp.~112--118, 1991.

\bibitem{pavlov2003tri}
M.~V. Pavlov and S.~P. Tsarev, ``Tri-hamiltonian structures of egorov systems
  of hydrodynamic type,'' {\em Functional Analysis and Its Applications},
  vol.~37, pp.~32--45, 2003.

\bibitem{pavlov1994exact}
M.~V. Pavlov, ``Exact integrability of a system of benney equations,'' in {\em
  Doklady Akademii Nauk}, vol.~339, pp.~311--313, Russian Academy of Sciences,
  1994.

\bibitem{pavlov1994multi}
M.~V. Pavlov, ``Multi-hamiltonian structures of the whitham equations,'' in
  {\em Doklady Akademii Nauk}, vol.~338, pp.~165--167, Russian Academy of
  Sciences, 1994.

\bibitem{dubrovin2011linearly}
B.~A. Dubrovin, M.~V. Pavlov, and S.~Zykov, ``Linearly degenerate {H}amiltonian
  {PDEs} and a new class of solutions to the {WDVV} associativity equations,''
  {\em Functional Analysis and Its Applications}, vol.~45, no.~4, pp.~278--290,
  2011.

\bibitem{pavlov2012generalized}
M.~V. Pavlov, V.~B. Taranov, and G.~A. El, ``Generalized hydrodynamic
  reductions of the kinetic equation for a soliton gas,'' {\em Theoretical and
  Mathematical Physics}, vol.~171, pp.~675--682, 2012.

\bibitem{ferapontov2022kinetic}
E.~Ferapontov and M.~Pavlov, ``Kinetic equation for soliton gas: integrable
  reductions,'' {\em Journal of Nonlinear Science}, vol.~32, no.~2, p.~26,
  2022.

\bibitem{pavlov2003integrable}
M.~V. Pavlov, ``Integrable hydrodynamic chains,'' {\em Journal of Mathematical
  Physics}, vol.~44, no.~9, pp.~4134--4156, 2003.

\bibitem{spohn2020collision}
H.~Spohn, ``Collision rate ansatz for the classical {T}oda lattice,'' {\em
  Physical Review E}, vol.~101, no.~6, p.~060103, 2020.

\bibitem{yoshimura2020collision}
T.~Yoshimura and H.~Spohn, ``Collision rate ansatz for quantum integrable
  systems,'' {\em SciPost Physics}, vol.~9, no.~3, p.~040, 2020.

\bibitem{mokhov1990non}
O.~I. Mokhov and E.~V. Ferapontov, ``Non-local {H}amiltonian operators of
  hydrodynamic type related to metrics of constant curvature,'' {\em Uspekhi
  Matematicheskikh Nauk}, vol.~45, no.~3, pp.~191--192, 1990.

\bibitem{ferapontov1991differential}
E.~V. Ferapontov, ``Differential geometry of nonlocal {H}amiltonian operators
  of hydrodynamic type,'' {\em Functional Analysis and Its Applications},
  vol.~25, no.~3, pp.~195--204, 1991.

\bibitem{takahashibook}
M.~Takahashi, {\em Thermodynamics of One-Dimensional Solvable Models}.
\newblock Cambridge University Press, 1999.

\bibitem{doyon2017large}
B.~Doyon, J.~Dubail, R.~Konik, and T.~Yoshimura, ``Large-scale description of
  interacting one-dimensional {B}ose gases: generalized hydrodynamics
  supersedes conventional hydrodynamics,'' {\em Physical review letters},
  vol.~119, no.~19, p.~195301, 2017.

\bibitem{thiffeault1998invariants}
J.-L. Thiffeault and P.~Morrison, ``Invariants and labels in {Lie-Poisson}
  systems,'' {\em Annals of the New York Academy of Sciences}, vol.~867, no.~1,
  pp.~109--119, 1998.

\bibitem{morrison1980maxwell}
P.~J. Morrison, ``The {Maxwell-Vlasov} equations as a continuous hamiltonian
  system,'' {\em Physics Letters A}, vol.~80, no.~5-6, pp.~383--386, 1980.

\bibitem{marsden1982hamiltonian}
J.~E. Marsden and A.~Weinstein, ``The {H}amiltonian structure of the
  {Maxwell-Vlasov} equations,'' {\em Physica D: nonlinear phenomena}, vol.~4,
  no.~3, pp.~394--406, 1982.

\bibitem{weinstein1981comments}
A.~Weinstein and P.~J. Morrison, ``Comments on: {T}he {Maxwell-Vlasov}
  equations as a continuous {H}amiltonian system,'' {\em Physics Letters A},
  vol.~86, no.~4, pp.~235--236, 1981.

\bibitem{zakharov1981benney}
V.~Zakharov, ``On the {B}enney equations,'' {\em Physica D: Nonlinear
  Phenomena}, vol.~3, no.~1-2, pp.~193--202, 1981.

\bibitem{kupershmidt1987hydrodynamical}
B.~Kupershmidt, ``Hydrodynamical {P}oisson brackets and local {L}ie algebras,''
  {\em Physics Letters A}, vol.~121, no.~4, pp.~167--174, 1987.

\bibitem{chesnokov2017stability}
A.~Chesnokov, G.~El, S.~L. Gavrilyuk, and M.~V. Pavlov, ``Stability of shear
  shallow water flows with free surface,'' {\em SIAM Journal on Applied
  Mathematics}, vol.~77, no.~3, pp.~1068--1087, 2017.

\bibitem{piterbarg1995poisson}
L.~Piterbarg, ``The {P}oisson bracket for 2{D} hydrodynamics reduces to the
  {G}ardner bracket,'' {\em Physics Letters A}, vol.~205, no.~2-3,
  pp.~149--157, 1995.

\bibitem{jayawardana2022clebsch}
B.~Jayawardana, P.~J. Morrison, and T.~Ohsawa, ``Clebsch canonization of
  {Lie--Poisson} systems,'' {\em Journal of Geometric Mechanics}, vol.~14,
  no.~4, pp.~635--658, 2022.

\bibitem{luesink2024casimir}
E.~Luesink, S.~Ephrati, P.~Cifani, and B.~Geurts, ``Casimir preserving
  stochastic {Lie--Poisson} integrators,'' {\em Advances in Continuous and
  Discrete Models}, vol.~2024, no.~1, p.~1, 2024.

\bibitem{mokhov1988dubrovin}
O.~I. Mokhov, ``{Dubrovin-Novikov} type {P}oisson brackets ({DN}-brackets),''
  {\em Functional Analysis and Its Applications}, vol.~22, no.~4, pp.~336--338,
  1988.

\bibitem{mokhov2008classification}
O.~I. Mokhov, ``The classification of nonsingular multidimensional
  {Dubrovin-Novikov} brackets,'' {\em Functional Analysis and Its
  Applications}, vol.~42, no.~1, pp.~33--44, 2008.

\bibitem{mokhov1998symplectic}
O.~I. Mokhov, ``Symplectic and {P}oisson structures on loop spaces of smooth
  manifolds, and integrable systems,'' {\em Russian Mathematical Surveys},
  vol.~53, no.~3, p.~515, 1998.

\bibitem{vergallo2021homogeneous}
P.~Vergallo and R.~Vitolo, ``Homogeneous {H}amiltonian operators and the theory
  of coverings,'' {\em Differential Geometry and its Applications}, vol.~75,
  p.~101713, 2021.

\bibitem{ferapontov2014projective}
E.~V. Ferapontov, M.~V. Pavlov, and R.~F. Vitolo, ``Projective-geometric
  aspects of homogeneous third-order hamiltonian operators,'' {\em Journal of
  Geometry and Physics}, vol.~85, pp.~16--28, 2014.

\bibitem{ferapontov2016towards}
E.~Ferapontov, M.~V. Pavlov, and R.~Vitolo, ``Towards the classification of
  homogeneous third-order hamiltonian operators,'' {\em International
  Mathematics Research Notices}, vol.~2016, no.~22, pp.~6829--6855, 2016.

\bibitem{vergallo2023projective}
P.~Vergallo and R.~Vitolo, ``Projective geometry of homogeneous second-order
  hamiltonian operators,'' {\em Nonlinearity}, vol.~36, no.~10, p.~5311, 2023.

\bibitem{dharetal1993}
A.~Dhar, G.~Mandal, and S.~R. Wadia, ``W{$\infty$} coherent states and
  path-integral derivation of bosonization of non-relativistic fermions in one
  dimension,'' {\em Modern Physics Letters A}, vol.~08, no.~37, pp.~3557--3568,
  1993.

\bibitem{lobb2009lagrangian}
S.~Lobb and F.~Nijhoff, ``Lagrangian multiforms and multidimensional
  consistency,'' {\em Journal of Physics A: Mathematical and Theoretical},
  vol.~42, no.~45, p.~454013, 2009.

\bibitem{suris2016lagrangian}
Y.~B. Suris and M.~Vermeeren, {\em On the Lagrangian structure of integrable
  hierarchies}.
\newblock Springer Berlin Heidelberg, 2016.

\bibitem{sleigh2019variational}
D.~Sleigh, F.~Nijhoff, and V.~Caudrelier, ``A variational approach to {L}ax
  representations,'' {\em Journal of Geometry and Physics}, vol.~142,
  pp.~66--79, 2019.

\bibitem{sleigh2020variational}
D.~Sleigh, F.~Nijhoff, and V.~Caudrelier, ``Variational symmetries and
  lagrangian multiforms,'' {\em Letters in Mathematical Physics}, vol.~110,
  pp.~805--826, 2020.

\bibitem{caudrelier2021multiform}
V.~Caudrelier and M.~Stoppato, ``Multiform description of the {AKNS} hierarchy
  and classical r-matrix,'' {\em Journal of Physics A: Mathematical and
  Theoretical}, vol.~54, no.~23, p.~235204, 2021.

\bibitem{caudrelier2024classical}
V.~Caudrelier, M.~Stoppato, and B.~Vicedo, ``Classical {Yang--Baxter} equation,
  {L}agrangian multiforms and ultralocal integrable hierarchies,'' {\em
  Communications in Mathematical Physics}, vol.~405, no.~1, p.~12, 2024.

\bibitem{sleigh2023lagrangian}
D.~Sleigh, F.~W. Nijhoff, and V.~Caudrelier, ``Lagrangian multiforms for
  {Kadomtsev--Petviashvili} ({KP}) and the {Gelfand--Dickey} hierarchy,'' {\em
  International Mathematics Research Notices}, vol.~2023, no.~2,
  pp.~1420--1460, 2023.

\bibitem{biagettin2024therma}
L.~Biagetti, G.~Cecile, and {J. De Nardis}, ``Three-stage thermalisation of a
  quasi-integrable system,'' 2024.

\bibitem{gaudin2014bethe}
M.~Gaudin, {\em The Bethe wavefunction}.
\newblock Cambridge University Press, 2014.

\bibitem{ruggiero2020quantum}
P.~Ruggiero, P.~Calabrese, B.~Doyon, and J.~Dubail, ``Quantum generalized
  hydrodynamics,'' {\em Physical review letters}, vol.~124, no.~14, p.~140603,
  2020.

\bibitem{abraham2012manifolds}
R.~Abraham, J.~E. Marsden, and T.~Ratiu, {\em Manifolds, tensor analysis, and
  applications}, vol.~75.
\newblock Springer Science \& Business Media, 2012.

\end{thebibliography}

\end{document}